\documentclass[twocolumn,preprintnumbers,amsmath,amssymb,aps,prb,longbibliography]{revtex4-1}
\usepackage{graphicx}
\begin{document}

\title{Braiding Majorana Fermions and Creating Quantum Logic Gates with Vortices On
a Periodic Pinning Structure }
\author{
X. Ma$^{1,2}$,
C. J. O. Reichhardt$^{1}$, and  C. Reichhardt$^{1}$
}
\affiliation{
$^1$ Theoretical Division,
Los Alamos National Laboratory, Los Alamos, New Mexico 87545 USA\\
$^2$ Department of Physics, University of Notre Dame, Notre Dame, Indiana 46556 USA\\
}

\date{\today}
\begin{abstract}
We show how vortices that support Majorana fermions
when placed on a periodic pinning array can be used for vortex exchange and
independent braiding
by performing a series of specific moves with a probe tip.
Using these braiding operations, we demonstrate
realizations of a Hadamard and a CNOT gate.
We specifically consider the first matching field
at which there is one vortex per pinning site,
and we show that there
are two basic dynamic operations,
move and exchange,
from which basic braiding operations can
be constructed in order to create
specific logic gates.
The periodic pinning array permits both
control of the world lines of the
vortices
and freedom for vortex manipulation using a
set of specific moves of the probe
during which
the probe tip strength
and height remain unchanged.
We
measure the robustness of the
different moves against thermal effects and show
that the three different operations produce
distinct force signatures on the moving tip.
\end{abstract}
\maketitle

\section{Introduction}

Manipulation of individual vortices in type-II superconductors can now
be achieved with
a variety of
methods, including
local magnetic fields \cite{Gardner02},
magnetic force microscopes (MFMs) \cite{Straver08,Auslaender09,Luan09,Shapira15},
mechanical forces
\cite{Kremen16},
scanning tunneling tips \cite{Ge16},
and optically \cite{Veshchunov16}.
It is possible for the
vortices to be moved over certain distances
\cite{Gardner02,Straver08,Kremen16}, entangled \cite{Reichhardt04},
and arranged in special
positions \cite{Gardner02,Auslaender09,Kremen16,Ge16,Veshchunov16}.
The forces induced on the tip by the motion of the
vortex can also serve as a probe of the pinning
properties \cite{Straver08,Luan09,Kafri06,Kafri07,Reichhardt09a},
the dynamics of individual vortices coupled to pinning
\cite{Auslaender09,Reichhardt09a,Reichhardt08,Reichhardt10c,Ma18}, or
the creation of vortices \cite{Ge17,Dremov19}. As advances in nanoscale
fabrication continue,
it will likely become possible
to develop even more precise control of the vortex motion
and also to manipulate multiple vortices at the same time.
One promising application of
vortex manipulation is to perform the braiding of
Majorana fermions for quantum computing
in materials for which
Majorana fermions are localized in the vortex core.

Majorana fermions were first introduced by Ettore Majorana, and
they have the interesting property of being their own
antiparticles \cite{Majorana37}.
Currently it is unclear whether certain elementary particles in high
energy physics are Majorana fermions;
however, Majorana fermions in the form of
quasiparticles in condensed
matter systems
has been a rapidly growing field,
and there is now evidence
that such states indeed
occur in numerous
systems \cite{Rokhinson12,Mourik12,Beenakker13,Deng14,NadjPerge14,Banerjee16,He17,Wang18,Das12}.
Another reason that
such states are of interest in condensed matter
is that, due to their intrinsic non-Abelian statistics,
Majorana fermions can be
used to support topologically protected states for quantum
computation \cite{Beenakker13,DasSarma15,Nayak08}.

Majorana fermions in condensed matter are non-Abelian anyons
\cite{Stern10} with non-trivial exchange operations which do not commute.
Instead of generating a phase $2\pi$ for bosons, $\pi$ for fermions, and arbitrary phase for Abelian anyons, the exchange
of Majorana fermions leads to a unitary transformation within the degenerate
ground state manifold which does not depend on the method or
details of its execution \cite{Leijnse12,Lutchyn18}.
The inherent non-Abelian statistics can be used to support topologically-protected qubits
for quantum computation \cite{Freedman06}.
Non-Abelian anyons were first predicted
by Moore and Read to occur in the fractional quantum Hall state \cite{Moore91},
and later, Read and Green established a close connection
between a two-dimensional (2D) spinless $p+ip$ superconductor and the
Moore-Read quantum Hall state \cite{Read00},
where non-Abelian statistics must be shared in the $p$-wave superconductors.
Kitaev showed that non-Abelian statistics can also emerge in
spinless one-dimensional (1D) superconductors \cite{Kitaev01}.
These superconductors can contain topological phases which
support exotic excitations at their boundaries and
inside their topological
defects \cite{Read00,Kitaev01,Volovik03,Alicea12}.
In particular,
Majorana fermions can be localized at the ends of
1D superconductors \cite{Kitaev01}
and can be bonded to superconducting vortex cores in 2D materials \cite{Stone06}.
When vortices that are bonded to Majorana fermions are exchanged adiabatically,
the Majorana fermions will exhibit non-Abelian statistics \cite{Ivanov01}.

In 2008,
Fu and
Kane
proposed a physical realization of $p+ip$ superconductivity
at the interface of an $s$-wave superconductor and
a topological insulator \cite{Fu08}.
Recent experiments using spin selective Andreev reflection verified
the existence of Majorana fermions at a superconducting vortex core
\cite{Sun16,Sun17}.
By manipulating the vortices,
it is possible
to manipulate the Majorana fermions (MF) trapped inside,
thereby achieving
transformations of the quantum states encoded by the MFs.
The manipulation of MFs trapped in non-interacting
(distantly separated) vortices in $p$-wave superconductors was
studied recently via self-consistent Bogolioubov-de Gennes calculations
\cite{Wu17}, which showed that MF states are robust against the movement
of the vortices.
More recently, there have
been proposals for the manipulation of vortex states in superconducting
structures which would allow the
braiding of individual vortices  \cite{November19}
or vortex ensembles \cite{Polshyn19,Posske19} as well as
other operations
\cite{Liang12,Beenakker19,Stern19,Beenakker19a,Roising19}
that could be applied to
quantum computing.
Since there are many ways to create different types of pinning lattice structures
for vortices in superconductors
\cite{Baert95,Harada96,Martin97,Berdiyorov06,Kemmler06,deSouzaSilva07,Goldberg09,Kemmler09,Gutierrez09,Libal09,Hoffmann12,Swiecicki12,Latimer13,Wang13,Ray14b,Trastoy14,Sadovskyy17,Zechner17,Xue18,Ge18,Wang18a}
as well as numerous methods
for achieving individual vortex manipulation
\cite{Gardner02,Straver08,Kremen16,Ge16,Veshchunov16,Ge17,Dremov19},
a natural direction
to study is what type of vortex pinning array
would allow the performance of vortex exchanges that
could realize the different logic gates
required for topological quantum computing.

In this work we examine vortex manipulation
in a
topological superconductor, consisting of the interface between
an $s$-wave superconductor and a topological insulator,
that contains a square lattice
of pinning sites in the form of blind holes.
The vortex manipulation is achieved
using a moving MFM probe.
We propose basic operations that can
independently realize vortex exchange and braiding
without incorporating the world lines of other vortices.
We analyze the robustness of these operations against noise
and propose using the periodic potential force signals exerted on the
moving probe
to detect
the microscopic behavior of the vortices
during the different basic motions.

In addition to performing vortex exchanges,
we also propose a method to braid the world lines of vortices
in which the vortices end up at the
same positions as in their initial state,
which  provides more freedom for vortex manipulation.
Based on the wave function of quasiparticles in Moore-Read states,
Georgiev \cite{Georgiev06,Georgiev08} proposed braid matrices that relate braiding operations to
transformations of the quantum state,
making it possible to
construct braiding operations that realize quantum gates, including a Hadamard gate
and a controlled-NOT (CNOT) gate.  Since the topological
equivalence between Moore-Read states and 2D
$p$-wave superconducting states has already been established
\cite{Read00,Ivanov01},
we follow the braiding schemes in Refs.~\cite{Georgiev06,Georgiev08} to
demonstrate our method for realizing quantum gates using vortices.
We also discuss how our technique could be used in a similar scheme for
skyrmion systems,
based on
proposals for the stabilization of bound Majorana states in
skyrmions \cite{Yang16,Gungordu18,Rex19}, the pinning of skyrmions on periodic substrates \cite{Reichhardt18},
and the manipulation of individual skyrmions \cite{Hanneken16}.
There are also systems in which skyrmions and superconducting
vortices are coupled
\cite{Menezes19}.

\section{System}

\begin{figure}
\includegraphics[width=3.5in]{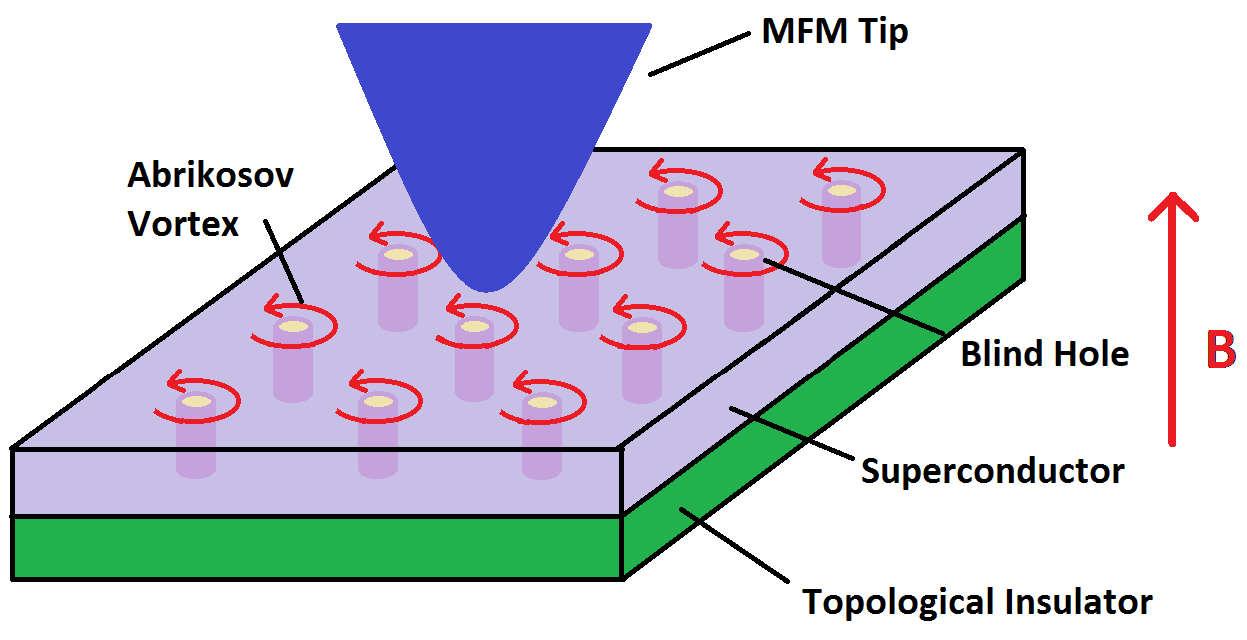}
\caption{
Schematic of the system, consisting of a superconductor (pink) coupled to a
topological insulator (green).
A magnetic field ${\bf B}$ is applied  perpendicular to the layers.
The superconducting layer contains
a square array of blind holes (yellow) that each capture one
superconducting vortex (pink columns), and a Majorana fermion is
localized inside each vortex.
An MFM tip is used to manipulate individual vortices.
}
\label{fig:1}
\end{figure}

In Fig.~\ref{fig:1} we show a schematic of our system which
consists of
a superconductor coupled to a topological insulator.
The superconductor contains a square array of blind holes that act as
pinning sites capable of capturing at most one vortex.
A magnetic field ${\bf B}$
is applied in the $z$-direction, perpendicular to the superconducting plane,
with a value that corresponds to the first matching field
$B_{\phi}$ at which the number of vortices $N_v$ is equal to the number
of pinning sites $N_p$.
It is known from previous work that at the matching field,
the vortices fill all of the pinning sites to
form a commensurate structure \cite{Harada96}.
A probe such as an MFM tip is used to manipulate individual vortices.

We consider a sample of size $L \times L$ with
$L=40\lambda$,
where all lengths are measured in terms of the London penetration depth $\lambda$.
The pinning array contains $N_p=400$
pinning sites arranged in a square array with a lattice
constant of $a=2\lambda$.
As discussed in Refs.~\cite{Posske19,Cheng09,Cheng10},
the hybridization strength of two vortex Majorana fermions
is very small, so in our setup it can be neglected.
We use the same molecular dynamics simulation technique employed
in previous work on vortices in periodic pinning arrays \cite{Ma18},
and
model the pinning sites
as finite range parabolic traps with a pinning radius of
$r_{p} = 0.3\lambda$ and a maximum pinning force of $F_{tr}=0.3$.
The probe tip is also represented as
a finite range parabolic trap
with a
maximum trapping force of $F_{tr} = 0.65$
and a radius of $R_{tr}=0.65\lambda$
that is translating at a velocity $V_{tr}=0.1$.
The probe tip is moved slowly enough that the system
remains in the adiabatic limit.
At higher drives, the vortex can slip out of the probe tip,
and the pinning parameters for the nonadiabatic
regime have been characterized in previous work \cite{Ma18}.

The dynamics of vortex $i$ arise from the following
overdamped equation of motion:
\begin{equation}
 \eta\frac{d {\bf r}_{i}}{dt} = {\bf F}^{vv}_{i} + {\bf F}^{vp}_{i} + {\bf F}^{tr}_{i} .
\end{equation}
Here
$\eta=1$ is the damping constant
and ${\bf r}_{i}$ is the position  of vortex $i$.
The vortex-vortex interaction force is
${\bf F}^{vv}_{i} = \sum^{N_{v}}_{j=1}K_{1}(r_{ij}/\lambda){\hat {\bf r}_{ij}}$,
where $K_{1}$ is the
modified Bessel function of the second kind, $r_{ij} = |{\bf r}_{i} - {\bf r}_{j}|$,
${\hat {\bf r}}_{ij} = ({\bf r}_{i} - {\bf r}_{j})/r_{ij}$, and $r_{j}$ is the position of vortex $j$.
We measure all forces in units of $f_{0} = \phi^{2}_{0}/(2\pi\mu_{0}\lambda^3)$ where $\phi_{0} = h/2e$ is the flux quantum.
The pinning force
from the harmonic traps is given by
 ${\bf F}^{vp}_{i} = -\sum_{k}^{N_p}(F_{p}/r_{p})({\bf r}_{i} - {\bf r}_{k}^{(p)})\Theta(r_{p} - |{\bf r}_{i} - {\bf r}_k^{(p)}|)$,
where $F_{p}$ is the maximum pinning strength
and ${\bf r}^{(p)}_k$ is the
location of pinning site $k$.
The force ${\bf F}^{tr}$
from the probe tip
has the same form as the pinning force but the
maximum probe trapping force is $F_{tr}$
and the probe trap radius $R_{tr} = 0.65\lambda$
is larger than the pinning radius.
The probe tip speed is $V_{tr}=0.1$.
We use a simulation time step of
$\Delta t = 0.02$ so that the typical time required for the probe
tip to move a distance
$2\lambda$, or one
pinning lattice constant,
is $1000$ simulation time steps.
In general, after
each move of the probe tip,
we wait $100$ simulation time steps before beginning the next move.
This increases the ability of the probe tip to hold a trapped vortex.

We note that in previous work on this system \cite{Ma18}
we considered
different types of pinning potentials
such as a Gaussian
trap with
potential $U(r) = U_{p}\exp(-\kappa R_p^2)$.
We found that the general behavior was almost the same
for either harmonic or Gaussian traps
with only minor changes.
In general, we expect that our results will
be robust as long as the pinning sites have a well defined, uniform length scale and
pinning force.

\section{Basic Operations}

\begin{figure}
\includegraphics[width=3.5in]{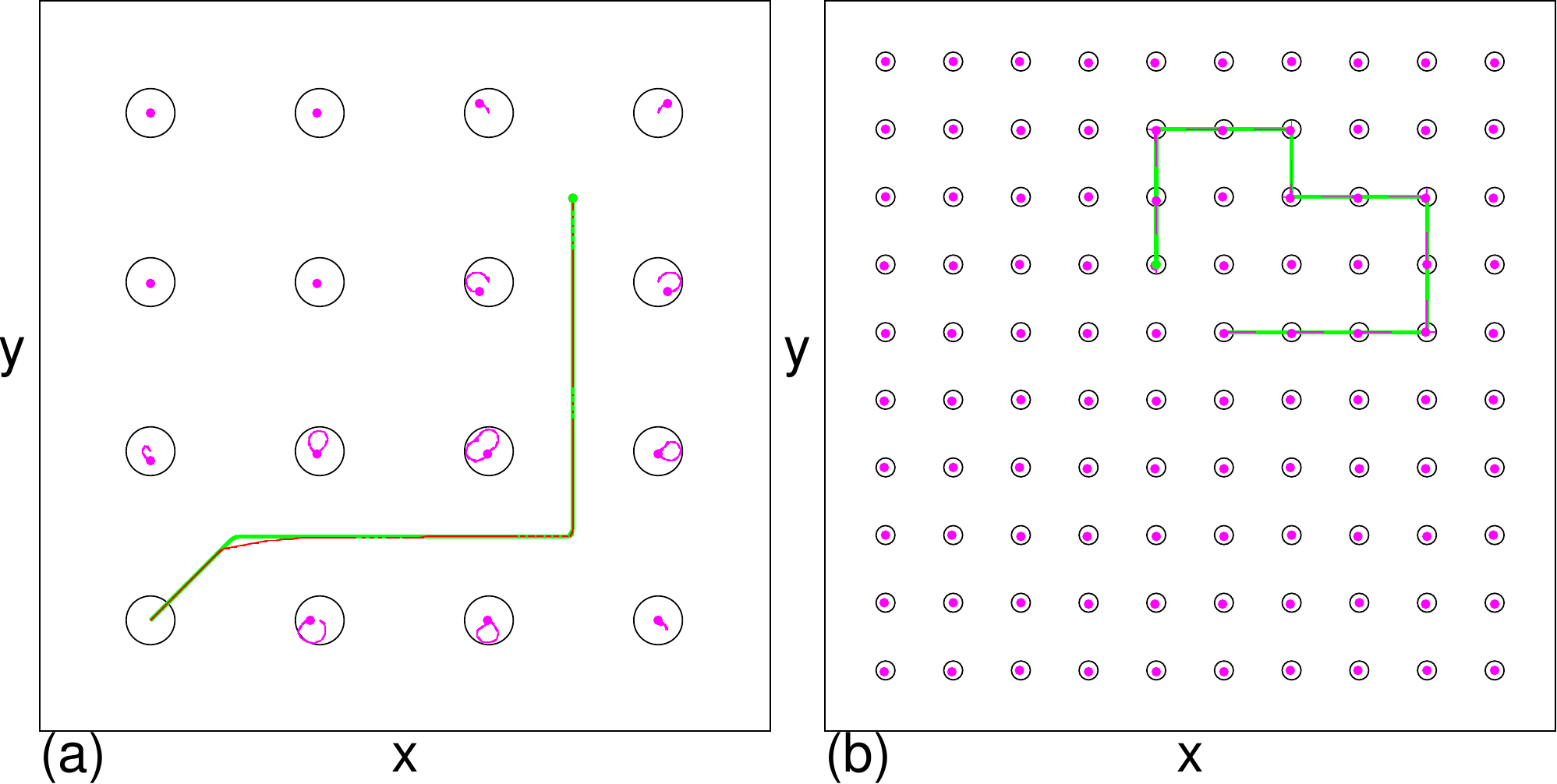}
\caption{
  Pinning site locations (open circles),
  vortex locations (filled pink circles), vortex trajectories (magenta lines),
  trapped vortex location (filled green circle), trapped vortex trajectory (red line),
  and probe tip trajectory (green line) for
 the fundamental operations.
 (a) The {\it capture} operation in which the probe tip moves along
 the $\langle 1 1\rangle$ direction
 over the pinning site and the vortex inside the pin becomes
 trapped by the probe tip, followed by the {\it move} operation in which
 the vortex is carried by the probe tip to a new location.
In the illustrated trajectory, the probe tip moves the captured vortex
a distance $2a$ in the $x$ direction and $2a$ in the
$y$ direction, where $a$ is the pinning lattice constant.
Only the vortex trapped by the probe tip is depinned.
(b) The {\it reposition} operation where the empty probe tip moves along
the $\langle 1 0\rangle$ direction over the pinning sites and does
not capture a vortex.
Individual vortices are temporarily pushed a short distance
  out of their pinning sites by the probe tip, but
  fall back into their original pinning sites as the probe tip moves away from the pin.
  In the illustrated trajectory, the probe is moving counterclockwise.
}
\label{fig:2}
\end{figure}

\subsection{Fundamental Operations}
There are three fundamental operations that can be combined in order
to create different logic gates.
They are distinguished by whether the probe tip moves across a pinning
site along the $\langle 1 0\rangle$ direction (parallel with the $x$ or $y$
axis),
if the tip moves
along the $\langle 1 1\rangle$ direction (at a $45^\circ$ angle from the
$x$ or $y$ axis), or if the tip does not move over a pinning site at all.
Due to the symmetry of the commensurate vortex lattice,
when the probe tip moves over a pin along $\langle 1 0\rangle$, it moves the
vortex in the pinning site toward its nearest neighbor
a distance $a$ away.  This neighbor exerts a sufficiently strong repulsive force
that the vortex falls out of the
probe tip and returns to its original pinned location.
If instead the probe tip moves over the pin along $\langle 1 1\rangle$,
the vortex in the pin moves toward
a neighbor that is a distance $\sqrt{2}a$
away.  The repulsion from this neighbor
is weak enough that the probe tip is able to capture the
vortex successfully and pull it out of its pinning site.
The key to our proposed logic operation technique is the difference between
capturing and not capturing the pinned vortices
depending on whether the probe tip moves
along $\langle 1 0\rangle$ or along $\langle 1 1\rangle$ as it crosses the pinning site.

The fundamental operations can be described as follows:
(1) {\it Capture}.
By moving the probe along $\langle 1 1\rangle$ as it passes over a pinning site,
the probe tip captures the vortex that was in the pinning site and drags it
out of the pin.
(2) {\it Move}.  When the probe tip contains a vortex and does not pass over
a pinning site, it can move the trapped vortex
over a fixed distance through the interstitial region
between pinning sites.
The efficiency of this operation is quantified by the distance between the position of the vortex which is supposed to move with the tip and the position of the tip at the end of these operations.
(3) {\it Reposition}. By moving the probe along $\langle 1 0\rangle$
over a pinning
site, the empty probe tip can be translated
to a new position without knocking any vortices out of their pinning sites.
This operation is quantified by the distance between the positions of vortices after all these operations and the positions of their original pinning sites.

In Fig.~\ref{fig:2}(a), we illustrate the capture operation followed
by the move operation.
Here a single vortex in the bottom left pinning site is captured
by the tip which moves across the pin along the $\langle 1 1\rangle$ direction.
The trapped vortex is then dragged a distance $2a$ in the $x$ direction followed by
a distance $2a$ in the $y$ direction through the interstitial area between
the pinning sites.
The dragged vortex closely follows the tip trajectory
while the other vortices remain in their pinning sites,
exhibiting
some rotational movement induced by the vortex-vortex interactions
as the dragged vortex moves past.
Figure~\ref{fig:2}(b) shows the reposition operation, where the empty
trap moves over pinning sites along the $\langle 1 0\rangle$ direction
without depinning any vortices.
Individual vortices are dragged by the tip over a short distance, but fall out
of the tip due to the repulsion from the neighboring vortex and return to their
original pinning sites, as indicated by the short linear vortex trajectories.

\begin{figure}
  \begin{minipage}[c]{3.5in}
  \begin{minipage}[c]{0.5\textwidth}
    \includegraphics[width=\textwidth]{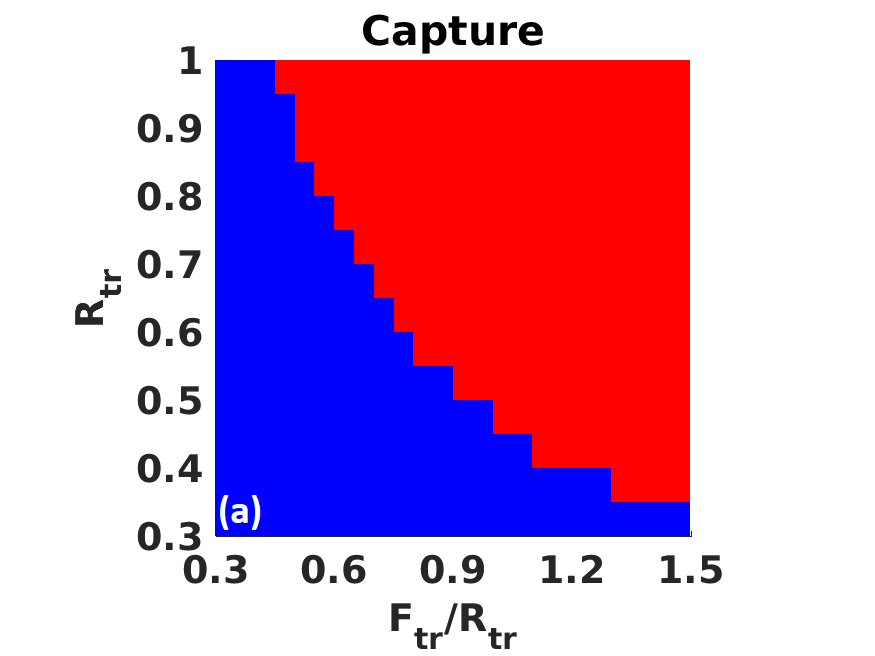}
  \end{minipage}%
  \begin{minipage}[c]{0.5\textwidth}
    \includegraphics[width=\textwidth]{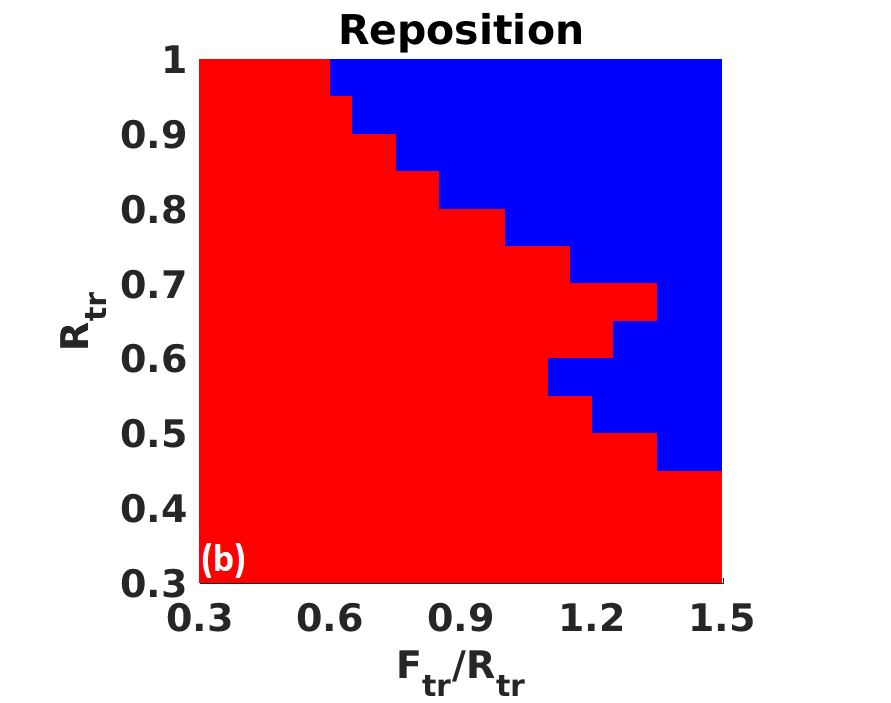}
  \end{minipage}%
  \end{minipage}
  \begin{minipage}[c]{3.5in}
    \begin{center}
      \includegraphics[width=0.5\textwidth]{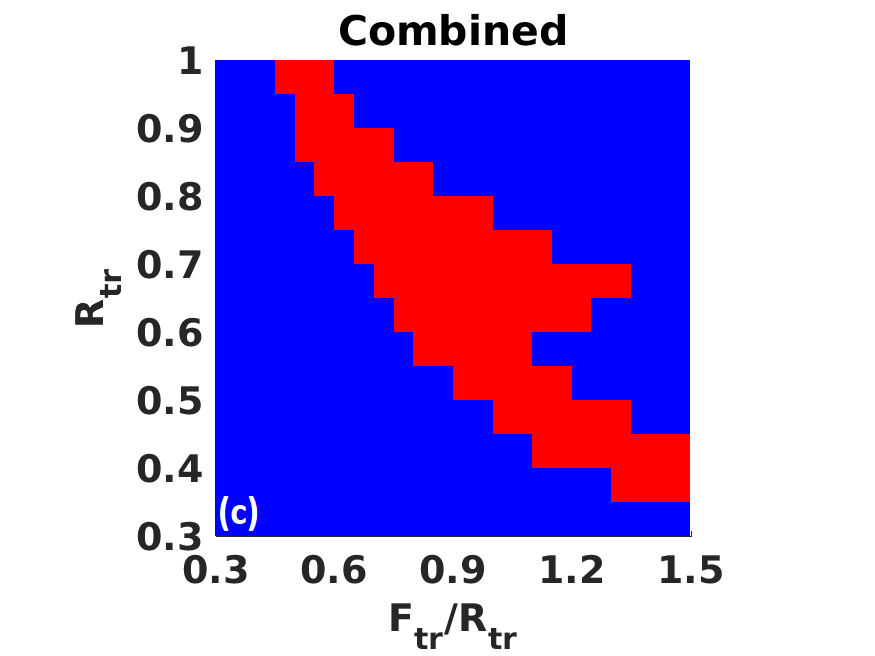}
      \end{center}
  \end{minipage}
  \caption{
    Plots of the parameter regimes indicating where the fundamental operations are
    performed successfully (red) or unsuccessfully (blue) as a function of probe tip
    radius $R_{tr}$ vs probe tip spring constant $F_{tr}/R_{tr}$ in a sample with lattice
    constant $a=2\lambda$ and pinning strength $F_p=0.3$.
    (a) The capture operation in which the tip crosses the pin along $\langle 1 1\rangle$.
    (b) The reposition operation in which the tip crosses the pins along $\langle 1 0\rangle$.
    (c) The combination of capture and reposition, with red indicating the region of
    parameter space in which both operations can be achieved simultaneously.
}
\label{fig:3}
\end{figure}

Experimentally, for a fixed pinning force and lattice constant $a$,
the attractive strength of the probe tip can be tuned such that the
capture and reposition
operations can both be performed.
In Fig.~\ref{fig:3}(a) we indicate the portion of the probe tip radius $R_{tr}$ and
probe tip spring constant $F_{tr}/R_{tr}$ parameter space in which the capture operation is
successful.
The measure of these operations are converted to 0 (failure) and 1 (success) according to the quantification mentioned above with a threshold of 0.5.
As the probe tip radius becomes smaller, the
minimum strength of the probe required to permit capture to occur increases.
For the reposition operation,
Fig.~\ref{fig:3}(b) shows
an opposite trend in which a decrease in the size of the probe tip radius
decreases the maximum strength of the probe required to permit repositioning to occur.
In Fig.~\ref{fig:3}(c), we show that there is a finite window of parameter space in
which both operations can simultaneously be achieved.
The widest range of tip strengths falls at a tip radius that is approximately
twice the size of the pinning radius.
For the remainder of this work,
we consider the optimal regime
with $R_{tr}=0.65$ and $F_{tr}/R_{tr}=1.0$.

\subsection{Exchange Operations}

\begin{figure}
\includegraphics[width=3.5in]{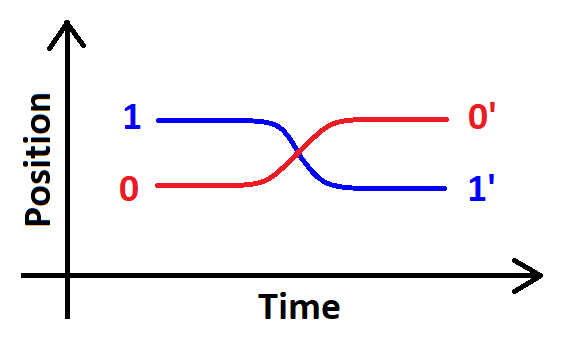}
\caption{ The world lines as a function of position and time
  for an exchange operation of vortex 0 (red) with vortex 1 (blue).
}
\label{fig:4}
\end{figure}

\begin{figure}
\includegraphics[width=3.5in]{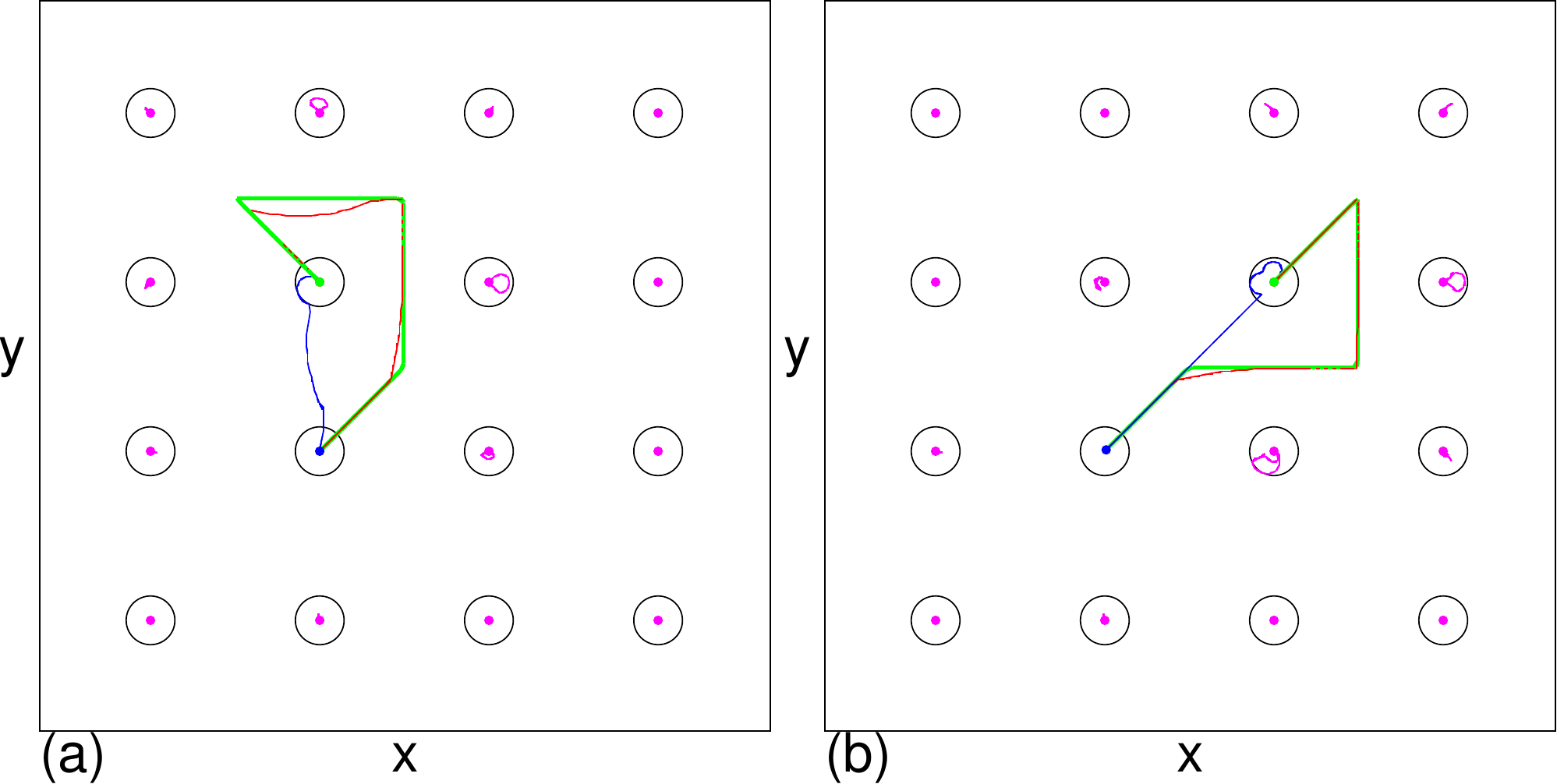}
\caption{
  Pinning site locations (open circles), vortex locations (filled circles), vortex
  trajectories (lines), and probe tip trajectory (green line) for two-vortex exchange
  operations between vortex 0 (red circle and trajectory) and vortex 1 (blue circle
  and trajectory).
  (a) A $\langle 1 0\rangle$ exchange.
  (b) A $\langle 1 1\rangle$ exchange.
}
\label{fig:5}
\end{figure}

Now that we have established the fundamental operations
and identified optimal parameters for these operations, we consider
the higher order operations needed for exchange and braiding.
In Fig.~\ref{fig:4} we show the world lines
for the simple two-vortex exchange operation of vortex 0 and vortex 1.
The simplest exchange is a $\langle 1 0\rangle$ exchange between
two vortices that are one lattice constant apart along either the $x$ or $y$
direction.  As illustrated in Fig.~\ref{fig:5}(a),
the probe tip first captures vortex 0 from the pin in the lower row, moves
it a distance $a$ in the $+y$ direction, and then moves it a distance $a$ in
the $-x$ direction. During the motion in $-x$ direction, vortex 1 was pushed down towards the empty pinning site below. Next the probe performs a
reverse capture operation in which vortex 0 is pulled into the pinning site originally
occupied by vortex 1 while vortex 1 is pushed
by the repulsion from the vortex trapped by the probe tip
into the empty pinning site that was vacated by vortex 0.
Throughout this exchange operation, the other vortices
remain pinned.
We can characterize the path of the tip by a series
of $(\theta,r/a)$ instructions where $\theta$ is
the
direction of motion and $r/a$ is the distance traveled in this direction in
units of the lattice constant $a$.
The procedure for the motion in Fig.~\ref{fig:5}(a) is
($[1 1]$, $\sqrt{2}/2$), ($[0 1]$, 1), ($[\overline{1} 0]$, 1),
($[1 \overline{1}]$, $\sqrt{2}/2$),
which, if written in terms of angles of motion from the $x$ axis, is the same as
($45^\circ$, $\sqrt{2}/2$), ($90^\circ$, 1), ($180^{\circ}$, 1),
($315^\circ$, $\sqrt{2}/2$).
For our parameters,
the move along $[11]$ or 45$^\circ$ takes 707 simulation time steps.

\begin{figure}
\includegraphics[width=3.5in]{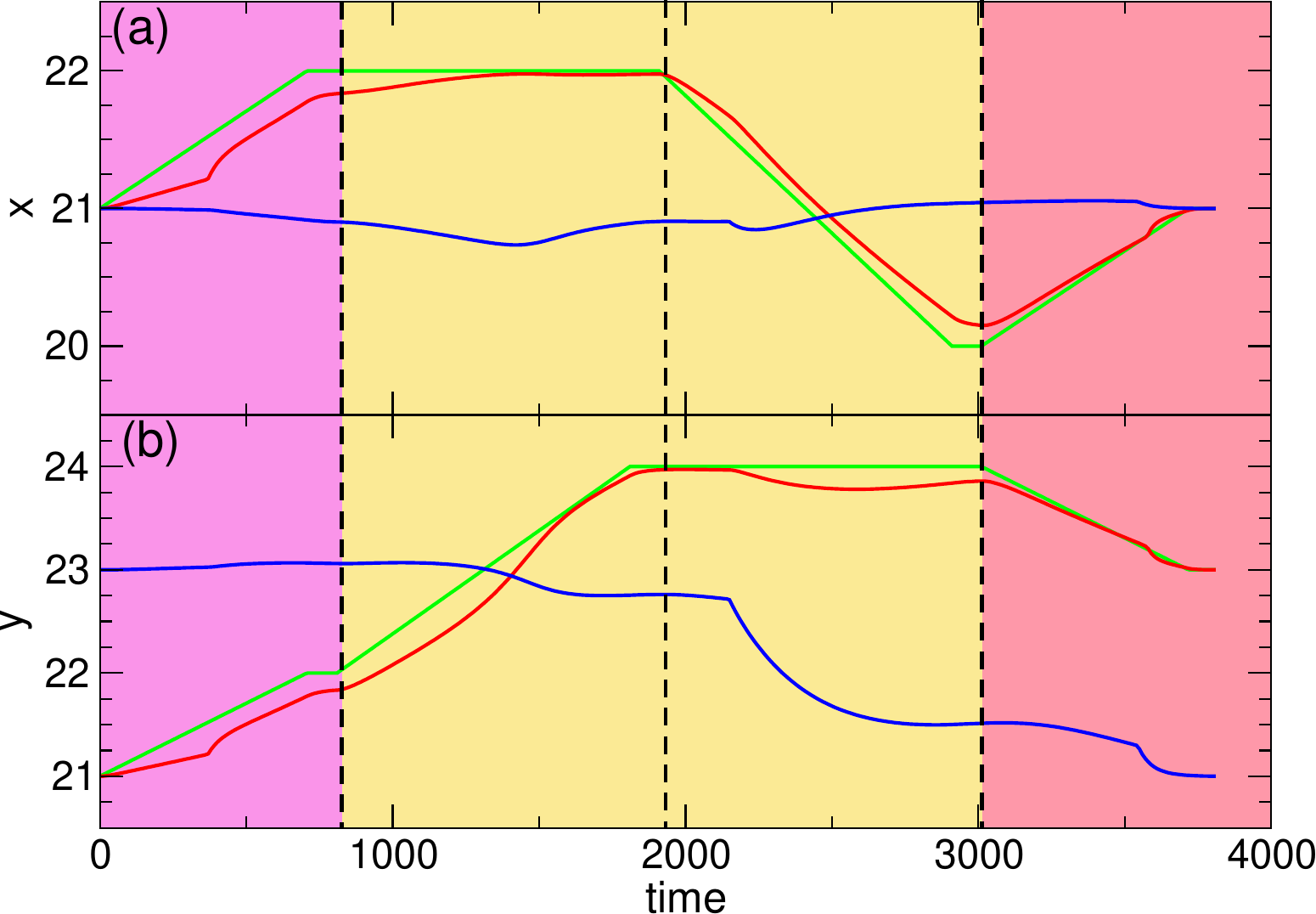}
\caption{The $x$ (a) and $y$ (b) positions vs time in simulation steps
  for vortex 0 (red), vortex 1 (blue), and the probe tip (green)
  for the $\langle 1 0\rangle$
  exchange illustrated in Fig.~\ref{fig:5}(a).
  The vertical dashed lines
  delineate the four
  stages of probe tip motion.
  Magenta: capture; yellow: $\langle 1 0\rangle$ moves; pink: reverse capture.
  }
\label{fig:6}
\end{figure}

The exchange operation can also be characterized
by the displacements in $x$ and $y$
of vortices 0 and 1 along with the probe tip,
as shown in Fig.~\ref{fig:6}(a) and (b).
Vertical dashed lines indicate the four stages of probe tip motion.
Here,
vortex 0
closely follows the probe tip.
Since the exchange is in the $y$-direction, the
$x$-positions of the vortices have the same value at the beginning and end of
the operation, while
the $y$-positions of vortex 0 and vortex 1 switch places by the end of the
operation.

\begin{figure}
\includegraphics[width=3.5in]{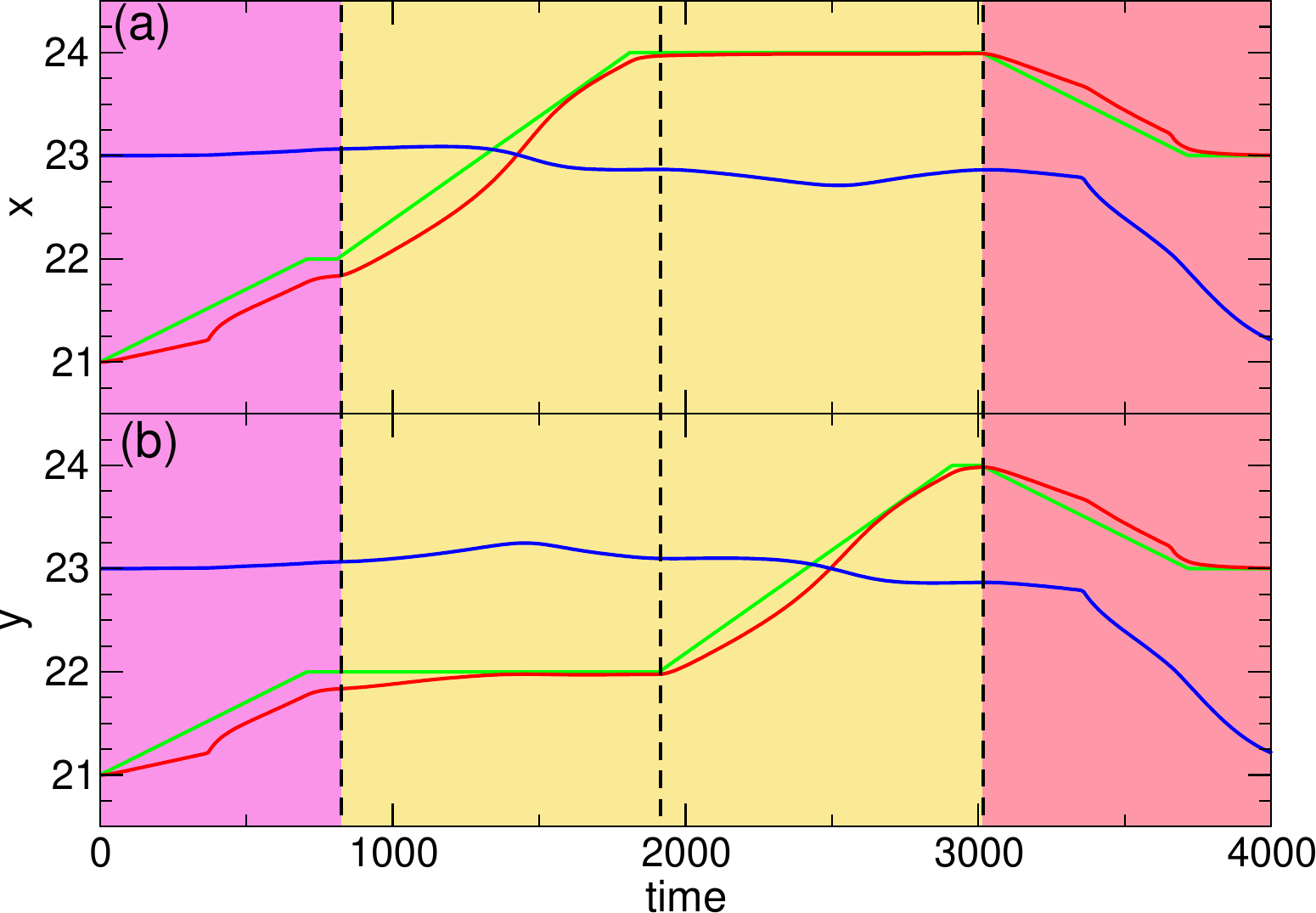}
\caption{The $x$ (a) and $y$ (b) positions vs time in simulation steps
  for vortex 0 (red), vortex 1 (blue), and the probe tip (green)
  for the $\langle 1 1\rangle$
  exchange illustrated in Fig.~\ref{fig:5}(b).
  The vertical dashed lines
  delineate the four stages of probe tip motion.  Magenta: capture; yellow: $\langle 1 0\rangle$ move; pink: reverse capture.
}
\label{fig:6a}
\end{figure}

We can also perform exchange operations along the $\langle 1 1\rangle$ direction,
as illustrated in
Fig.~\ref{fig:5}(b) where the probe tip again follows four stages of movement.
Vortex 0 is first captured by the tip and is next moved a distance $a$ in the $+x$
direction followed by a distance $a$ in the $+y$ direction.
The probe tip then performs a reverse capture in which
vortex 0 is dragged into the pinning site occupied by vortex 1, which is
ejected and travels into the pinning site vacated by vortex 0.
The procedure for this exchange is
([11], $\sqrt{2}/2$), ([10], 1), ([01], 1), ([$\overline{1}\overline{1}$], $\sqrt{2}/2$).
We note that in the last step we wait $400$ simulation time steps
to increase the stability of the operation
since the $\langle 1 1\rangle$
exchange is more susceptible to fluctuations,
as we describe in Sec.~V.
In Fig.~\ref{fig:6a}(a,b) we plot
the $x$ and $y$-displacements of vortex 0, vortex 1 and the tip
as a function of time for the $\langle 1 1\rangle$
exchange.
Vortex 0 follows the probe tip,
and the final positions of vortices 0 and 1 are swapped in both the $x$ and $y$
directions by the end of the operation.

\subsection{Braiding Operations}

\begin{figure}
\includegraphics[width=3.5in]{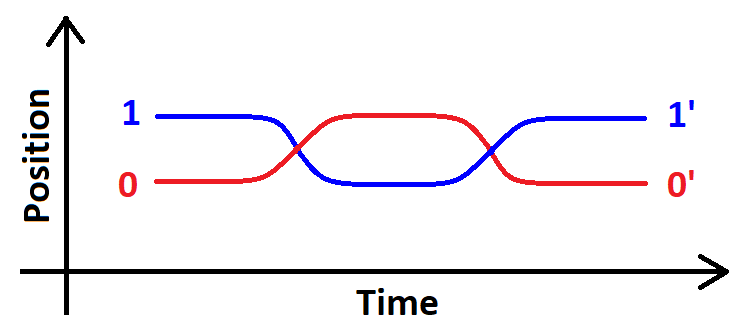}
\caption{ The world lines
  as a function of position and time for a braiding operation of vortex 0
  (red) and vortex 1 (blue).
}
\label{fig:7}
\end{figure}

\begin{figure}
\includegraphics[width=3.5in]{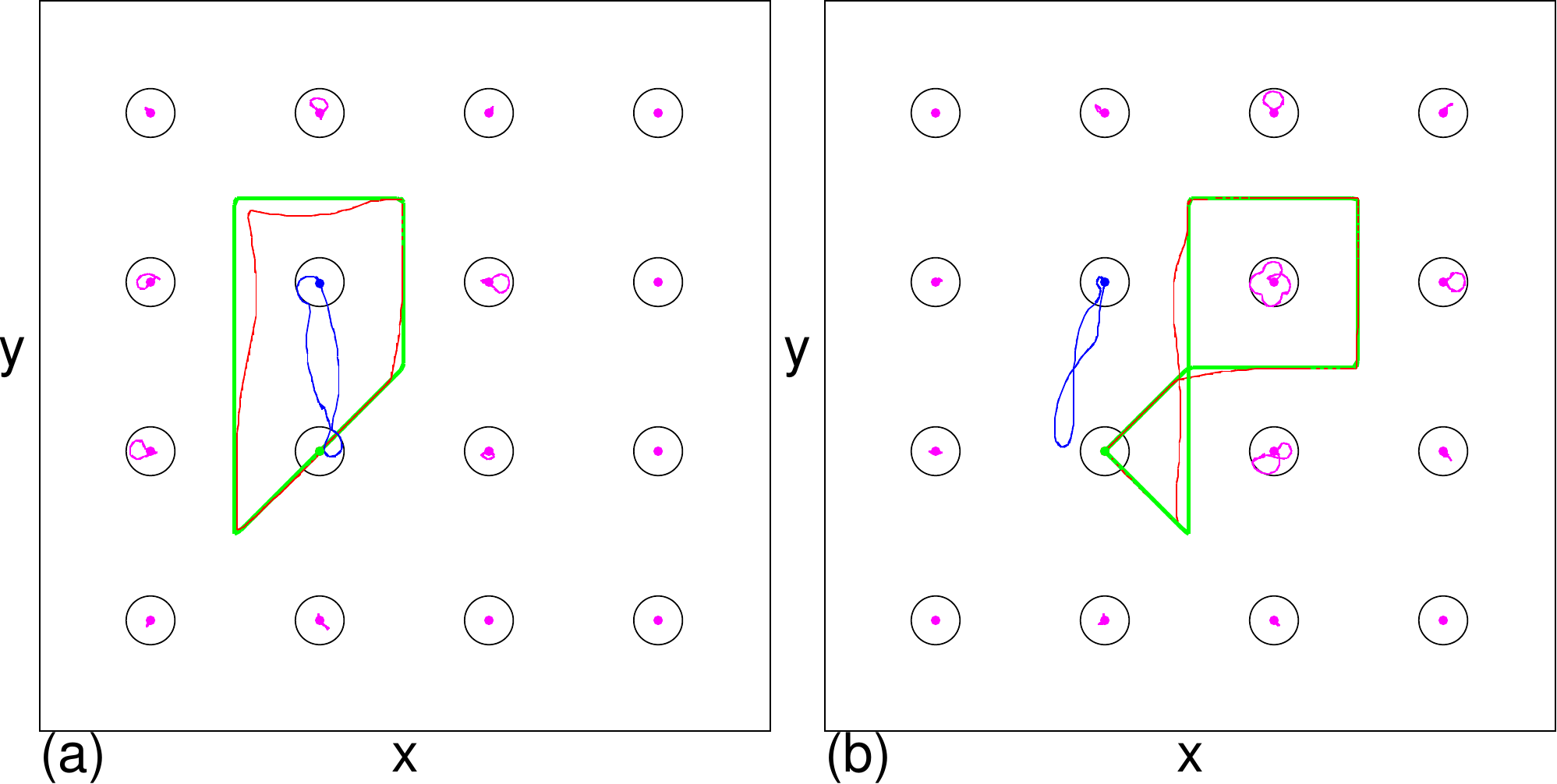}
\caption{Pinning site locations (open circles), vortex locations (filled circles),
  vortex trajectories (lines), and probe tip trajectory (green line) for
  braiding operations between vortex 0 (red circle and trajectory) and vortex
  1 (blue circle and trajectory.
  (a) A $\langle 10\rangle$ braid. (b) A $\langle 11\rangle$ braid.
}
\label{fig:8}
\end{figure}

\begin{figure}
\includegraphics[width=3.5in]{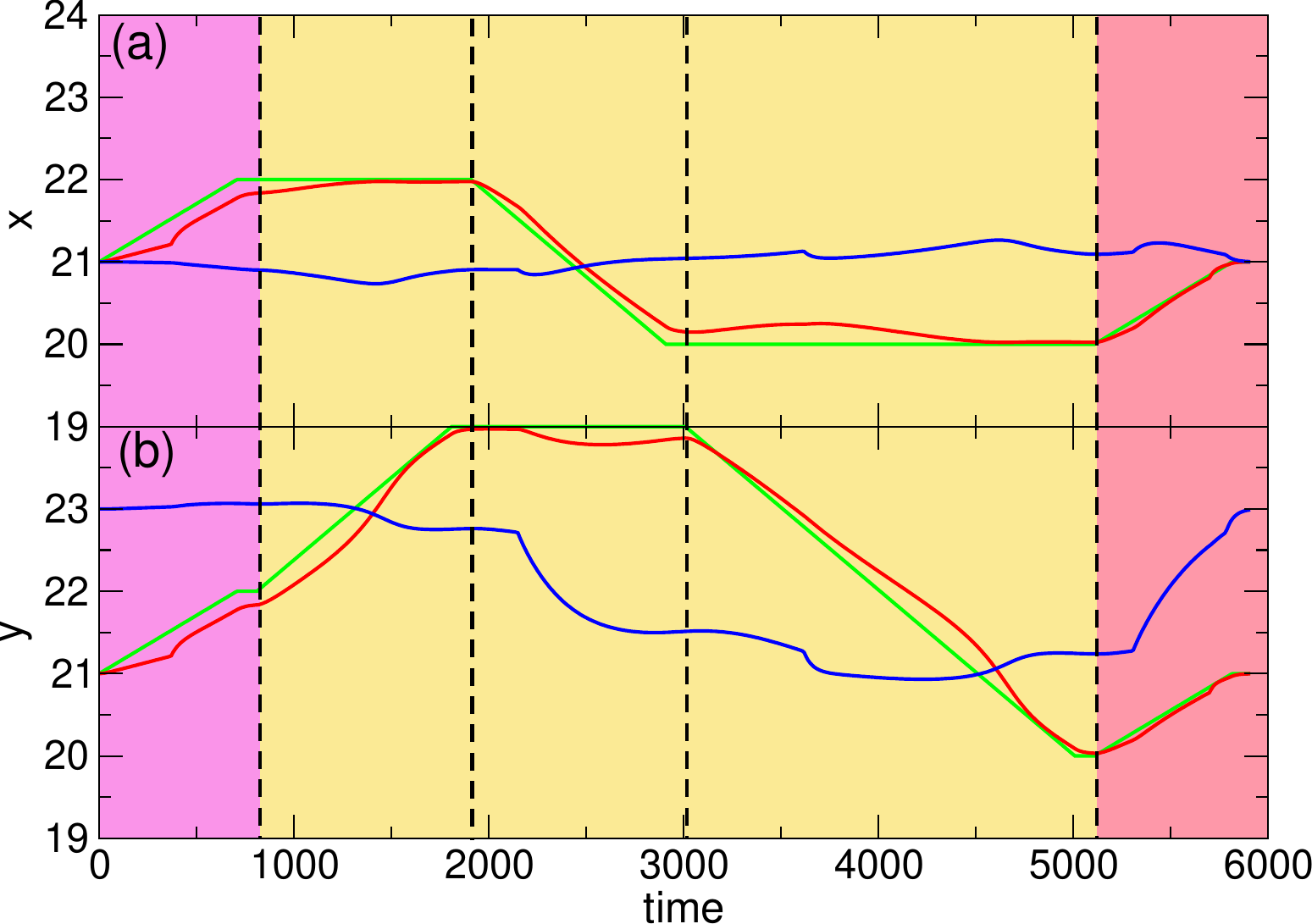}
\caption{The $x$ (a) and $y$  (b) positions vs time
  in simulation steps for vortex 0 (red), vortex 1 (blue), and
  the probe tip (green) for the $\langle 1 0\rangle$
  braid operation illustrated in Fig.~\ref{fig:8}(a).
  The vertical dashed lines denote the five stages of probe tip motion.
  Magenta: capture; yellow: $\langle 10\rangle$ moves; pink: reverse capture.
}
\label{fig:9}
\end{figure}

To perform a braid,
the probe tip carries one vortex around one or more other vortices and then
returns the original vortex to its starting position,
as illustrated in the world line
diagram in Fig.~\ref{fig:7}.
During this operation, the vortices that are not involved in the braiding
may show small perturbations in their positions but they
remain localized in the pinning sites.
In some cases, one of these background
vortices can temporarily move out of its pinning site before returning
to the same site;
however, there is no exchange of these
additional vortices,
and their world lines
do not interact with
the world lines of the dragged or braided vortices.
In Fig.~\ref{fig:8}(a) we illustrate the vortex positions and trajectories
for a local $\langle 1 0\rangle$
braiding operation which
is achieved with 5 probe tip motion stages.
Figure~\ref{fig:9}(a,b) shows the $x$ and $y$ positions of
vortex 0, vortex 1, and the probe tip during
the $\langle 1 0\rangle$ braiding operation.
In this case, the capture and reverse capture operations are performed at the
same pinning site location in order to return vortex 0 to its original position.
The procedure for this braiding operation is
([11], $\sqrt{2}/2$), ([01], 1), ([$\overline{1}$0], 1), ([0$\overline{1}$], 2),
([11], $\sqrt{2}/2$).
As indicated by Fig.~\ref{fig:8}(a),
vortex 1 is actually forced out of its pinning site during the third stage,
but it returns to its original pinning site during the fifth stage.
All of the vortices return to their original positions
at the end of the operation, in agreement with
the world lines illustrated in Fig.~\ref{fig:7}.
We can also perform a $\langle 11\rangle$ braiding operation using
six stages of probe tip motion, as shown
in Fig.~\ref{fig:8}(b).
The procedure can be written as
([11], $\sqrt{2}/2$), ([10], 1), ([01], 1), ([$\overline{1}$0], 1), ([0$\overline{1}$], 2),
([$\overline{1}$1], $\sqrt{2}/2$).
In this case, a vortex that is not involved in the braiding operation makes a
temporary excursion out of its pinning site before returning,
but its world line does not interact with
the world lines of vortices 0 or 1.

\begin{figure}
\includegraphics[width=3.5in]{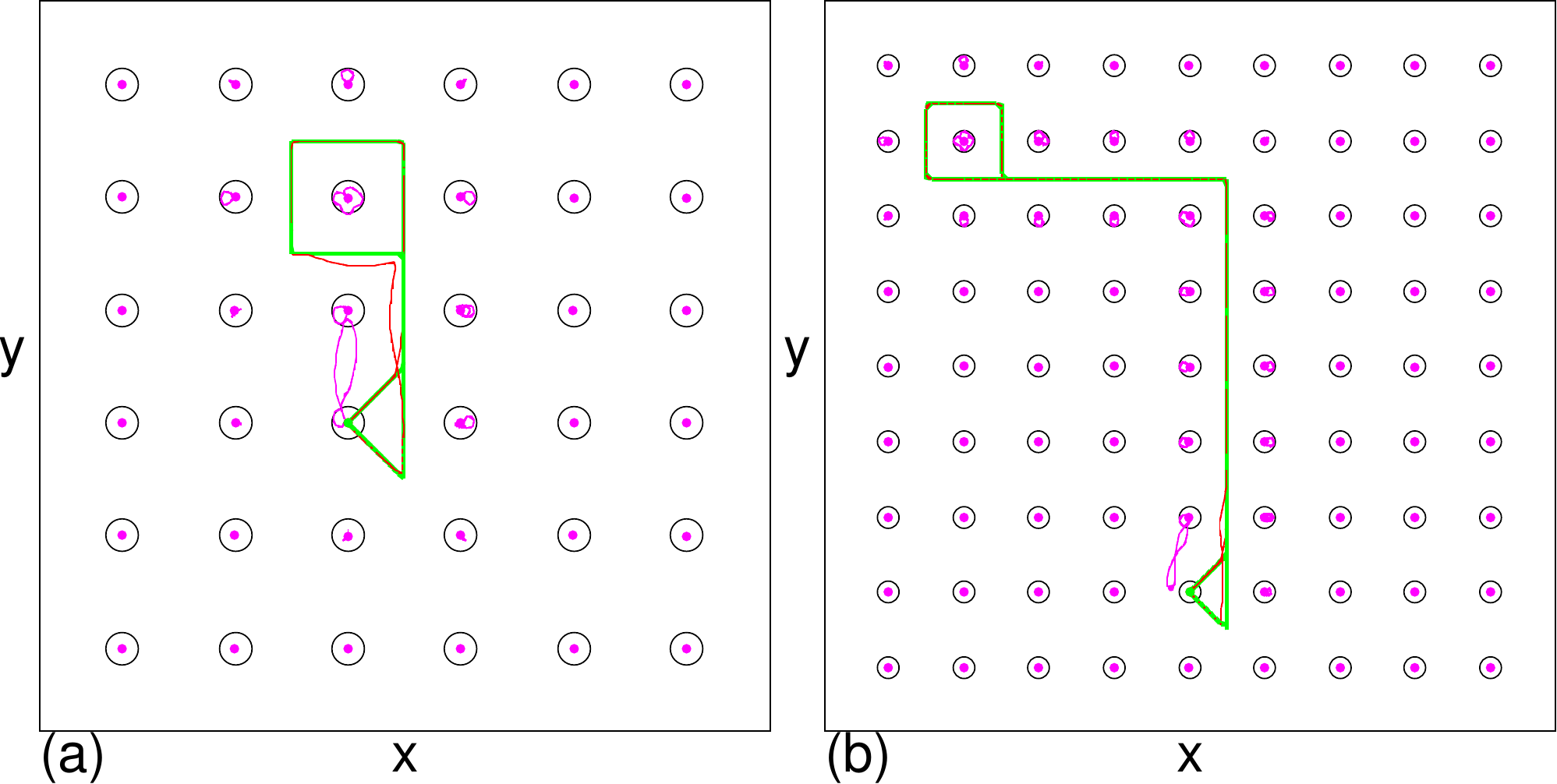}
\caption{
  Pinning site locations (open circles), vortex locations (filled circles), vortex
  trajectories (lines), and probe tip trajectory (green line) for braiding operations
  between vortex 0 (red circle and trajectory) and non-neighboring vortices.
  (a) Braiding of vortices that are separated by two lattice constants.
  (b) Braiding of distant vortices.
 }
\label{fig:10}
\end{figure}

The braiding operation can be extended to vortices that are not nearest
neighbors,
as shown in Fig.~\ref{fig:10}(a) where vortices that are two lattice constants away from
each other are braided in a seven stage operation.
The braid can be extended out to arbitrary distances as long as the coherence time
of the Majorana fermions is not exceeded by the operation time.
A nine step braiding operation applied to two distant vortices is illustrated
in Fig.~\ref{fig:10}(b).

\section{Gate Operations}

\begin{figure}
\includegraphics[width=3.5in]{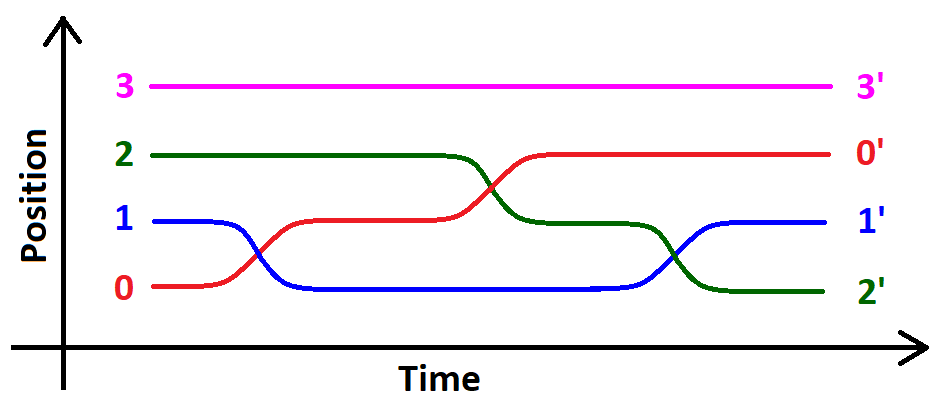}
\caption{The world lines as a function of position and time
  for a Hadamard gate involving vortex 0 (red), 1 (blue), and 2 (green), with
the world line of an uninvolved vortex 3 (pink) included for comparison.
}
\label{fig:11}
\end{figure}

\begin{figure}
\includegraphics[width=3.5in]{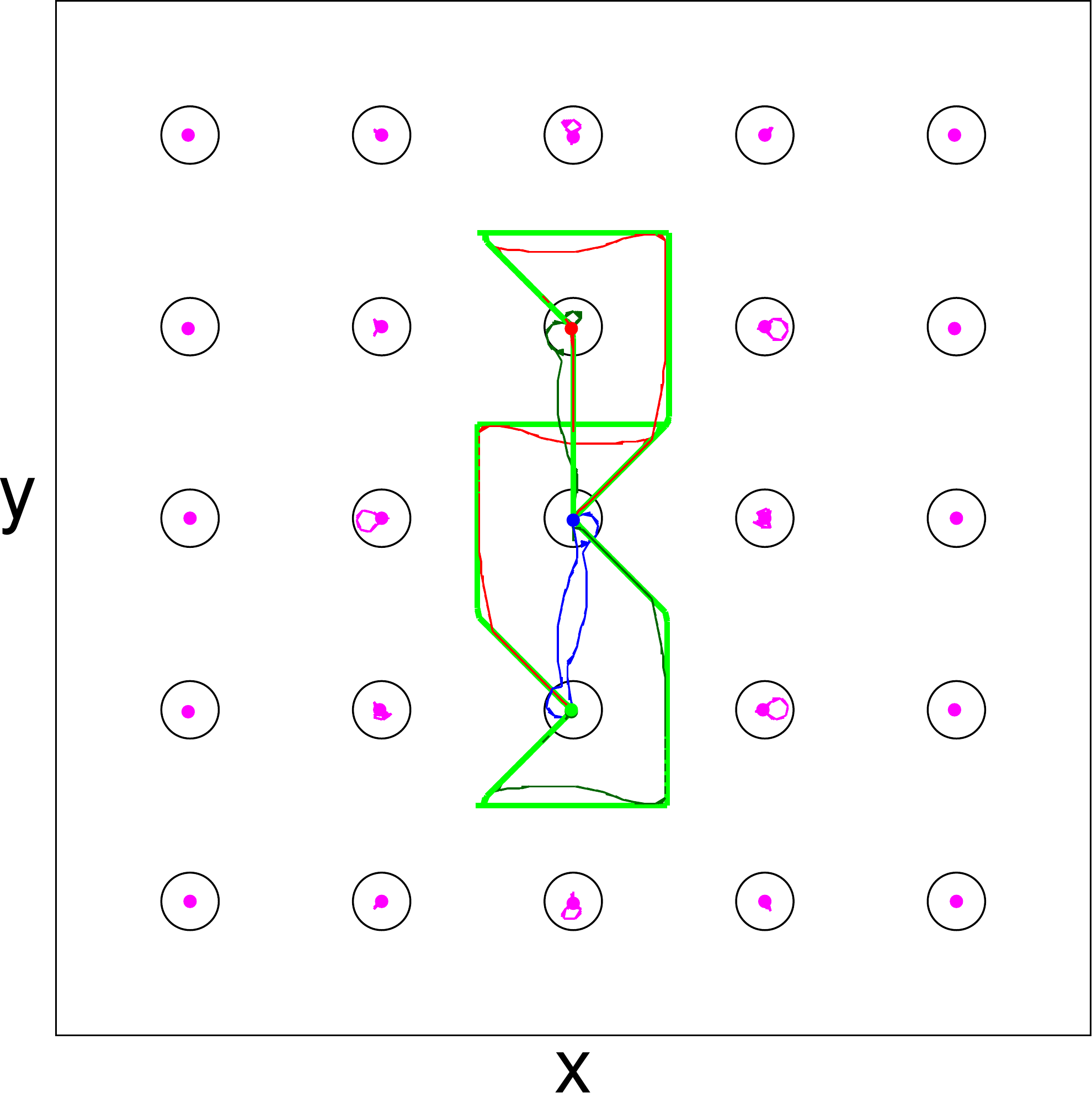}
\caption{
  Pinning site locations (open circles), vortex locations (filled circles), vortex
  trajectories (lines), and probe tip trajectory (green line) for
  a Hadamard gate involving
  vortex 0 (red circle and trajectory), vortex 1 (blue circle and trajectory), and
  vortex 2 (dark green circle and trajectory).
  }
\label{fig:12}
\end{figure}

\begin{figure}
\includegraphics[width=3.5in]{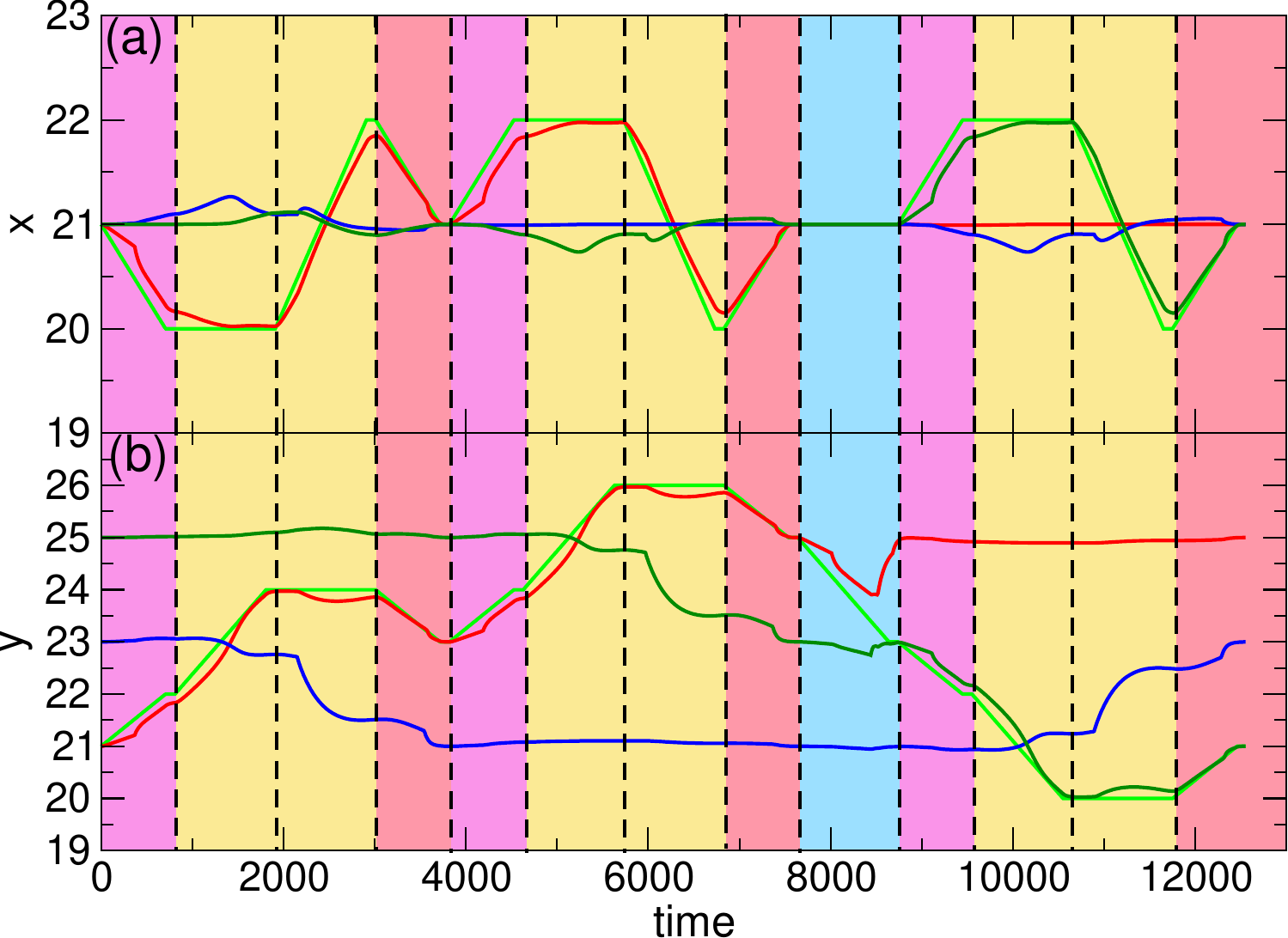}
\caption{
The $x$ (a) and $y$ (b) positions vs time in simulation steps for
vortex 0 (red), vortex 1 (blue), vortex 2 (dark green), and the
probe tip (light green)
for the Hadamard gate operation
illustrated in Fig.~\ref{fig:12}.
The vertical dashed lines indicate the 13 stages of probe tip motion.
Magenta: capture; yellow: $\langle 1 0\rangle$ moves; blue: repositioning of probe;
pink: reverse capture.
}
\label{fig:13}
\end{figure}

Now that we have demonstrated
the exchange and braiding operations,
we
show
how to combine these operations in order to
create
Hadamard and CNOT gates.
In Fig.~\ref{fig:11} we illustrate the world lines for a
Hadamard gate involving three vortices, while
in Fig.~\ref{fig:12} we plot the vortex and probe tip trajectories from
a simulation of the Hadamard gate operation.
The Hadamard gate creates a quantum superposition of the Majorana fermions,
and our realization of this gate
requires 13 stages of probe tip motion.
In the first 4 stages, we perform a $\langle 1 0\rangle$ exchange clockwise
between vortices 0 and 1:
1. ([$\overline{1}$1], $\sqrt{2}/2$), 2. ([01], 1), 3. ([10], 1),
4. ([$\overline{1}\overline{1}$], $\sqrt{2}/2$).
The probe captures vortex 0 and moves it to the
pinning site occupied by vortex 1, which is ejected and travels to the pinning
site originally occupied by vortex 0.
In the next 4 stages, we perform a $\langle 1 0\rangle$
counterclockwise
exchange
between vortices 0 and 2:
5. ([11], $\sqrt{2}/2$), 6. ([01], 1), 7. ([$\overline{1}$0], 1),
8. ([1$\overline{1}$], $\sqrt{2}/2$).
The probe captures vortex 0 and moves it to the pinning site occupied by vortex 2,
which is ejected and travels to the pinning site originally occupied by vortex 1.
In the next stage, the probe tip is repositioned
to the pinning site originally occupied by vortex 1 without moving a vortex:
9. ([0$\overline{1}$], 1).
In the final 4 stages, we perform a $\langle 1 0\rangle$ exchange
between vortices 2 and 1 clockwise:
10. ([1$\overline{1}$], $\sqrt{2}/2$),
11. ([0$\overline{1}$], 1),
12. ([$\overline{1}$0], 1),
13. ([11], $\sqrt{2}/2$).
The probe captures vortex 2 and moves it to the pinning site originally occupied
by vortex 0.  Vortex 1 is ejected from this pinning site and returns to the pinning site
it originally occupied.
In Fig.~\ref{fig:13} we plot the $x$ and $y$
positions versus time in simulation steps for
vortices 0, 1, and 2 along with the probe tip for
the successful simulated operation of the Hadamard gate.

\begin{figure}
\includegraphics[width=3.5in]{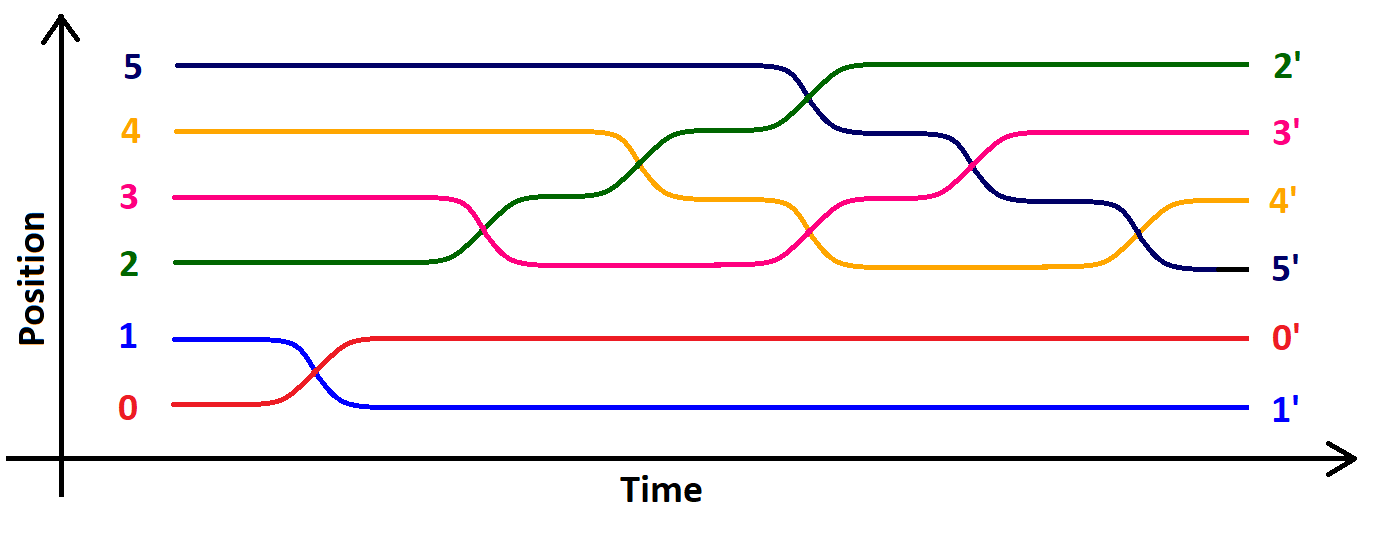}
\caption{
  The world lines as a function of position and time
  for a CNOT gate involving
vortex 0 (red), 1 (blue), 2 (dark green), 3 (pink), 4 (orange), and 5 (violet).
}
\label{fig:14}
\end{figure}

\begin{figure}
\includegraphics[width=3.5in]{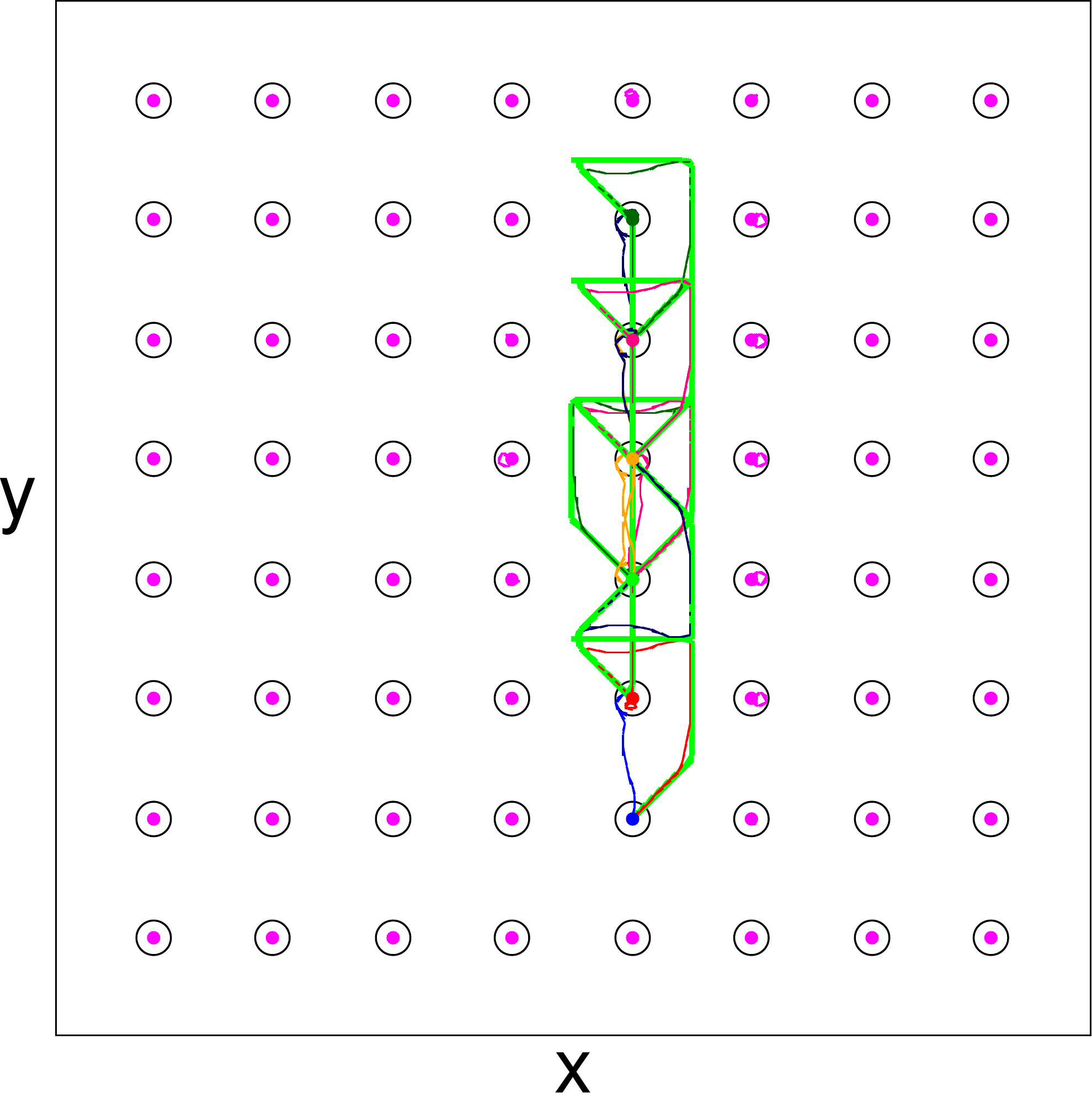}
\caption{
  Pinning site locations (open circles), vortex locations (filled circles),
  vortex trajectories (lines), and probe tip trajectory (green line) for a CNOT
  gate involving vortex
  0 (red circle and trajectory),
  1 (blue circle and trajectory),
  2 (green circle and trajectory),
  3 (pink circle and trajectory),
  4 (orange circle and trajectory), and
  5 (violet circle and trajectory).
}
\label{fig:14a}
\end{figure}

\begin{figure}
\includegraphics[width=3.5in]{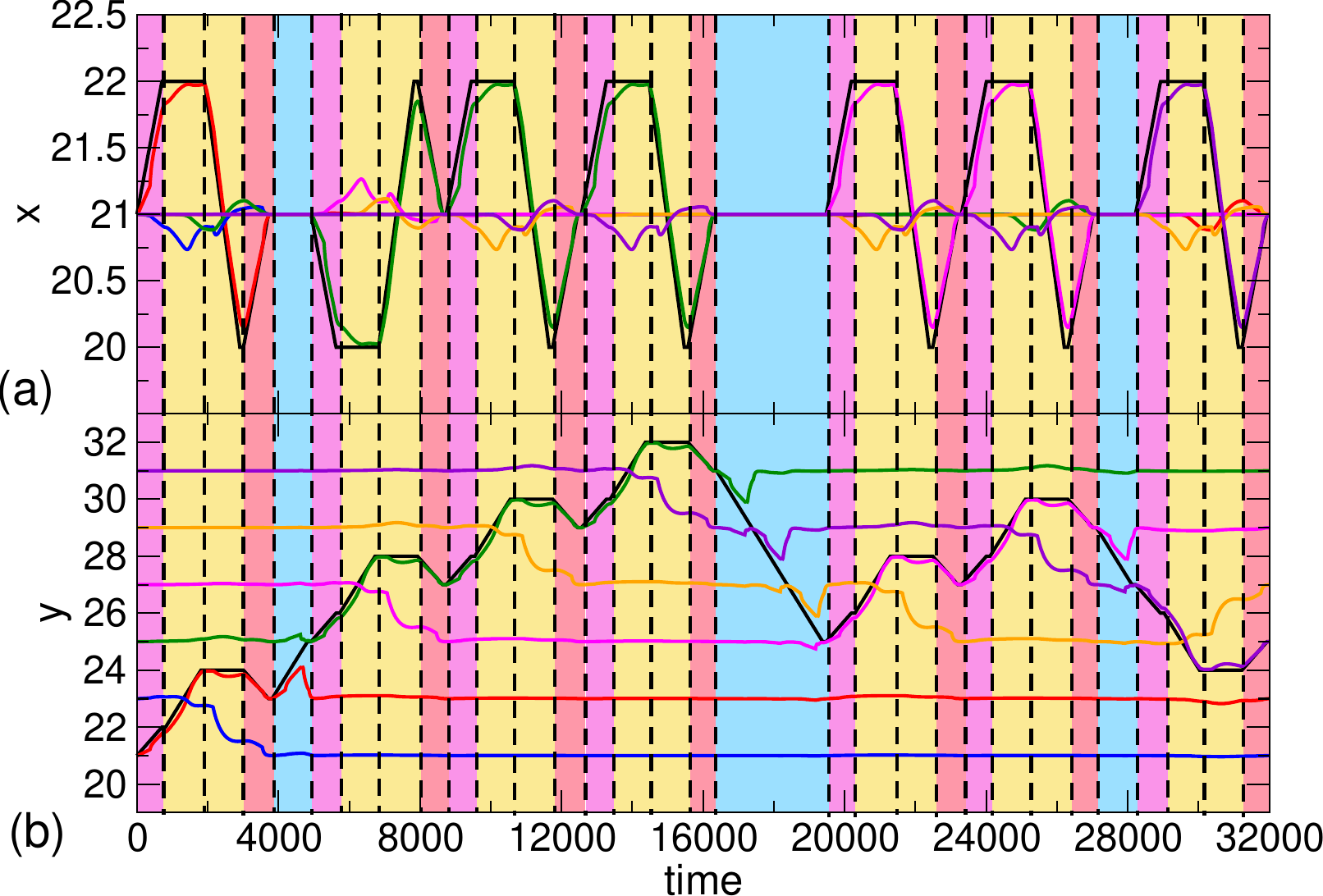}
\caption{The $x$ (a) and $y$ (b) positions vs time in simulation steps
  for vortex 0 (red), vortex 1 (blue), vortex 2 (dark green), vortex 3 (pink),
  vortex 4 (orange), vortex 5 (violet), and the probe tip (black) for the CNOT
  gate operation illustrated in Fig.~\ref{fig:14a}.
  The vertical dashed lines indicate the 31 stages of probe tip motion.
  Magenta: capture; yellow: $\langle 10\rangle$ moves; blue: repositioning of
  probe; pink: reverse capture.
 }
\label{fig:15}
\end{figure}

Now that we have shown the
achievement of a single qubit gate by
a vortex braiding operation, we
demonstrate a CNOT gate as an example of a 2-qubit gate.
In Fig.~\ref{fig:14}(a) we show the world lines for
a CNOT gate involving six vortices.
We select 6 vortices that are in a vertical line, as shown in Fig.~\ref{fig:14a} where we
illustrate the motion of the vortices and the probe tip.
The CNOT operation requires 31 stages of probe motion.
In the first 4 stages, we perform a $\langle 1 0 \rangle$ exchange
between vortices 0 and 1:
1. ([11], $\sqrt{2}/2$), 2. ([01], 1), 3. ([$\overline{1}$0], 1),
4. ([1$\overline{1}$], $\sqrt{2}/2$).
The probe captures vortex 0 and moves it to the pinning site occupied by
vortex 1, which is ejected and travels to the pinning site originally occupied
by vortex 0.
In the next stage, the probe tip is repositioned to the pinning site occupied by
vortex 2 without moving a vortex:
5. ([01], 1).
During the next 4 stages, we perform a $\langle 1 0\rangle$ exchange
between vortices 2 and 3:
6. ([$\overline{1}$1], $\sqrt{2}/2$), 7. ([01], 1), 8. ([10], 1), 9. ([$\overline{1}\overline{1}$],
$\sqrt{2}/2$).
The probe captures vortex 2 and moves it to the pinning site occupied by vortex 3,
which is ejected and travels to the pinning site originally occupied by vortex 2.
In the next 4 stages we perform a $\langle 1 0\rangle$ exchange between
vortices 2 and 4:
10. ([11], $\sqrt{2}/2$), 11. ([01], 1), 12. ([$\overline{1}$0], 1),
13. ([1$\overline{1}$], $\sqrt{2}/2$).
The probe captures vortex 2 and moves it to the pinning site occupied by vortex 4,
which is ejected and travels to the pinning site originally occupied by vortex 3.
The next 4 stages are a $\langle 1 0\rangle$ exchange between
vortices 2 and 5:
14. ([11], $\sqrt{2}/2$), 15. ([01], 1), 16. ([$\overline{1}$0], 1),
17. ([1$\overline{1}$], $\sqrt{2}/2$).
The probe captures vortex 2 and moves it to the pinning site occupied by
vortex 5, which is ejected and travels to the pinning site originally occupied by
vortex 4.
In the next stage, the probe tip is repositioned to the pinning site originally
occupied by vortex 2 without moving a vortex:
18. ([0$\overline{1}$], 3).
The next 4 stages are a $\langle 1 0\rangle$ exchange between vortices 3 and 4:
19. ([11], $\sqrt{2}/2$), 20. ([01], 1), 21. ([$\overline{1}$0], 1),
22. ([1$\overline{1}$], $\sqrt{2}/2$).
The probe captures vortex 3 and moves it to the pinning site originally occupied
by vortex 3.  This returns vortex 3 to its starting location and causes the ejection
of vortex 4, which travels to the pinning site originally occupied by vortex 2.
The following 4 stages are a $\langle 1 0\rangle$ exchange between vortices
3 and 5:
23. ([11], $\sqrt{2}/2$), 24. ([01], 1), 25. ([$\overline{1}$0], 1),
26. ([1$\overline{1}$], $\sqrt{2}/2$).
The probe captures vortex 3 and moves it to the pinning site originally occupied
by vortex 4.  This causes the ejection of vortex 5, which travels to the pinning site
originally occupied by vortex 3.
In the next stage, the probe tip is repositioned to the pinning site originally
occupied by vortex 3 without moving a vortex:
27. ([0$\overline{1}$], 1).
During the final four stages, a $\langle 1 0\rangle$ exchange is performed
between vortices 4 and 5:
28. ([1$\overline{1}$], $\sqrt{2}/2$), 29. ([0$\overline{1}$], 1), 30. ([$\overline{1}$0], 1),
31. ([11], $\sqrt{2}/2$).
The probe captures vortex 5 and moves it to the pinning site originally occupied
by vortex 2.  This causes the ejection of vortex 4, which travels to the pinning site
originally occupied by vortex 3.
In Fig.~\ref{fig:15} we plot the $x$ and $y$ positions of
vortices 0 through 5 and the probe tip during the simulated operation of
the CNOT gate during a time span of 32000 simulation time steps.

We note that other quantum gates can be realized following similar
vortex braiding methods, and thus complicated logic operations can be achieved.
In principle it would be possible to perform the CNOT and Hadamard operations
using
only a 1D chain of pinned vortices.
In that case, however, it is possible that other unconfined
vortices outside of the chain could occupy random positions that might interfere
with the logic operations.  By trapping all of the vortices in a periodic pinning array,
we can guarantee that a minimum distance is maintained between the passive
background vortices and the active vortices that are involved in the logic
operation.

\section{Robustness}

\begin{figure}
\includegraphics[width=3.5in]{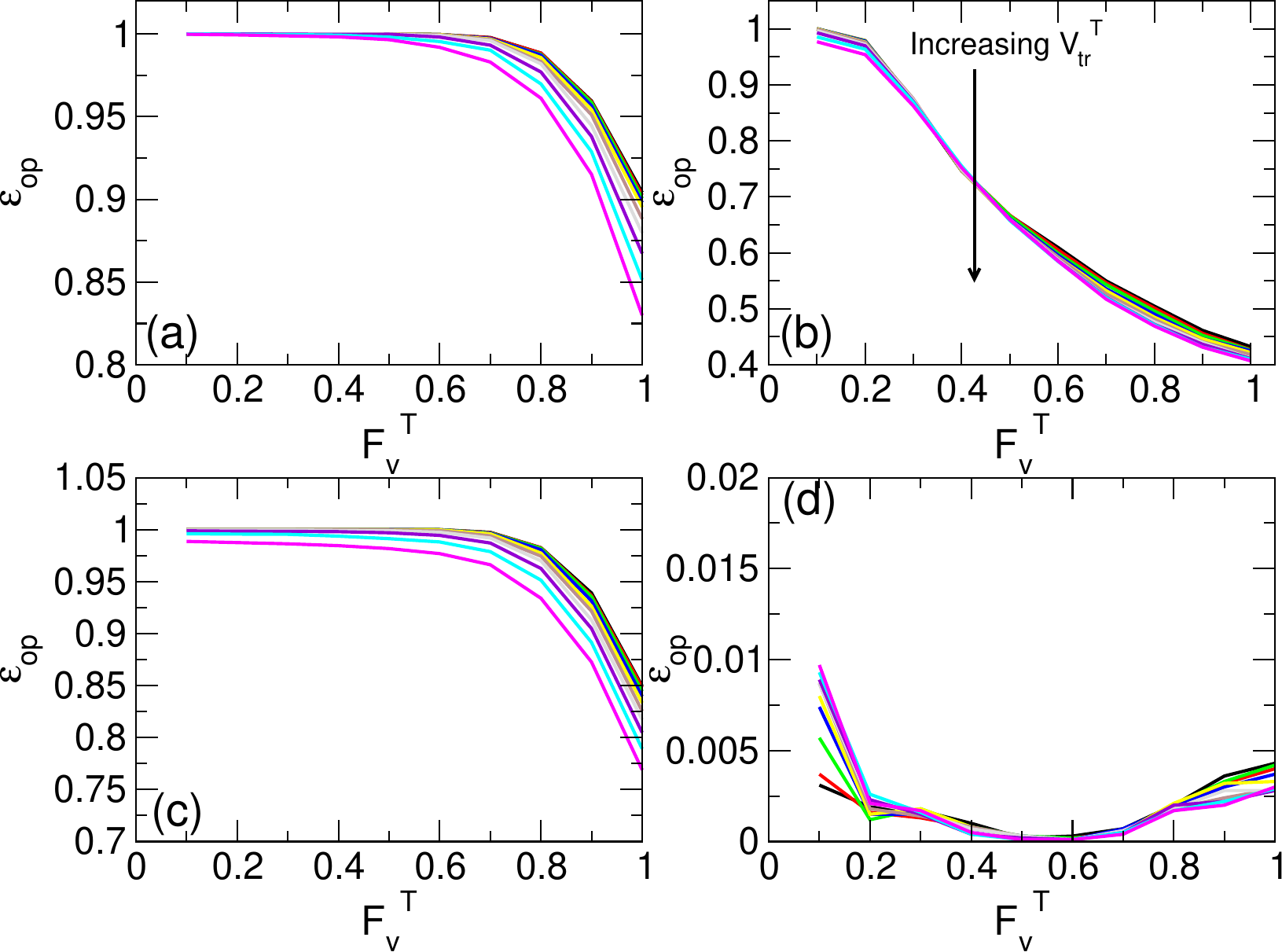}
\caption{ The fidelity $\epsilon_{\rm op}$ of the operations
  vs the magnitude $F_v^T$ of the thermal noise on the
  vortices for varied probe tip fluctuation magnitudes
  $V_{tr}^T=$ 0.02 (black), 0.04, 0.06, 0.08, 0.10, 0.12, 0.14, 0.16, 0.18,
  and 0.20 (magenta).
  (a) The capture operation.
  (b) The reposition operation.
  (c) The $\langle 1 0\rangle$ exchange.
(d) The $\langle 1 1\rangle$ exchange, which shows a strong sensitivity to noise.
 }
\label{fig:16}
\end{figure}

We next analyze the robustness of our logic
operations against noise.
To achieve this, we add
thermal
fluctuations to the motion of the vortices and compare this with the
addition of
thermal motions to the probe tip itself.
In all cases, the probe tip moves at
an average velocity of $V_{tr}=0.1$ and we fix $F_{tr} = 0.3$.
The thermal fluctuations $F^T_i$ have the properties
$\langle F^T_i\rangle=0$ and $\langle F^T_i(t)F_j^T(t^\prime)\rangle=2\eta k_B T\delta_{ij}\delta(t-t^\prime)$.
We consider fluctuations ranging from
$V_{tr}^T=0.02$ to $V_{tr}^T=0.2$ in intervals of $\Delta V_{tr}^T=0.02$ for the probe tip,
and fluctuations ranging from $F_v^T=0.1$ to $F_v^T=1.0$ in intervals
of $\Delta F_v^T=0.1$ for the vortices.
We perform each logic operation
$O_L=10000$ times in the presence of the fluctuations and
compute the fidelity $\epsilon_{\rm op}=O_s/O_L$, given by the ratio of the number
of successful operations $O_s$ to the total number of attempted
operations $O_L$.
In Fig.~\ref{fig:16}(a) we plot $\epsilon_{\rm op}$ versus vortex
thermal noise magnitude $F_v^T$ for the capture operation performed at different
values of the tip noise $V_{tr}^T$.
The fidelity remains close to $\epsilon_{\rm op}=1$ for $F_v^T<0.8$ and begins to
drop as the vortex thermal noise increases above this value, while the tip noise
has only a very weak effect on the fidelity.
In Fig.~\ref{fig:16}(b) we show $\epsilon_{\rm op}$ versus $F_v^T$ for the
reposition operation,
which depends more sensitively on the vortex thermal noise.
For $F_v^T<0.2$, $\epsilon_{\rm op} \approx 1$;
however, by the time $F_v^T=0.6$, the fidelity has dropped to a value of less than
50\%.
The reposition operation is also only weakly affected by tip fluctuations.
Figure~\ref{fig:16}(c) shows the fidelity versus $F_v^T$
for the $\langle 1 0\rangle$ exchange operation, which has $\epsilon_{\rm op} \approx 1$
when $F_v^T<0.8$.
In Fig.~\ref{fig:16}(d), the plot of $\epsilon_{\rm op}$ versus $F_v^T$ for the
$\langle 1 1\rangle$ exchange operation, which functions
well at zero temperature, shows that the fidelity drops nearly to zero as soon as
vortex thermal fluctuations are added.
This extreme sensitivity to noise makes the $\langle 1 1 \rangle$ exchange operation
poorly suited for use in logic gates, and this is why we did not employ this operation
in our proposed gates.

\begin{figure}
\includegraphics[width=3.5in]{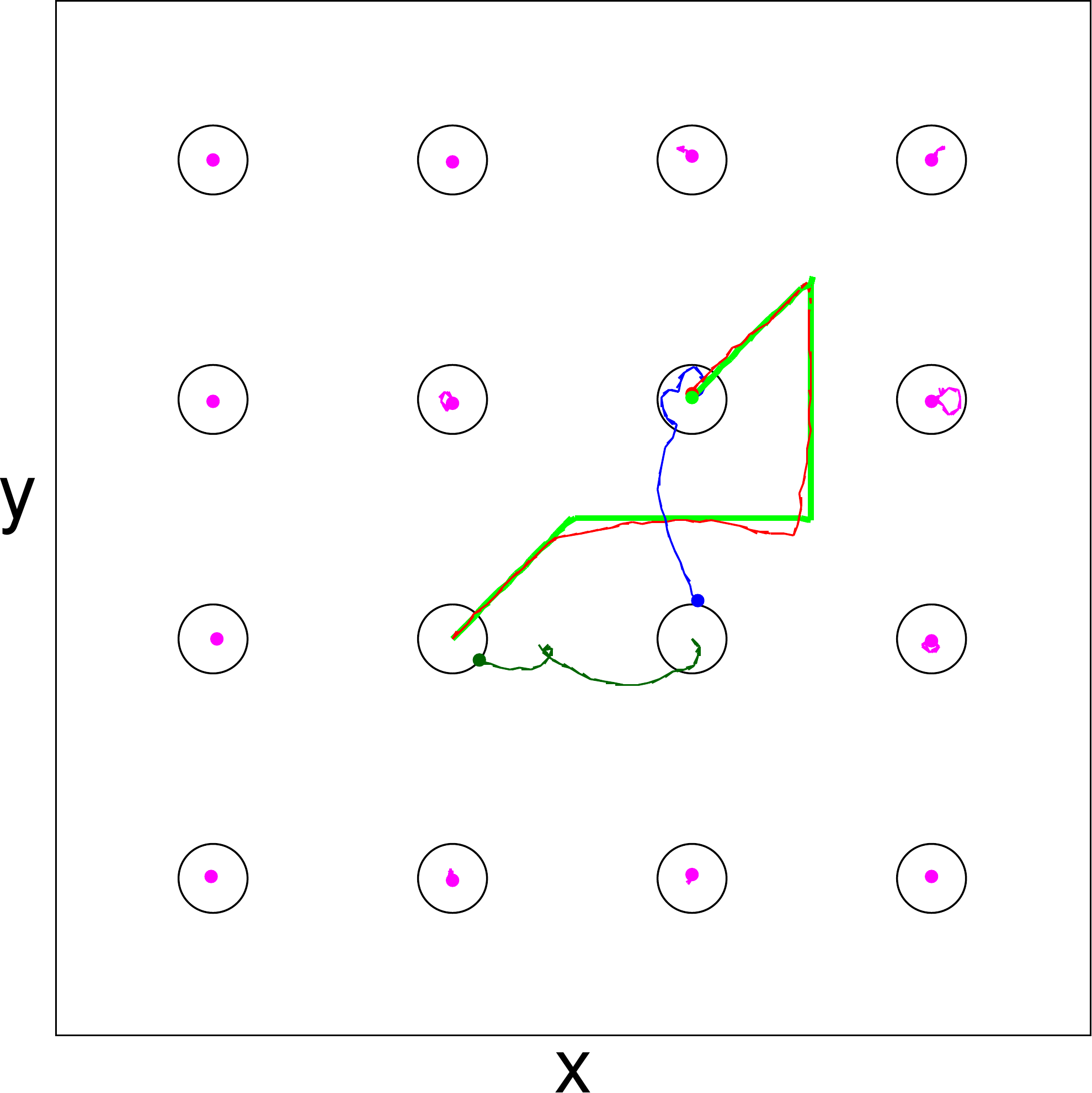}
\caption{
  Pinning site locations (open circles), vortex locations (filled circles), vortex
  trajectories (lines), and probe tip trajectory (green line) for
  the $\langle 1 1\rangle$ exchange operation performed at a finite
  temperature of $F_v^T=0.1.$  Vortex 0 (red circle and trajectory) and vortex 1
  (blue circle and trajectory) should be the only two vortices participating in the
  exchange; however, due to the thermal fluctuations, vortex 2 (dark green circle
  and trajectory) is able to depin and interfere with the operation.
 }
\label{fig:17}
\end{figure}

The origin of the extreme noise sensitivity of the
$\langle 1 1 \rangle$ exchange operation is illustrated
in Fig.~\ref{fig:17},
where we highlight the trajectories
at $F_v^T=0.1$ of the probe tip, vortices 0 and 1, and
a third vortex that is supposed to remain in the passive background of the
operation.
When $F_v^T = 0$, as in Fig.~\ref{fig:5}(b),  the third vortex does not
interfere with the operation;
however, once thermal noise is added,
the motion of vortex 1 can cause the third vortex to depin and move into the
pinning site originally occupied by vortex 0, resulting in a failure of the exchange.
We find that a similar effect occurs for the
$\langle 1 1\rangle$ braiding operation as
well.
This suggests that only the $\langle 1 0\rangle$ exchange and braiding operations,
which are not sensitive to the weak shear mode of the vortex lattice, should be used
for gate creation if thermal noise will be significant.

\section{Force Signals}

\begin{figure}
\includegraphics[width=3.5in]{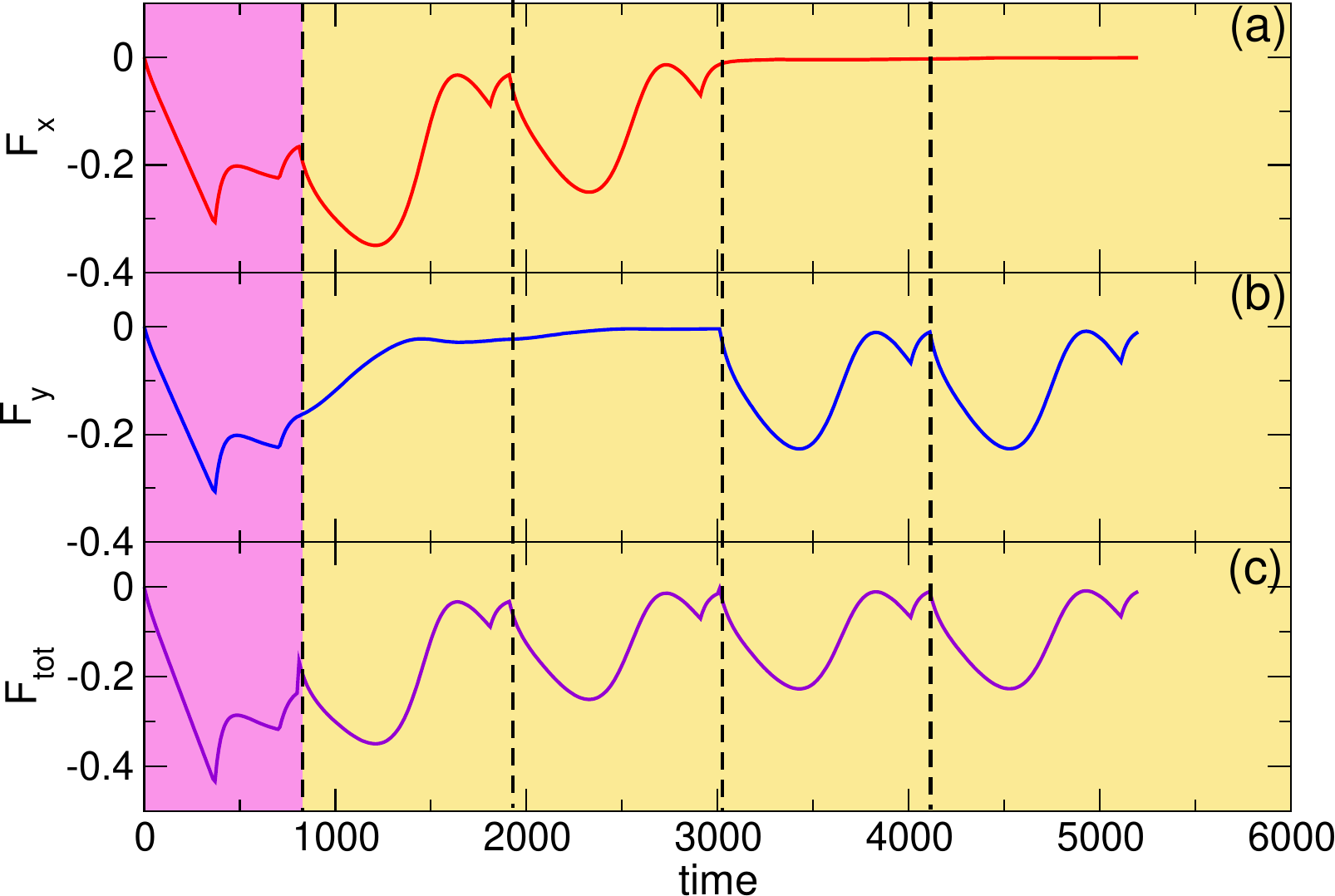}
\caption{Probe tip force signatures $F_x$ (a),
  $F_y$ (b), and $F_{\rm tot}$ (c) vs time in simulation steps
  for the capture (magenta) and $\langle 1 0\rangle$ move (yellow) operations
  illustrated in Fig.~\ref{fig:2}(a).
}
\label{fig:18}
\end{figure}

\begin{figure}
\includegraphics[width=3.5in]{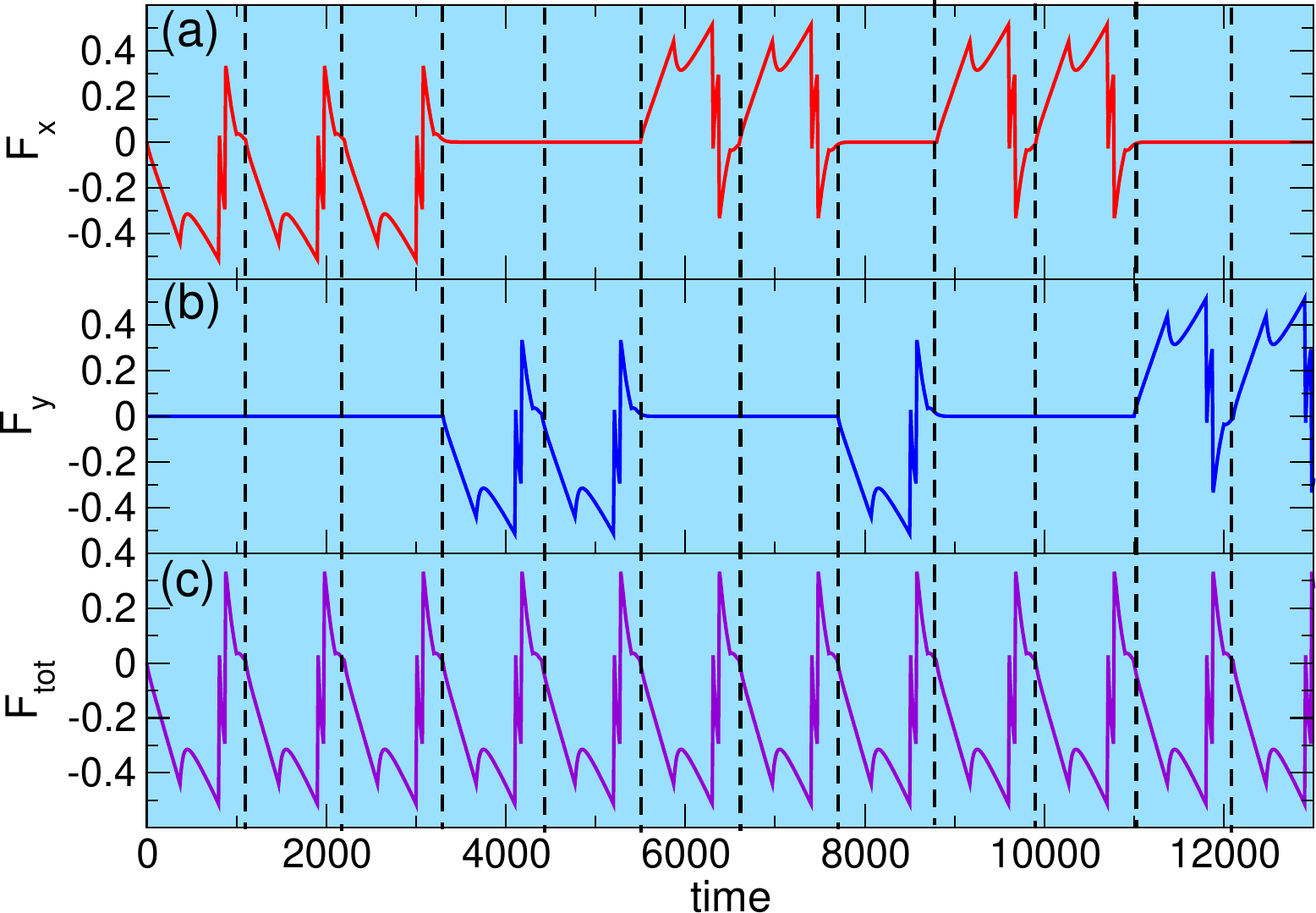}
\caption{Probe tip force signatures $F_x$ (a),
  $F_y$ (b), and $F_{\rm tot}$ (c) vs time in simulation steps
  for the reposition (blue) operations
  illustrated in Fig.~\ref{fig:2}(b).
}
\label{fig:18a}
\end{figure}

In our simulations, we can trace the complete vortex trajectory during the
operations;
however, in experiments performed using an MFM or other tip,
it would be
valuable to be able to determine whether the operation was successful by simply
measuring the force fluctuations experienced by the tip.
We find that the operations produce
a specific set of force patterns on the probe tip that can be divided
into five groups.
In the first group,
the probe tip pulls a vortex out of a pinning site.
In the second group, the probe tip transports a vortex symmetrically between
two pinned vortices, which remain pinned.
The third group is for asymmetric transport of a vortex by the probe tip
between two pinned vortices, where one of the pinned vortices has only a single
pinned neighboring vortex perpendicular to the direction in which the probe tip is
moving.  This vortex is depinned by the passage of the vortex trapped by the probe tip.
In the fourth group, the probe tip deposits its trapped vortex in a pinning site.
The fifth group is when the probe tip is repositioned and does not carry a trapped
vortex.
The characteristic force signals of these five groups of patterns can be used to determine
whether the attempted operation was successfully completed.
Additionally, a measurement of the $x$ and $y$ components of the forces make it
possible to reconstruct the trajectory of the probe tip in the experiment.
In Fig.~\ref{fig:18}(a,b,c) we plot $F_x$, $F_y$, and the total force $F_{\rm tot}$ on the
probe tip versus time for the capture and move operations illustrated
in Fig.~\ref{fig:2}(a), while in
Fig.~\ref{fig:18a}(a,b,c) we show the same for the reposition operations
from Fig.~\ref{fig:2}(b).
The capture signal, highlighted in pink in Fig.~\ref{fig:18}, is an abrupt
downward spike in all three quantities, while the move
signal, highlighted in yellow in Fig.~\ref{fig:18}, has a sinusoidal form in
the direction of motion.
The repositioning signal, shown in blue in Fig.~\ref{fig:18a}, has a double spike feature.

\begin{figure}
\includegraphics[width=3.5in]{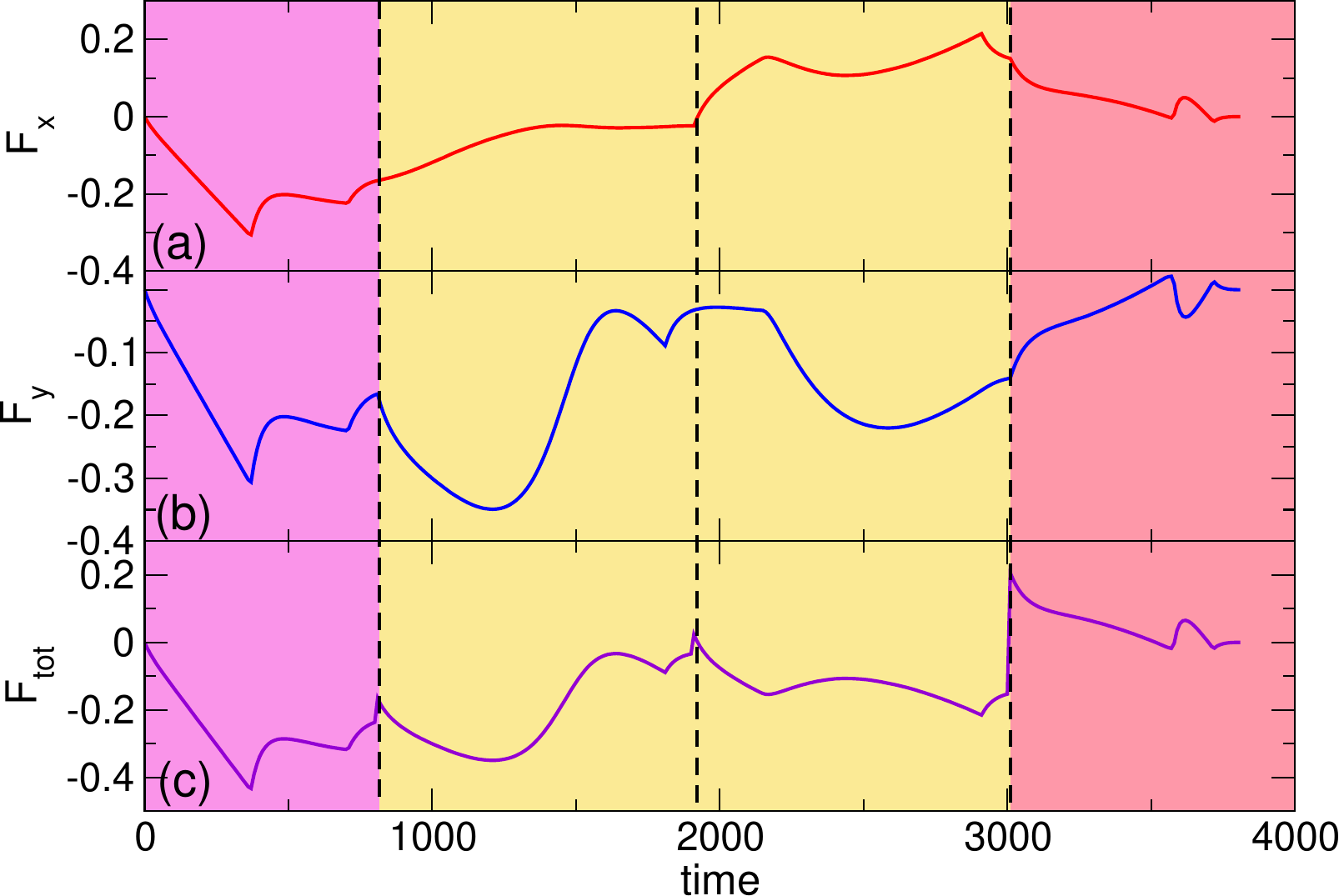}
\caption{Probe tip force signatures $F_x$ (a),
  $F_y$ (b), and $F_{\rm tot}$ (c) vs time in simulation steps
  for the $\langle 1 0\rangle$ exchange operation illustrated
  in Figs.~\ref{fig:5}(a) and \ref{fig:6}.
  Magenta: capture; yellow: $\langle 10\rangle$ moves; pink: reverse capture.
}
\label{fig:19}
\end{figure}

\begin{figure}
\includegraphics[width=3.5in]{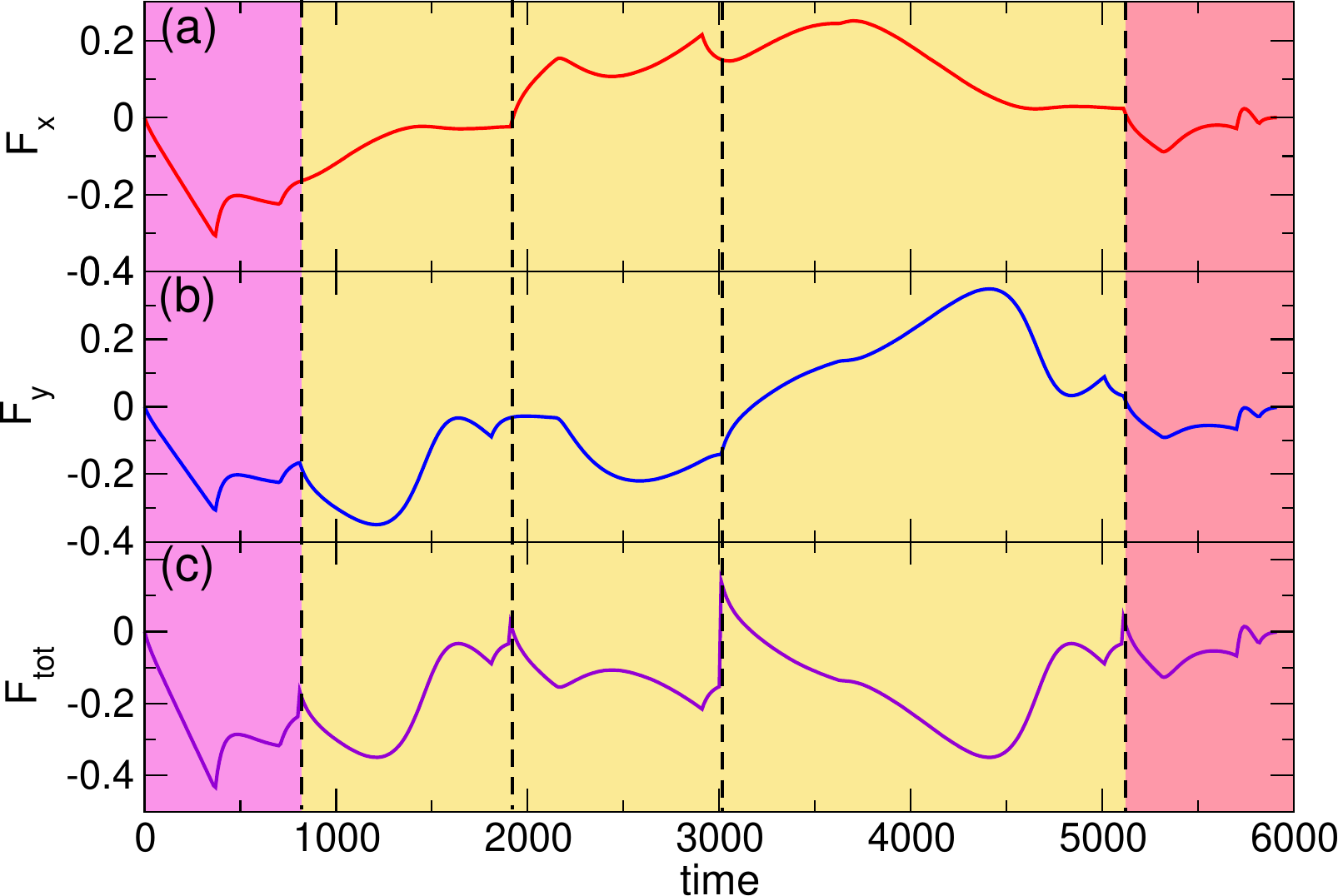}
\caption{Probe tip force signatures $F_x$ (a),
  $F_y$ (b), and $F_{\rm tot}$ (c) vs time in simulation steps
  for the $\langle 1 0\rangle$ braiding operation illustrated
  in Figs.~\ref{fig:8}(a) and \ref{fig:9}.
  Magenta: capture; yellow: $\langle 10\rangle$ moves; pink: reverse capture.
}
\label{fig:19a}
\end{figure}

In Fig.~\ref{fig:19} we plot $F_x$, $F_y$, and $F_{\rm tot}$ versus time
for the $\langle 1 0\rangle$ exchange operation, and in
Fig.~\ref{fig:19a} we plot the same quantities for the
$\langle 1 0\rangle$ braid operation.  Each of these operations shows
the reverse capture signature at the end; additionally, in the fourth
stage of the $\langle 1 0\rangle$ braid, the signature associated with the
indirect depinning of a vortex from a pinning site that is not immediately
underneath the probe tip can be seen.

\begin{figure}
\includegraphics[width=3.5in]{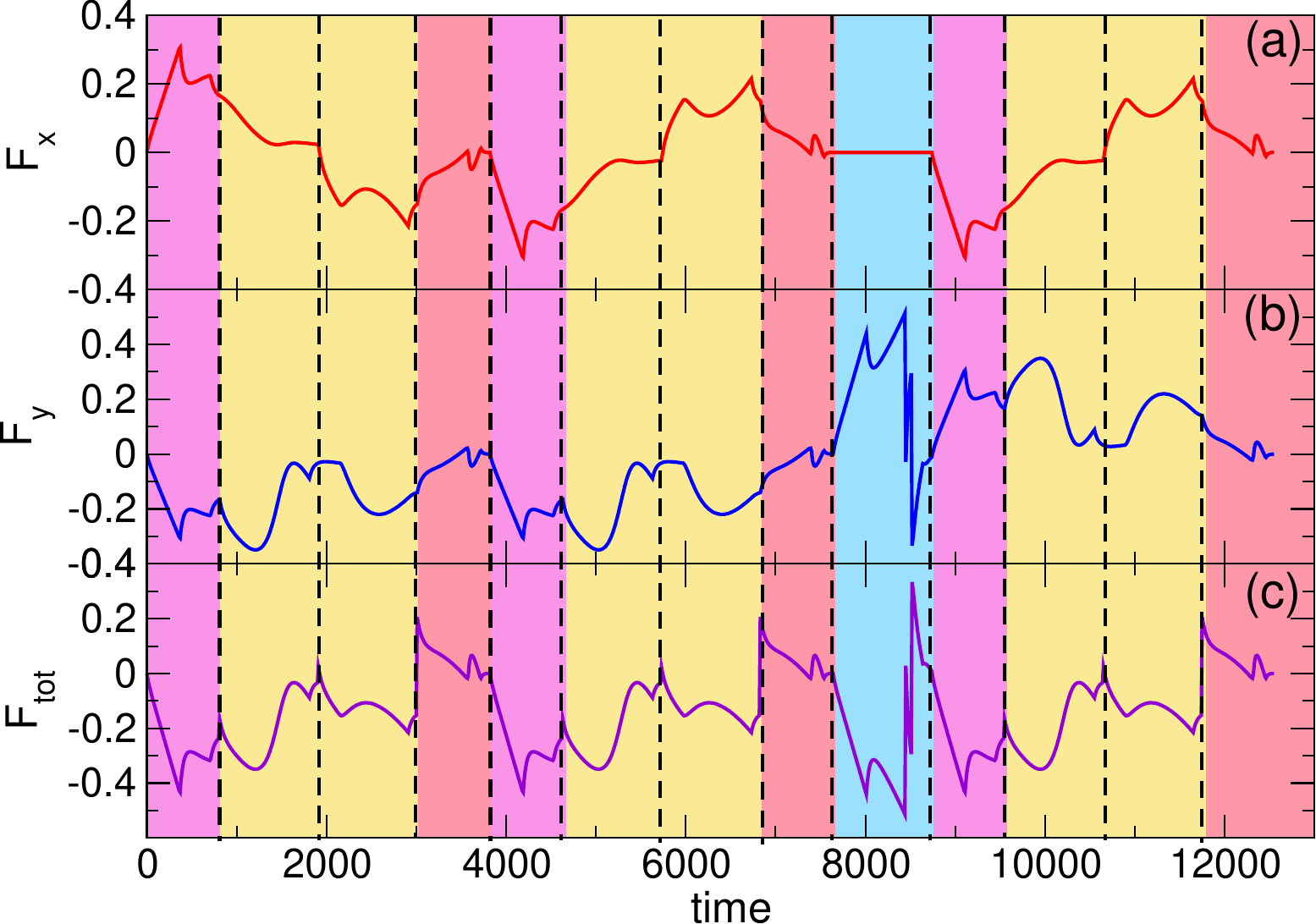}
\caption{Probe tip force signatures $F_x$ (a),
  $F_y$ (b), and $F_{\rm tot}$ (c) vs time in simulation steps
  for the Hadamard gate operation illustrated in Figs.~\ref{fig:12} and \ref{fig:13}.
  Magenta: capture; yellow: $\langle 10\rangle$ moves; pink: reverse capture;
  blue: repositioning of probe.
}
\label{fig:20}
\end{figure}

\begin{figure}
\includegraphics[width=3.5in]{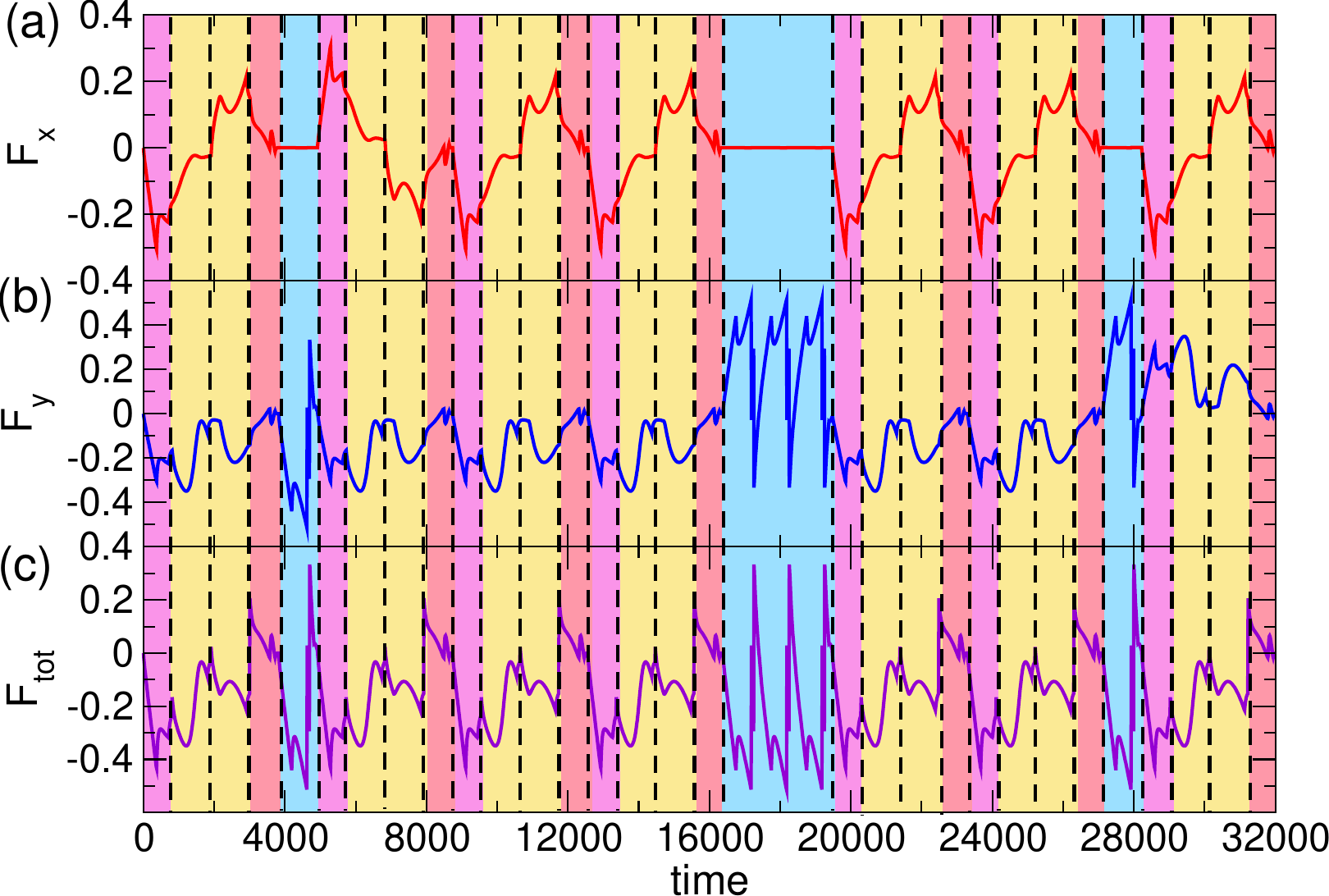}
\caption{Probe tip force signatures $F_x$ (a),
  $F_y$ (b), and $F_{\rm tot}$ (c) vs time in simulation steps
  for the CNOT gate operation illustrated in Figs.~\ref{fig:14a} and \ref{fig:15}.
  Magenta: capture; yellow: $\langle 10\rangle$ moves; pink: reverse capture;
  blue: repositioning of probe.
}
\label{fig:21}
\end{figure}

Figure~\ref{fig:20} shows the
probe tip force signals for the Hadamard gate and
Fig.~\ref{fig:21} shows the same for the CNOT gate.
In both cases, signatures of the independent components of the
gate moves can be seen.
In actual
experiments,
the exact form of the pinning potential
and probe tip interaction will likely differ from what we have assumed;
however, there
should still be distinct force signatures
for the different categories of motion.
In principle
it would be possible to construct
a library of the different force signatures that could be compared
with the experimentally measured signals in order to
confirm whether the operation was carried out successfully.

\section{Discussion}
An important advantage of using a tip to move the vortices
is that the same tip can also be used to measure the  response of the Majorana
fermion, as suggested
in Ref.~\cite{November19} for Majorana fermions contained inside
superconducting vortices.
A similar technique could be used to manipulate skyrmions
containing Majorana fermion bound states \cite{Yang16,Gungordu18,Rex19};
however, in this case, it is necessary to consider the dynamics carefully
due to the strong Magnus component of the skyrmion motion, which is
likely to affect how the skyrmions
move under the influence of the tip and interact with pinning sites
\cite{Reichhardt18}.
In this work, we considered moving the vortices with an MFM tip;
however, there are also proposals for creating vortex logic
devices using
applied currents
and specially structured pinning geometries \cite{Hastings03,Milosevic07}.
Alternative geometries such as these could also be used to achieve
braiding operations.

\section{Summary}
We have shown how to braid Majorana fermions in vortices on a periodic pinning array
using an MFM tip.
We specifically examine a superconductor coupled to a topological insulator with a square array of pinning sites in the form of blind holes at a magnetic field for
which there is one vortex per pinning site.
After demonstrating the fundamental operations of vortex capture,
vortex motion, and probe tip relocation without motion of a vortex,
we show how to perform vortex exchange
and basic braiding operations by following a series of specific steps of probe
tip motion.
Based on
the braiding operations, we
construct Hadamard and CNOT gates,
and show using numerical simulations that
these gates can be operated successfully.
In the presence of thermal noise,
braids and motions involving the $\langle 11\rangle$ direction
are not robust, so we utilize only the $\langle 1 0\rangle$ direction for
maneuvering the vortices, while reserving $\langle 11\rangle$ motions for
capture and reverse capture of vortices from the pinning sites.
We show that the basic moves produce
specific sequences of $x$ and $y$ force components on the probe tip,
and that these sequences can be used to determine
whether a gate operation has been completed successfully in experiment.
Our results could be
generalized
for different pinning geometries,
which could further optimize the robustness of the operations.

\acknowledgments
This work was supported by the US Department of Energy through
the Los Alamos National Laboratory.  Los Alamos National Laboratory is
operated by Triad National Security, LLC, for the National Nuclear Security
Administration of the U. S. Department of Energy (Contract No. 892333218NCA000001).

\bibliography{mybib}

\begin{thebibliography}{85}%
\makeatletter
\providecommand \@ifxundefined [1]{%
 \@ifx{#1\undefined}
}%
\providecommand \@ifnum [1]{%
 \ifnum #1\expandafter \@firstoftwo
 \else \expandafter \@secondoftwo
 \fi
}%
\providecommand \@ifx [1]{%
 \ifx #1\expandafter \@firstoftwo
 \else \expandafter \@secondoftwo
 \fi
}%
\providecommand \natexlab [1]{#1}%
\providecommand \enquote  [1]{``#1''}%
\providecommand \bibnamefont  [1]{#1}%
\providecommand \bibfnamefont [1]{#1}%
\providecommand \citenamefont [1]{#1}%
\providecommand \href@noop [0]{\@secondoftwo}%
\providecommand \href [0]{\begingroup \@sanitize@url \@href}%
\providecommand \@href[1]{\@@startlink{#1}\@@href}%
\providecommand \@@href[1]{\endgroup#1\@@endlink}%
\providecommand \@sanitize@url [0]{\catcode `\\12\catcode `\$12\catcode
  `\&12\catcode `\#12\catcode `\^12\catcode `\_12\catcode `\%12\relax}%
\providecommand \@@startlink[1]{}%
\providecommand \@@endlink[0]{}%
\providecommand \url  [0]{\begingroup\@sanitize@url \@url }%
\providecommand \@url [1]{\endgroup\@href {#1}{\urlprefix }}%
\providecommand \urlprefix  [0]{URL }%
\providecommand \Eprint [0]{\href }%
\providecommand \doibase [0]{http://dx.doi.org/}%
\providecommand \selectlanguage [0]{\@gobble}%
\providecommand \bibinfo  [0]{\@secondoftwo}%
\providecommand \bibfield  [0]{\@secondoftwo}%
\providecommand \translation [1]{[#1]}%
\providecommand \BibitemOpen [0]{}%
\providecommand \bibitemStop [0]{}%
\providecommand \bibitemNoStop [0]{.\EOS\space}%
\providecommand \EOS [0]{\spacefactor3000\relax}%
\providecommand \BibitemShut  [1]{\csname bibitem#1\endcsname}%
\let\auto@bib@innerbib\@empty
\bibitem [{\citenamefont {Gardner}\ \emph {et~al.}(2002)\citenamefont
  {Gardner}, \citenamefont {Wynn}, \citenamefont {Bonn}, \citenamefont {Liang},
  \citenamefont {Hardy}, \citenamefont {Kirtley}, \citenamefont {Kogan},\ and\
  \citenamefont {Moler}}]{Gardner02}%
  \BibitemOpen
  \bibfield  {author} {\bibinfo {author} {\bibfnamefont {B.~W.}\ \bibnamefont
  {Gardner}}, \bibinfo {author} {\bibfnamefont {J.~C.}\ \bibnamefont {Wynn}},
  \bibinfo {author} {\bibfnamefont {D.~A.}\ \bibnamefont {Bonn}}, \bibinfo
  {author} {\bibfnamefont {R.~X.}\ \bibnamefont {Liang}}, \bibinfo {author}
  {\bibfnamefont {W.~N.}\ \bibnamefont {Hardy}}, \bibinfo {author}
  {\bibfnamefont {J.~R.}\ \bibnamefont {Kirtley}}, \bibinfo {author}
  {\bibfnamefont {V.~G.}\ \bibnamefont {Kogan}}, \ and\ \bibinfo {author}
  {\bibfnamefont {K.~A.}\ \bibnamefont {Moler}},\ }\bibfield  {title} {\enquote
  {\bibinfo {title} {Manipulation of single vortices in
  {YBa$_2$Cu$_3$O$_{6.354}$} with a locally applied magnetic field},}\ }\href
  {\doibase 10.1063/1.1445468} {\bibfield  {journal} {\bibinfo  {journal}
  {Appl. Phys. Lett.}\ }\textbf {\bibinfo {volume} {80}},\ \bibinfo {pages}
  {1010--1012} (\bibinfo {year} {2002})}\BibitemShut {NoStop}%
\bibitem [{\citenamefont {Straver}\ \emph {et~al.}(2008)\citenamefont
  {Straver}, \citenamefont {Hoffman}, \citenamefont {Auslaender}, \citenamefont
  {Rugar},\ and\ \citenamefont {Moler}}]{Straver08}%
  \BibitemOpen
  \bibfield  {author} {\bibinfo {author} {\bibfnamefont {E.~W.~J.}\
  \bibnamefont {Straver}}, \bibinfo {author} {\bibfnamefont {J.~E.}\
  \bibnamefont {Hoffman}}, \bibinfo {author} {\bibfnamefont {O.~M.}\
  \bibnamefont {Auslaender}}, \bibinfo {author} {\bibfnamefont
  {D.}~\bibnamefont {Rugar}}, \ and\ \bibinfo {author} {\bibfnamefont
  {Kathryn~A.}\ \bibnamefont {Moler}},\ }\bibfield  {title} {\enquote {\bibinfo
  {title} {Controlled manipulation of individual vortices in a
  superconductor},}\ }\href {\doibase 10.1063/1.3000963} {\bibfield  {journal}
  {\bibinfo  {journal} {Appl. Phys. Lett.}\ }\textbf {\bibinfo {volume} {93}},\
  \bibinfo {pages} {172514} (\bibinfo {year} {2008})}\BibitemShut {NoStop}%
\bibitem [{\citenamefont {Auslaender}\ \emph {et~al.}(2009)\citenamefont
  {Auslaender}, \citenamefont {Luan}, \citenamefont {Straver}, \citenamefont
  {Hoffman}, \citenamefont {Koshnick}, \citenamefont {Zeldov}, \citenamefont
  {Bonn}, \citenamefont {Liang}, \citenamefont {Hardy},\ and\ \citenamefont
  {Moler}}]{Auslaender09}%
  \BibitemOpen
  \bibfield  {author} {\bibinfo {author} {\bibfnamefont {O.~M.}\ \bibnamefont
  {Auslaender}}, \bibinfo {author} {\bibfnamefont {L.}~\bibnamefont {Luan}},
  \bibinfo {author} {\bibfnamefont {E.~W.~J.}\ \bibnamefont {Straver}},
  \bibinfo {author} {\bibfnamefont {J.~E.}\ \bibnamefont {Hoffman}}, \bibinfo
  {author} {\bibfnamefont {N.~C.}\ \bibnamefont {Koshnick}}, \bibinfo {author}
  {\bibfnamefont {E.}~\bibnamefont {Zeldov}}, \bibinfo {author} {\bibfnamefont
  {D.~A.}\ \bibnamefont {Bonn}}, \bibinfo {author} {\bibfnamefont
  {R.}~\bibnamefont {Liang}}, \bibinfo {author} {\bibfnamefont {W.~N.}\
  \bibnamefont {Hardy}}, \ and\ \bibinfo {author} {\bibfnamefont {K.~A.}\
  \bibnamefont {Moler}},\ }\bibfield  {title} {\enquote {\bibinfo {title}
  {Mechanics of individual isolated vortices in a cuprate superconductor},}\
  }\href {\doibase 10.1038/NPHYS1127} {\bibfield  {journal} {\bibinfo
  {journal} {Nature Phys.}\ }\textbf {\bibinfo {volume} {5}},\ \bibinfo {pages}
  {35--39} (\bibinfo {year} {2009})}\BibitemShut {NoStop}%
\bibitem [{\citenamefont {Luan}\ \emph {et~al.}(2009)\citenamefont {Luan},
  \citenamefont {Auslaender}, \citenamefont {Bonn}, \citenamefont {Liang},
  \citenamefont {Hardy},\ and\ \citenamefont {Moler}}]{Luan09}%
  \BibitemOpen
  \bibfield  {author} {\bibinfo {author} {\bibfnamefont {L.}~\bibnamefont
  {Luan}}, \bibinfo {author} {\bibfnamefont {O.~M.}\ \bibnamefont
  {Auslaender}}, \bibinfo {author} {\bibfnamefont {D.~A.}\ \bibnamefont
  {Bonn}}, \bibinfo {author} {\bibfnamefont {R.}~\bibnamefont {Liang}},
  \bibinfo {author} {\bibfnamefont {W.~N.}\ \bibnamefont {Hardy}}, \ and\
  \bibinfo {author} {\bibfnamefont {K.~A.}\ \bibnamefont {Moler}},\ }\bibfield
  {title} {\enquote {\bibinfo {title} {Magnetic force microscopy study of
  interlayer kinks in individual vortices in the underdoped cuprate
  superconductor {YBa$_2$Cu$_3$O$_{6+x}$}},}\ }\href {\doibase
  10.1103/PhysRevB.79.214530} {\bibfield  {journal} {\bibinfo  {journal} {Phys.
  Rev. B}\ }\textbf {\bibinfo {volume} {79}},\ \bibinfo {pages} {214530}
  (\bibinfo {year} {2009})}\BibitemShut {NoStop}%
\bibitem [{\citenamefont {Shapira}\ \emph {et~al.}(2015)\citenamefont
  {Shapira}, \citenamefont {Lamhot}, \citenamefont {Shpielberg}, \citenamefont
  {Kafri}, \citenamefont {Ramshaw}, \citenamefont {Bonn}, \citenamefont
  {Liang}, \citenamefont {Hardy},\ and\ \citenamefont
  {Auslaender}}]{Shapira15}%
  \BibitemOpen
  \bibfield  {author} {\bibinfo {author} {\bibfnamefont {N.}~\bibnamefont
  {Shapira}}, \bibinfo {author} {\bibfnamefont {Y.}~\bibnamefont {Lamhot}},
  \bibinfo {author} {\bibfnamefont {O.}~\bibnamefont {Shpielberg}}, \bibinfo
  {author} {\bibfnamefont {Y.}~\bibnamefont {Kafri}}, \bibinfo {author}
  {\bibfnamefont {B.~J.}\ \bibnamefont {Ramshaw}}, \bibinfo {author}
  {\bibfnamefont {D.~A.}\ \bibnamefont {Bonn}}, \bibinfo {author}
  {\bibfnamefont {R.}~\bibnamefont {Liang}}, \bibinfo {author} {\bibfnamefont
  {W.~N.}\ \bibnamefont {Hardy}}, \ and\ \bibinfo {author} {\bibfnamefont
  {O.~M.}\ \bibnamefont {Auslaender}},\ }\bibfield  {title} {\enquote {\bibinfo
  {title} {Disorder-induced power-law response of a superconducting vortex on a
  plane},}\ }\href {\doibase 10.1103/PhysRevB.92.100501} {\bibfield  {journal}
  {\bibinfo  {journal} {Phys. Rev. B}\ }\textbf {\bibinfo {volume} {92}},\
  \bibinfo {pages} {100501} (\bibinfo {year} {2015})}\BibitemShut {NoStop}%
\bibitem [{\citenamefont {Kremen}\ \emph {et~al.}(2016)\citenamefont {Kremen},
  \citenamefont {Wissberg}, \citenamefont {Haham}, \citenamefont {Persky},
  \citenamefont {Frenkel},\ and\ \citenamefont {Kalisky}}]{Kremen16}%
  \BibitemOpen
  \bibfield  {author} {\bibinfo {author} {\bibfnamefont {A.}~\bibnamefont
  {Kremen}}, \bibinfo {author} {\bibfnamefont {S.}~\bibnamefont {Wissberg}},
  \bibinfo {author} {\bibfnamefont {N.}~\bibnamefont {Haham}}, \bibinfo
  {author} {\bibfnamefont {E.}~\bibnamefont {Persky}}, \bibinfo {author}
  {\bibfnamefont {Y.}~\bibnamefont {Frenkel}}, \ and\ \bibinfo {author}
  {\bibfnamefont {B.}~\bibnamefont {Kalisky}},\ }\bibfield  {title} {\enquote
  {\bibinfo {title} {Mechanical control of individual superconducting
  vortices},}\ }\href {\doibase 10.1021/acs.nanolett.5b04444} {\bibfield
  {journal} {\bibinfo  {journal} {Nano Lett.}\ }\textbf {\bibinfo {volume}
  {16}},\ \bibinfo {pages} {1626--1630} (\bibinfo {year} {2016})}\BibitemShut
  {NoStop}%
\bibitem [{\citenamefont {Ge}\ \emph {et~al.}(2016)\citenamefont {Ge},
  \citenamefont {Gladilin}, \citenamefont {Tempere}, \citenamefont {Xue},
  \citenamefont {Devreese}, \citenamefont {Van~de Vondel}, \citenamefont
  {Zhou},\ and\ \citenamefont {Moshchalkov}}]{Ge16}%
  \BibitemOpen
  \bibfield  {author} {\bibinfo {author} {\bibfnamefont {J.-Y.}\ \bibnamefont
  {Ge}}, \bibinfo {author} {\bibfnamefont {V.~N.}\ \bibnamefont {Gladilin}},
  \bibinfo {author} {\bibfnamefont {J.}~\bibnamefont {Tempere}}, \bibinfo
  {author} {\bibfnamefont {C.}~\bibnamefont {Xue}}, \bibinfo {author}
  {\bibfnamefont {J.~T.}\ \bibnamefont {Devreese}}, \bibinfo {author}
  {\bibfnamefont {J.}~\bibnamefont {Van~de Vondel}}, \bibinfo {author}
  {\bibfnamefont {Y.}~\bibnamefont {Zhou}}, \ and\ \bibinfo {author}
  {\bibfnamefont {V.~V.}\ \bibnamefont {Moshchalkov}},\ }\bibfield  {title}
  {\enquote {\bibinfo {title} {Nanoscale assembly of superconducting vortices
  with scanning tunnelling microscope tip},}\ }\href {\doibase
  10.1038/ncomms13880} {\bibfield  {journal} {\bibinfo  {journal} {Nature
  Commun.}\ }\textbf {\bibinfo {volume} {7}},\ \bibinfo {pages} {13880}
  (\bibinfo {year} {2016})}\BibitemShut {NoStop}%
\bibitem [{\citenamefont {Veshchunov}\ \emph {et~al.}(2016)\citenamefont
  {Veshchunov}, \citenamefont {Magrini}, \citenamefont {Mironov}, \citenamefont
  {Godin}, \citenamefont {Trebbia}, \citenamefont {Buzdin}, \citenamefont
  {Tamarat},\ and\ \citenamefont {Lounis}}]{Veshchunov16}%
  \BibitemOpen
  \bibfield  {author} {\bibinfo {author} {\bibfnamefont {I.~S.}\ \bibnamefont
  {Veshchunov}}, \bibinfo {author} {\bibfnamefont {W.}~\bibnamefont {Magrini}},
  \bibinfo {author} {\bibfnamefont {S.~V.}\ \bibnamefont {Mironov}}, \bibinfo
  {author} {\bibfnamefont {A.~G.}\ \bibnamefont {Godin}}, \bibinfo {author}
  {\bibfnamefont {J.~B.}\ \bibnamefont {Trebbia}}, \bibinfo {author}
  {\bibfnamefont {A.~I.}\ \bibnamefont {Buzdin}}, \bibinfo {author}
  {\bibfnamefont {Ph.}\ \bibnamefont {Tamarat}}, \ and\ \bibinfo {author}
  {\bibfnamefont {B.}~\bibnamefont {Lounis}},\ }\bibfield  {title} {\enquote
  {\bibinfo {title} {Optical manipulation of single flux quanta},}\ }\href
  {\doibase 10.1038/ncomms12801} {\bibfield  {journal} {\bibinfo  {journal}
  {Nature Commun.}\ }\textbf {\bibinfo {volume} {7}},\ \bibinfo {pages} {12801}
  (\bibinfo {year} {2016})}\BibitemShut {NoStop}%
\bibitem [{\citenamefont {Olson~Reichhardt}\ and\ \citenamefont
  {Hastings}(2004)}]{Reichhardt04}%
  \BibitemOpen
  \bibfield  {author} {\bibinfo {author} {\bibfnamefont {C.~J.}\ \bibnamefont
  {Olson~Reichhardt}}\ and\ \bibinfo {author} {\bibfnamefont {M.~B.}\
  \bibnamefont {Hastings}},\ }\bibfield  {title} {\enquote {\bibinfo {title}
  {Do vortices entangle?}}\ }\href {\doibase 10.1103/PhysRevLett.92.157002}
  {\bibfield  {journal} {\bibinfo  {journal} {Phys. Rev. Lett.}\ }\textbf
  {\bibinfo {volume} {92}},\ \bibinfo {pages} {157002} (\bibinfo {year}
  {2004})}\BibitemShut {NoStop}%
\bibitem [{\citenamefont {Kafri}\ \emph {et~al.}(2006)\citenamefont {Kafri},
  \citenamefont {Nelson},\ and\ \citenamefont {Polkovnikov}}]{Kafri06}%
  \BibitemOpen
  \bibfield  {author} {\bibinfo {author} {\bibfnamefont {Y.}~\bibnamefont
  {Kafri}}, \bibinfo {author} {\bibfnamefont {D.~R.}\ \bibnamefont {Nelson}}, \
  and\ \bibinfo {author} {\bibfnamefont {A.}~\bibnamefont {Polkovnikov}},\
  }\bibfield  {title} {\enquote {\bibinfo {title} {Unzipping flux lines from
  extended defects in type-{II} superconductors},}\ }\href {\doibase
  10.1209/epl/i2005-10390-9} {\bibfield  {journal} {\bibinfo  {journal}
  {Europhys. Lett.}\ }\textbf {\bibinfo {volume} {73}},\ \bibinfo {pages}
  {253--259} (\bibinfo {year} {2006})}\BibitemShut {NoStop}%
\bibitem [{\citenamefont {Kafri}\ \emph {et~al.}(2007)\citenamefont {Kafri},
  \citenamefont {Nelson},\ and\ \citenamefont {Polkovnikov}}]{Kafri07}%
  \BibitemOpen
  \bibfield  {author} {\bibinfo {author} {\bibfnamefont {Y.}~\bibnamefont
  {Kafri}}, \bibinfo {author} {\bibfnamefont {D.~R.}\ \bibnamefont {Nelson}}, \
  and\ \bibinfo {author} {\bibfnamefont {A.}~\bibnamefont {Polkovnikov}},\
  }\bibfield  {title} {\enquote {\bibinfo {title} {Unzipping vortices in
  type-{II} superconductors},}\ }\href {\doibase 10.1103/PhysRevB.76.144501}
  {\bibfield  {journal} {\bibinfo  {journal} {Phys. Rev. B}\ }\textbf {\bibinfo
  {volume} {76}},\ \bibinfo {pages} {144501} (\bibinfo {year}
  {2007})}\BibitemShut {NoStop}%
\bibitem [{\citenamefont {Reichhardt}(2009)}]{Reichhardt09a}%
  \BibitemOpen
  \bibfield  {author} {\bibinfo {author} {\bibfnamefont {C.}~\bibnamefont
  {Reichhardt}},\ }\bibfield  {title} {\enquote {\bibinfo {title} {Vortices
  wiggled and dragged},}\ }\href {\doibase 10.1038/nphys1169} {\bibfield
  {journal} {\bibinfo  {journal} {Nature Phys.}\ }\textbf {\bibinfo {volume}
  {5}},\ \bibinfo {pages} {15--16} (\bibinfo {year} {2009})}\BibitemShut
  {NoStop}%
\bibitem [{\citenamefont {Olson~Reichhardt}\ and\ \citenamefont
  {Reichhardt}(2008)}]{Reichhardt08}%
  \BibitemOpen
  \bibfield  {author} {\bibinfo {author} {\bibfnamefont {C.~J.}\ \bibnamefont
  {Olson~Reichhardt}}\ and\ \bibinfo {author} {\bibfnamefont {C.}~\bibnamefont
  {Reichhardt}},\ }\bibfield  {title} {\enquote {\bibinfo {title} {Viscous
  decoupling transitions for individually dragged particles in systems with
  quenched disorder},}\ }\href {\doibase 10.1103/PhysRevE.78.011402} {\bibfield
   {journal} {\bibinfo  {journal} {Phys. Rev. E}\ }\textbf {\bibinfo {volume}
  {78}},\ \bibinfo {pages} {011402} (\bibinfo {year} {2008})}\BibitemShut
  {NoStop}%
\bibitem [{\citenamefont {Reichhardt}\ and\ \citenamefont
  {Reichhardt}(2010)}]{Reichhardt10c}%
  \BibitemOpen
  \bibfield  {author} {\bibinfo {author} {\bibfnamefont {C.~J.~Olson}\
  \bibnamefont {Reichhardt}}\ and\ \bibinfo {author} {\bibfnamefont
  {C.}~\bibnamefont {Reichhardt}},\ }\bibfield  {title} {\enquote {\bibinfo
  {title} {Driving an individual vortex in the presence of a periodic pinning
  array},}\ }\href {\doibase 10.1016/j.physc.2010.02.068} {\bibfield  {journal}
  {\bibinfo  {journal} {Physica C}\ }\textbf {\bibinfo {volume} {470}},\
  \bibinfo {pages} {779--781} (\bibinfo {year} {2010})}\BibitemShut {NoStop}%
\bibitem [{\citenamefont {Ma}\ \emph {et~al.}(2018)\citenamefont {Ma},
  \citenamefont {Reichhardt},\ and\ \citenamefont {Reichhardt}}]{Ma18}%
  \BibitemOpen
  \bibfield  {author} {\bibinfo {author} {\bibfnamefont {X.}~\bibnamefont
  {Ma}}, \bibinfo {author} {\bibfnamefont {C.~J.~O.}\ \bibnamefont
  {Reichhardt}}, \ and\ \bibinfo {author} {\bibfnamefont {C.}~\bibnamefont
  {Reichhardt}},\ }\bibfield  {title} {\enquote {\bibinfo {title} {Manipulation
  of individual superconducting vortices and stick-slip motion in periodic
  pinning arrays},}\ }\href {\doibase 10.1103/PhysRevB.97.214521} {\bibfield
  {journal} {\bibinfo  {journal} {Phys. Rev. B}\ }\textbf {\bibinfo {volume}
  {97}},\ \bibinfo {pages} {214521} (\bibinfo {year} {2018})}\BibitemShut
  {NoStop}%
\bibitem [{\citenamefont {Ge}\ \emph {et~al.}(2017)\citenamefont {Ge},
  \citenamefont {Gladilin}, \citenamefont {Tempere}, \citenamefont {Devreese},\
  and\ \citenamefont {Moshchalkov}}]{Ge17}%
  \BibitemOpen
  \bibfield  {author} {\bibinfo {author} {\bibfnamefont {J.-Y.}\ \bibnamefont
  {Ge}}, \bibinfo {author} {\bibfnamefont {V.~N.}\ \bibnamefont {Gladilin}},
  \bibinfo {author} {\bibfnamefont {J.}~\bibnamefont {Tempere}}, \bibinfo
  {author} {\bibfnamefont {J.}~\bibnamefont {Devreese}}, \ and\ \bibinfo
  {author} {\bibfnamefont {V.~V.}\ \bibnamefont {Moshchalkov}},\ }\bibfield
  {title} {\enquote {\bibinfo {title} {Controlled generation of quantized
  vortex-antivortex pairs in a superconducting condensate},}\ }\href {\doibase
  10.1021/acs.nanolett.7b02180} {\bibfield  {journal} {\bibinfo  {journal}
  {Nano Lett.}\ }\textbf {\bibinfo {volume} {17}},\ \bibinfo {pages}
  {5003--5007} (\bibinfo {year} {2017})}\BibitemShut {NoStop}%
\bibitem [{\citenamefont {Dremov}\ \emph {et~al.}(2019)\citenamefont {Dremov},
  \citenamefont {Grebenchuk}, \citenamefont {Shishkin}, \citenamefont
  {Baranov}, \citenamefont {Hovhannisyan}, \citenamefont {Skryabina},
  \citenamefont {Golovchanskiy}, \citenamefont {Chichkov}, \citenamefont
  {Brun}, \citenamefont {Cren}, \citenamefont {Krasnov}, \citenamefont
  {Golubov}, \citenamefont {Roditchev},\ and\ \citenamefont
  {Stolyarov}}]{Dremov19}%
  \BibitemOpen
  \bibfield  {author} {\bibinfo {author} {\bibfnamefont {V.~V.}\ \bibnamefont
  {Dremov}}, \bibinfo {author} {\bibfnamefont {S.~Yu.}\ \bibnamefont
  {Grebenchuk}}, \bibinfo {author} {\bibfnamefont {A.~G.}\ \bibnamefont
  {Shishkin}}, \bibinfo {author} {\bibfnamefont {D.~S.}\ \bibnamefont
  {Baranov}}, \bibinfo {author} {\bibfnamefont {R.~A.}\ \bibnamefont
  {Hovhannisyan}}, \bibinfo {author} {\bibfnamefont {O.~V.}\ \bibnamefont
  {Skryabina}}, \bibinfo {author} {\bibfnamefont {I.~A.}\ \bibnamefont
  {Golovchanskiy}}, \bibinfo {author} {\bibfnamefont {V.~I.}\ \bibnamefont
  {Chichkov}}, \bibinfo {author} {\bibfnamefont {C.}~\bibnamefont {Brun}},
  \bibinfo {author} {\bibfnamefont {T.}~\bibnamefont {Cren}}, \bibinfo {author}
  {\bibfnamefont {V.~M.}\ \bibnamefont {Krasnov}}, \bibinfo {author}
  {\bibfnamefont {A.~A.}\ \bibnamefont {Golubov}}, \bibinfo {author}
  {\bibfnamefont {D.}~\bibnamefont {Roditchev}}, \ and\ \bibinfo {author}
  {\bibfnamefont {V.~S.}\ \bibnamefont {Stolyarov}},\ }\bibfield  {title}
  {\enquote {\bibinfo {title} {Local {J}osephson vortex generation and
  manipulation with a magnetic force microscope},}\ }\href {\doibase
  10.1038/s41467-019-11924-0} {\bibfield  {journal} {\bibinfo  {journal}
  {Nature Commun.}\ }\textbf {\bibinfo {volume} {10}},\ \bibinfo {pages} {4009}
  (\bibinfo {year} {2019})}\BibitemShut {NoStop}%
\bibitem [{\citenamefont {Majorana}(1937)}]{Majorana37}%
  \BibitemOpen
  \bibfield  {author} {\bibinfo {author} {\bibfnamefont {E.}~\bibnamefont
  {Majorana}},\ }\bibfield  {title} {\enquote {\bibinfo {title} {Symmetrical
  theory of electrons and positrons},}\ }\href@noop {} {\bibfield  {journal}
  {\bibinfo  {journal} {Nuovo Cimento}\ }\textbf {\bibinfo {volume} {14}},\
  \bibinfo {pages} {171--184} (\bibinfo {year} {1937})}\BibitemShut {NoStop}%
\bibitem [{\citenamefont {Rokhinson}\ \emph {et~al.}(2012)\citenamefont
  {Rokhinson}, \citenamefont {Liu},\ and\ \citenamefont
  {Furdyna}}]{Rokhinson12}%
  \BibitemOpen
  \bibfield  {author} {\bibinfo {author} {\bibfnamefont {L.~P.}\ \bibnamefont
  {Rokhinson}}, \bibinfo {author} {\bibfnamefont {X.}~\bibnamefont {Liu}}, \
  and\ \bibinfo {author} {\bibfnamefont {J.~K.}\ \bibnamefont {Furdyna}},\
  }\bibfield  {title} {\enquote {\bibinfo {title} {The fractional a.c.
  josephson effect in a semiconductor-superconductor nanowire as a signature of
  {M}ajorana particles},}\ }\href {\doibase 10.1038/NPHYS2429} {\bibfield
  {journal} {\bibinfo  {journal} {Nature Phys.}\ }\textbf {\bibinfo {volume}
  {8}},\ \bibinfo {pages} {795--799} (\bibinfo {year} {2012})}\BibitemShut
  {NoStop}%
\bibitem [{\citenamefont {Mourik}\ \emph {et~al.}(2012)\citenamefont {Mourik},
  \citenamefont {Zuo}, \citenamefont {Frolov}, \citenamefont {Plissard},
  \citenamefont {Bakkers},\ and\ \citenamefont {Kouwenhoven}}]{Mourik12}%
  \BibitemOpen
  \bibfield  {author} {\bibinfo {author} {\bibfnamefont {V.}~\bibnamefont
  {Mourik}}, \bibinfo {author} {\bibfnamefont {K.}~\bibnamefont {Zuo}},
  \bibinfo {author} {\bibfnamefont {S.~M.}\ \bibnamefont {Frolov}}, \bibinfo
  {author} {\bibfnamefont {S.~R.}\ \bibnamefont {Plissard}}, \bibinfo {author}
  {\bibfnamefont {E.~P. A.~M.}\ \bibnamefont {Bakkers}}, \ and\ \bibinfo
  {author} {\bibfnamefont {L.~P.}\ \bibnamefont {Kouwenhoven}},\ }\bibfield
  {title} {\enquote {\bibinfo {title} {Signatures of {M}ajorana fermions in
  hybrid superconductor-semiconductor nanowire devices},}\ }\href {\doibase
  10.1126/science.1222360} {\bibfield  {journal} {\bibinfo  {journal}
  {Science}\ }\textbf {\bibinfo {volume} {336}},\ \bibinfo {pages} {1003--1007}
  (\bibinfo {year} {2012})}\BibitemShut {NoStop}%
\bibitem [{\citenamefont {Beenakker}(2013)}]{Beenakker13}%
  \BibitemOpen
  \bibfield  {author} {\bibinfo {author} {\bibfnamefont {C.~W.~J.}\
  \bibnamefont {Beenakker}},\ }\bibfield  {title} {\enquote {\bibinfo {title}
  {Search for {M}ajorana fermions in superconductors},}\ }in\ \href {\doibase
  10.1146/annurev-conmatphys-030212-184337} {\emph {\bibinfo {booktitle}
  {Annual Review of Condensed Matter Physics, Vol. 4}}},\ \bibinfo {series}
  {Annual Review of Condensed Matter Physics}, Vol.~\bibinfo {volume} {4},\
  \bibinfo {editor} {edited by\ \bibinfo {editor} {\bibfnamefont {J.~S.}\
  \bibnamefont {Langer}}}\ (\bibinfo {year} {2013})\ pp.\ \bibinfo {pages}
  {113--136}\BibitemShut {NoStop}%
\bibitem [{\citenamefont {Deng}\ \emph {et~al.}(2014)\citenamefont {Deng},
  \citenamefont {Yu}, \citenamefont {Huang}, \citenamefont {Larsson},
  \citenamefont {Caroff},\ and\ \citenamefont {Xu}}]{Deng14}%
  \BibitemOpen
  \bibfield  {author} {\bibinfo {author} {\bibfnamefont {M.~T.}\ \bibnamefont
  {Deng}}, \bibinfo {author} {\bibfnamefont {C.~L.}\ \bibnamefont {Yu}},
  \bibinfo {author} {\bibfnamefont {G.~Y.}\ \bibnamefont {Huang}}, \bibinfo
  {author} {\bibfnamefont {M.}~\bibnamefont {Larsson}}, \bibinfo {author}
  {\bibfnamefont {P.}~\bibnamefont {Caroff}}, \ and\ \bibinfo {author}
  {\bibfnamefont {H.~Q.}\ \bibnamefont {Xu}},\ }\bibfield  {title} {\enquote
  {\bibinfo {title} {Parity independence of the zero-bias conductance peak in a
  nanowire based topological superconductor-quantum dot hybrid device},}\
  }\href {\doibase 10.1038/srep07261} {\bibfield  {journal} {\bibinfo
  {journal} {Sci. Rep.}\ }\textbf {\bibinfo {volume} {4}},\ \bibinfo {pages}
  {7261} (\bibinfo {year} {2014})}\BibitemShut {NoStop}%
\bibitem [{\citenamefont {Nadj-Perge}\ \emph {et~al.}(2014)\citenamefont
  {Nadj-Perge}, \citenamefont {Drozdov}, \citenamefont {Li}, \citenamefont
  {Chen}, \citenamefont {Jeon}, \citenamefont {Seo}, \citenamefont {MacDonald},
  \citenamefont {Bernevig},\ and\ \citenamefont {Yazdani}}]{NadjPerge14}%
  \BibitemOpen
  \bibfield  {author} {\bibinfo {author} {\bibfnamefont {S.}~\bibnamefont
  {Nadj-Perge}}, \bibinfo {author} {\bibfnamefont {I.~K.}\ \bibnamefont
  {Drozdov}}, \bibinfo {author} {\bibfnamefont {J.}~\bibnamefont {Li}},
  \bibinfo {author} {\bibfnamefont {H.}~\bibnamefont {Chen}}, \bibinfo {author}
  {\bibfnamefont {S.}~\bibnamefont {Jeon}}, \bibinfo {author} {\bibfnamefont
  {J.}~\bibnamefont {Seo}}, \bibinfo {author} {\bibfnamefont {A.~H.}\
  \bibnamefont {MacDonald}}, \bibinfo {author} {\bibfnamefont {B.~A.}\
  \bibnamefont {Bernevig}}, \ and\ \bibinfo {author} {\bibfnamefont
  {A.}~\bibnamefont {Yazdani}},\ }\bibfield  {title} {\enquote {\bibinfo
  {title} {Observation of {M}ajorana fermions in ferromagnetic atomic chains on
  a superconductor},}\ }\href {\doibase 10.1126/science.1259327} {\bibfield
  {journal} {\bibinfo  {journal} {Science}\ }\textbf {\bibinfo {volume}
  {346}},\ \bibinfo {pages} {602--607} (\bibinfo {year} {2014})}\BibitemShut
  {NoStop}%
\bibitem [{\citenamefont {Banerjee}\ \emph {et~al.}(2016)\citenamefont
  {Banerjee}, \citenamefont {Bridges}, \citenamefont {Yan}, \citenamefont
  {Aczel}, \citenamefont {Li}, \citenamefont {Stone}, \citenamefont {Granroth},
  \citenamefont {Lumsden}, \citenamefont {Yiu}, \citenamefont {Knolle},
  \citenamefont {Bhattacharjee}, \citenamefont {Kovrizhin}, \citenamefont
  {Moessner}, \citenamefont {Tennant}, \citenamefont {Mandrus},\ and\
  \citenamefont {Nagler}}]{Banerjee16}%
  \BibitemOpen
  \bibfield  {author} {\bibinfo {author} {\bibfnamefont {A.}~\bibnamefont
  {Banerjee}}, \bibinfo {author} {\bibfnamefont {C.~A.}\ \bibnamefont
  {Bridges}}, \bibinfo {author} {\bibfnamefont {J.~Q.}\ \bibnamefont {Yan}},
  \bibinfo {author} {\bibfnamefont {A.~A.}\ \bibnamefont {Aczel}}, \bibinfo
  {author} {\bibfnamefont {L.}~\bibnamefont {Li}}, \bibinfo {author}
  {\bibfnamefont {M.~B.}\ \bibnamefont {Stone}}, \bibinfo {author}
  {\bibfnamefont {G.~E.}\ \bibnamefont {Granroth}}, \bibinfo {author}
  {\bibfnamefont {M.~D.}\ \bibnamefont {Lumsden}}, \bibinfo {author}
  {\bibfnamefont {Y.}~\bibnamefont {Yiu}}, \bibinfo {author} {\bibfnamefont
  {J.}~\bibnamefont {Knolle}}, \bibinfo {author} {\bibfnamefont
  {S.}~\bibnamefont {Bhattacharjee}}, \bibinfo {author} {\bibfnamefont {D.~L.}\
  \bibnamefont {Kovrizhin}}, \bibinfo {author} {\bibfnamefont {R.}~\bibnamefont
  {Moessner}}, \bibinfo {author} {\bibfnamefont {D.~A.}\ \bibnamefont
  {Tennant}}, \bibinfo {author} {\bibfnamefont {D.~G.}\ \bibnamefont
  {Mandrus}}, \ and\ \bibinfo {author} {\bibfnamefont {S.~E.}\ \bibnamefont
  {Nagler}},\ }\bibfield  {title} {\enquote {\bibinfo {title} {Proximate
  {K}itaev quantum spin liquid behaviour in a honeycomb magnet},}\ }\href
  {\doibase 10.1038/NMAT4604} {\bibfield  {journal} {\bibinfo  {journal}
  {Nature Mater.}\ }\textbf {\bibinfo {volume} {15}},\ \bibinfo {pages} {733}
  (\bibinfo {year} {2016})}\BibitemShut {NoStop}%
\bibitem [{\citenamefont {He}\ \emph {et~al.}(2017)\citenamefont {He},
  \citenamefont {Pan}, \citenamefont {Stern}, \citenamefont {Burks},
  \citenamefont {Che}, \citenamefont {Yin}, \citenamefont {Wang}, \citenamefont
  {Lian}, \citenamefont {Zhou}, \citenamefont {Choi}, \citenamefont {Murata},
  \citenamefont {Kou}, \citenamefont {Chen}, \citenamefont {Nie}, \citenamefont
  {Shao}, \citenamefont {Fan}, \citenamefont {Zhang}, \citenamefont {Liu},
  \citenamefont {Xia},\ and\ \citenamefont {Wang}}]{He17}%
  \BibitemOpen
  \bibfield  {author} {\bibinfo {author} {\bibfnamefont {Q.~L.}\ \bibnamefont
  {He}}, \bibinfo {author} {\bibfnamefont {L.}~\bibnamefont {Pan}}, \bibinfo
  {author} {\bibfnamefont {A.~L.}\ \bibnamefont {Stern}}, \bibinfo {author}
  {\bibfnamefont {E.~C.}\ \bibnamefont {Burks}}, \bibinfo {author}
  {\bibfnamefont {X.}~\bibnamefont {Che}}, \bibinfo {author} {\bibfnamefont
  {G.}~\bibnamefont {Yin}}, \bibinfo {author} {\bibfnamefont {J.}~\bibnamefont
  {Wang}}, \bibinfo {author} {\bibfnamefont {B.}~\bibnamefont {Lian}}, \bibinfo
  {author} {\bibfnamefont {Q.}~\bibnamefont {Zhou}}, \bibinfo {author}
  {\bibfnamefont {E.~S.}\ \bibnamefont {Choi}}, \bibinfo {author}
  {\bibfnamefont {K.}~\bibnamefont {Murata}}, \bibinfo {author} {\bibfnamefont
  {X.}~\bibnamefont {Kou}}, \bibinfo {author} {\bibfnamefont {Z.}~\bibnamefont
  {Chen}}, \bibinfo {author} {\bibfnamefont {T.}~\bibnamefont {Nie}}, \bibinfo
  {author} {\bibfnamefont {Q.}~\bibnamefont {Shao}}, \bibinfo {author}
  {\bibfnamefont {Y.}~\bibnamefont {Fan}}, \bibinfo {author} {\bibfnamefont
  {S.-C.}\ \bibnamefont {Zhang}}, \bibinfo {author} {\bibfnamefont
  {K.}~\bibnamefont {Liu}}, \bibinfo {author} {\bibfnamefont {J.}~\bibnamefont
  {Xia}}, \ and\ \bibinfo {author} {\bibfnamefont {K.~L.}\ \bibnamefont
  {Wang}},\ }\bibfield  {title} {\enquote {\bibinfo {title} {Chiral {M}ajorana
  fermion modes in a quantum anomalous {H}all insulator-superconductor
  structure},}\ }\href {\doibase 10.1126/science.aag2792} {\bibfield  {journal}
  {\bibinfo  {journal} {Science}\ }\textbf {\bibinfo {volume} {357}},\ \bibinfo
  {pages} {294--299} (\bibinfo {year} {2017})}\BibitemShut {NoStop}%
\bibitem [{\citenamefont {Wang}\ \emph
  {et~al.}(2018{\natexlab{a}})\citenamefont {Wang}, \citenamefont {Kong},
  \citenamefont {Fan}, \citenamefont {Chen}, \citenamefont {Zhu}, \citenamefont
  {Liu}, \citenamefont {Cao}, \citenamefont {Sun}, \citenamefont {Du},
  \citenamefont {Schneeloch}, \citenamefont {Zhong}, \citenamefont {Gu},
  \citenamefont {Fu}, \citenamefont {Ding},\ and\ \citenamefont
  {Gao}}]{Wang18}%
  \BibitemOpen
  \bibfield  {author} {\bibinfo {author} {\bibfnamefont {D.}~\bibnamefont
  {Wang}}, \bibinfo {author} {\bibfnamefont {L.}~\bibnamefont {Kong}}, \bibinfo
  {author} {\bibfnamefont {P.}~\bibnamefont {Fan}}, \bibinfo {author}
  {\bibfnamefont {H.}~\bibnamefont {Chen}}, \bibinfo {author} {\bibfnamefont
  {S.}~\bibnamefont {Zhu}}, \bibinfo {author} {\bibfnamefont {W.}~\bibnamefont
  {Liu}}, \bibinfo {author} {\bibfnamefont {L.}~\bibnamefont {Cao}}, \bibinfo
  {author} {\bibfnamefont {Y.}~\bibnamefont {Sun}}, \bibinfo {author}
  {\bibfnamefont {S.}~\bibnamefont {Du}}, \bibinfo {author} {\bibfnamefont
  {J.}~\bibnamefont {Schneeloch}}, \bibinfo {author} {\bibfnamefont
  {R.}~\bibnamefont {Zhong}}, \bibinfo {author} {\bibfnamefont
  {G.}~\bibnamefont {Gu}}, \bibinfo {author} {\bibfnamefont {L.}~\bibnamefont
  {Fu}}, \bibinfo {author} {\bibfnamefont {H.}~\bibnamefont {Ding}}, \ and\
  \bibinfo {author} {\bibfnamefont {H.-J.}\ \bibnamefont {Gao}},\ }\bibfield
  {title} {\enquote {\bibinfo {title} {Evidence for {M}ajorana bound states in
  an iron-based superconductor},}\ }\href {\doibase 10.1126/science.aao1797}
  {\bibfield  {journal} {\bibinfo  {journal} {Science}\ }\textbf {\bibinfo
  {volume} {362}},\ \bibinfo {pages} {333--335} (\bibinfo {year}
  {2018}{\natexlab{a}})}\BibitemShut {NoStop}%
\bibitem [{\citenamefont {Das}\ \emph {et~al.}(2012)\citenamefont {Das},
  \citenamefont {Ronen}, \citenamefont {Most}, \citenamefont {Oreg},
  \citenamefont {Heiblum},\ and\ \citenamefont {Shtrikman}}]{Das12}%
  \BibitemOpen
  \bibfield  {author} {\bibinfo {author} {\bibfnamefont {A.}~\bibnamefont
  {Das}}, \bibinfo {author} {\bibfnamefont {Y.}~\bibnamefont {Ronen}}, \bibinfo
  {author} {\bibfnamefont {Y.}~\bibnamefont {Most}}, \bibinfo {author}
  {\bibfnamefont {Y.}~\bibnamefont {Oreg}}, \bibinfo {author} {\bibfnamefont
  {M.}~\bibnamefont {Heiblum}}, \ and\ \bibinfo {author} {\bibfnamefont
  {H.}~\bibnamefont {Shtrikman}},\ }\bibfield  {title} {\enquote {\bibinfo
  {title} {Zero-bias peaks and splitting in an {Al-InAs} nanowire topological
  superconductor as a signature of {M}ajorana fermions},}\ }\href {\doibase
  10.1038/NPHYS2479} {\bibfield  {journal} {\bibinfo  {journal} {Nature Phys.}\
  }\textbf {\bibinfo {volume} {8}},\ \bibinfo {pages} {887--895} (\bibinfo
  {year} {2012})}\BibitemShut {NoStop}%
\bibitem [{\citenamefont {Das~Sarma}\ \emph {et~al.}(2015)\citenamefont
  {Das~Sarma}, \citenamefont {Freedman},\ and\ \citenamefont
  {Nayak}}]{DasSarma15}%
  \BibitemOpen
  \bibfield  {author} {\bibinfo {author} {\bibfnamefont {S.}~\bibnamefont
  {Das~Sarma}}, \bibinfo {author} {\bibfnamefont {M.}~\bibnamefont {Freedman}},
  \ and\ \bibinfo {author} {\bibfnamefont {C.}~\bibnamefont {Nayak}},\
  }\bibfield  {title} {\enquote {\bibinfo {title} {Majorana zero modes and
  topological quantum computation},}\ }\href {\doibase 10.1038/npjqi.2015.1}
  {\bibfield  {journal} {\bibinfo  {journal} {NPJ Quant. Inform.}\ }\textbf
  {\bibinfo {volume} {1}},\ \bibinfo {pages} {15001} (\bibinfo {year}
  {2015})}\BibitemShut {NoStop}%
\bibitem [{\citenamefont {Nayak}\ \emph {et~al.}(2008)\citenamefont {Nayak},
  \citenamefont {Simon}, \citenamefont {Stern}, \citenamefont {Freedman},\ and\
  \citenamefont {Das~Sarma}}]{Nayak08}%
  \BibitemOpen
  \bibfield  {author} {\bibinfo {author} {\bibfnamefont {C.}~\bibnamefont
  {Nayak}}, \bibinfo {author} {\bibfnamefont {S.~H.}\ \bibnamefont {Simon}},
  \bibinfo {author} {\bibfnamefont {A.}~\bibnamefont {Stern}}, \bibinfo
  {author} {\bibfnamefont {M.}~\bibnamefont {Freedman}}, \ and\ \bibinfo
  {author} {\bibfnamefont {S.}~\bibnamefont {Das~Sarma}},\ }\bibfield  {title}
  {\enquote {\bibinfo {title} {Non-{A}belian anyons and topological quantum
  computation},}\ }\href {\doibase 10.1103/RevModPhys.80.1083} {\bibfield
  {journal} {\bibinfo  {journal} {Rev. Mod. Phys.}\ }\textbf {\bibinfo {volume}
  {80}},\ \bibinfo {pages} {1083--1159} (\bibinfo {year} {2008})}\BibitemShut
  {NoStop}%
\bibitem [{\citenamefont {Stern}(2010)}]{Stern10}%
  \BibitemOpen
  \bibfield  {author} {\bibinfo {author} {\bibfnamefont {A.}~\bibnamefont
  {Stern}},\ }\bibfield  {title} {\enquote {\bibinfo {title} {Non-{A}belian
  states of matter},}\ }\href {\doibase 10.1038/nature08915} {\bibfield
  {journal} {\bibinfo  {journal} {Nature (London)}\ }\textbf {\bibinfo {volume}
  {464}},\ \bibinfo {pages} {187--193} (\bibinfo {year} {2010})}\BibitemShut
  {NoStop}%
\bibitem [{\citenamefont {Leijnse}\ and\ \citenamefont
  {Flensberg}(2012)}]{Leijnse12}%
  \BibitemOpen
  \bibfield  {author} {\bibinfo {author} {\bibfnamefont {M.}~\bibnamefont
  {Leijnse}}\ and\ \bibinfo {author} {\bibfnamefont {K.}~\bibnamefont
  {Flensberg}},\ }\bibfield  {title} {\enquote {\bibinfo {title} {Introduction
  to topological superconductivity and {M}ajorana fermions},}\ }\href {\doibase
  10.1088/0268-1242/27/12/124003} {\bibfield  {journal} {\bibinfo  {journal}
  {Semicond. Sci. Technol.}\ }\textbf {\bibinfo {volume} {27}},\ \bibinfo
  {pages} {124003} (\bibinfo {year} {2012})}\BibitemShut {NoStop}%
\bibitem [{\citenamefont {Lutchyn}\ \emph {et~al.}(2018)\citenamefont
  {Lutchyn}, \citenamefont {Bakkers}, \citenamefont {Kouwenhoven},
  \citenamefont {Krogstrup}, \citenamefont {Marcus},\ and\ \citenamefont
  {Oreg}}]{Lutchyn18}%
  \BibitemOpen
  \bibfield  {author} {\bibinfo {author} {\bibfnamefont {R.~M.}\ \bibnamefont
  {Lutchyn}}, \bibinfo {author} {\bibfnamefont {E.~P. A.~M.}\ \bibnamefont
  {Bakkers}}, \bibinfo {author} {\bibfnamefont {L.~P.}\ \bibnamefont
  {Kouwenhoven}}, \bibinfo {author} {\bibfnamefont {P.}~\bibnamefont
  {Krogstrup}}, \bibinfo {author} {\bibfnamefont {C.~M.}\ \bibnamefont
  {Marcus}}, \ and\ \bibinfo {author} {\bibfnamefont {Y.}~\bibnamefont
  {Oreg}},\ }\bibfield  {title} {\enquote {\bibinfo {title} {Majorana zero
  modes in superconductor-semiconductor heterostructures},}\ }\href {\doibase
  10.1038/s41578-018-0003-1} {\bibfield  {journal} {\bibinfo  {journal} {Nature
  Rev. Mater.}\ }\textbf {\bibinfo {volume} {3}},\ \bibinfo {pages} {52--68}
  (\bibinfo {year} {2018})}\BibitemShut {NoStop}%
\bibitem [{\citenamefont {Freedman}\ \emph {et~al.}(2006)\citenamefont
  {Freedman}, \citenamefont {Nayak},\ and\ \citenamefont
  {Walker}}]{Freedman06}%
  \BibitemOpen
  \bibfield  {author} {\bibinfo {author} {\bibfnamefont {M.}~\bibnamefont
  {Freedman}}, \bibinfo {author} {\bibfnamefont {C.}~\bibnamefont {Nayak}}, \
  and\ \bibinfo {author} {\bibfnamefont {K.}~\bibnamefont {Walker}},\
  }\bibfield  {title} {\enquote {\bibinfo {title} {Towards universal
  topological quantum computation in the $\nu=\frac{5}{2}$ fractional quantum
  {H}all state},}\ }\href {\doibase 10.1103/PhysRevB.73.245307} {\bibfield
  {journal} {\bibinfo  {journal} {Phys. Rev. B}\ }\textbf {\bibinfo {volume}
  {73}},\ \bibinfo {pages} {245307} (\bibinfo {year} {2006})}\BibitemShut
  {NoStop}%
\bibitem [{\citenamefont {Moore}\ and\ \citenamefont {Read}(1991)}]{Moore91}%
  \BibitemOpen
  \bibfield  {author} {\bibinfo {author} {\bibfnamefont {G.}~\bibnamefont
  {Moore}}\ and\ \bibinfo {author} {\bibfnamefont {N.}~\bibnamefont {Read}},\
  }\bibfield  {title} {\enquote {\bibinfo {title} {Nonabelions in the
  fractional quantum {H}all-effect},}\ }\href {\doibase
  10.1016/0550-3213(91)90407-O} {\bibfield  {journal} {\bibinfo  {journal}
  {Nucl. Phys. B}\ }\textbf {\bibinfo {volume} {360}},\ \bibinfo {pages}
  {362--396} (\bibinfo {year} {1991})}\BibitemShut {NoStop}%
\bibitem [{\citenamefont {Read}\ and\ \citenamefont {Green}(2000)}]{Read00}%
  \BibitemOpen
  \bibfield  {author} {\bibinfo {author} {\bibfnamefont {N.}~\bibnamefont
  {Read}}\ and\ \bibinfo {author} {\bibfnamefont {D.}~\bibnamefont {Green}},\
  }\bibfield  {title} {\enquote {\bibinfo {title} {Paired states of fermions in
  two dimensions with breaking of parity and time-reversal symmetries and the
  fractional quantum {H}all effect},}\ }\href {\doibase
  10.1103/PhysRevB.61.10267} {\bibfield  {journal} {\bibinfo  {journal} {Phys.
  Rev. B}\ }\textbf {\bibinfo {volume} {61}},\ \bibinfo {pages} {10267--10297}
  (\bibinfo {year} {2000})}\BibitemShut {NoStop}%
\bibitem [{\citenamefont {Kitaev}(2001)}]{Kitaev01}%
  \BibitemOpen
  \bibfield  {author} {\bibinfo {author} {\bibfnamefont {A.~Yu.}\ \bibnamefont
  {Kitaev}},\ }\bibfield  {title} {\enquote {\bibinfo {title} {Unpaired
  {M}ajorana fermions in quantum wires},}\ }\href {\doibase
  10.1070/1063-7869/44/10s/s29} {\bibfield  {journal} {\bibinfo  {journal}
  {Phys.-Usp.}\ }\textbf {\bibinfo {volume} {44}},\ \bibinfo {pages} {131--136}
  (\bibinfo {year} {2001})}\BibitemShut {NoStop}%
\bibitem [{\citenamefont {Volovik}(2003)}]{Volovik03}%
  \BibitemOpen
  \bibfield  {author} {\bibinfo {author} {\bibfnamefont {G.~E.}\ \bibnamefont
  {Volovik}},\ }\href@noop {} {\emph {\bibinfo {title} {The Universe in a
  Helium Droplet}}}\ (\bibinfo  {publisher} {Oxford University Press, Oxford},\
  \bibinfo {year} {2003})\BibitemShut {NoStop}%
\bibitem [{\citenamefont {Alicea}(2012)}]{Alicea12}%
  \BibitemOpen
  \bibfield  {author} {\bibinfo {author} {\bibfnamefont {J.}~\bibnamefont
  {Alicea}},\ }\bibfield  {title} {\enquote {\bibinfo {title} {New directions
  in the pursuit of {M}ajorana fermions in solid state systems},}\ }\href
  {\doibase 10.1088/0034-4885/75/7/076501} {\bibfield  {journal} {\bibinfo
  {journal} {Rep. Prog. Phys.}\ }\textbf {\bibinfo {volume} {75}},\ \bibinfo
  {pages} {076501} (\bibinfo {year} {2012})}\BibitemShut {NoStop}%
\bibitem [{\citenamefont {Stone}\ and\ \citenamefont {Chung}(2006)}]{Stone06}%
  \BibitemOpen
  \bibfield  {author} {\bibinfo {author} {\bibfnamefont {M.}~\bibnamefont
  {Stone}}\ and\ \bibinfo {author} {\bibfnamefont {S.-B.}\ \bibnamefont
  {Chung}},\ }\bibfield  {title} {\enquote {\bibinfo {title} {Fusion rules and
  vortices in ${p}_{x}+i{p}_{y}$ superconductors},}\ }\href {\doibase
  10.1103/PhysRevB.73.014505} {\bibfield  {journal} {\bibinfo  {journal} {Phys.
  Rev. B}\ }\textbf {\bibinfo {volume} {73}},\ \bibinfo {pages} {014505}
  (\bibinfo {year} {2006})}\BibitemShut {NoStop}%
\bibitem [{\citenamefont {Ivanov}(2001)}]{Ivanov01}%
  \BibitemOpen
  \bibfield  {author} {\bibinfo {author} {\bibfnamefont {D.~A.}\ \bibnamefont
  {Ivanov}},\ }\bibfield  {title} {\enquote {\bibinfo {title} {Non-{A}belian
  statistics of half-quantum vortices in $\mathit{p}$-wave superconductors},}\
  }\href {\doibase 10.1103/PhysRevLett.86.268} {\bibfield  {journal} {\bibinfo
  {journal} {Phys. Rev. Lett.}\ }\textbf {\bibinfo {volume} {86}},\ \bibinfo
  {pages} {268--271} (\bibinfo {year} {2001})}\BibitemShut {NoStop}%
\bibitem [{\citenamefont {Fu}\ and\ \citenamefont {Kane}(2008)}]{Fu08}%
  \BibitemOpen
  \bibfield  {author} {\bibinfo {author} {\bibfnamefont {L.}~\bibnamefont
  {Fu}}\ and\ \bibinfo {author} {\bibfnamefont {C.~L.}\ \bibnamefont {Kane}},\
  }\bibfield  {title} {\enquote {\bibinfo {title} {Superconducting proximity
  effect and {M}ajorana fermions at the surface of a topological insulator},}\
  }\href {\doibase 10.1103/PhysRevLett.100.096407} {\bibfield  {journal}
  {\bibinfo  {journal} {Phys. Rev. Lett.}\ }\textbf {\bibinfo {volume} {100}},\
  \bibinfo {pages} {096407} (\bibinfo {year} {2008})}\BibitemShut {NoStop}%
\bibitem [{\citenamefont {Sun}\ \emph {et~al.}(2016)\citenamefont {Sun},
  \citenamefont {Zhang}, \citenamefont {Hu}, \citenamefont {Li}, \citenamefont
  {Wang}, \citenamefont {Ma}, \citenamefont {Xu}, \citenamefont {Gao},
  \citenamefont {Guan}, \citenamefont {Li}, \citenamefont {Liu}, \citenamefont
  {Qian}, \citenamefont {Zhou}, \citenamefont {Fu}, \citenamefont {Li},
  \citenamefont {Zhang},\ and\ \citenamefont {Jia}}]{Sun16}%
  \BibitemOpen
  \bibfield  {author} {\bibinfo {author} {\bibfnamefont {H.-H.}\ \bibnamefont
  {Sun}}, \bibinfo {author} {\bibfnamefont {K.-W.}\ \bibnamefont {Zhang}},
  \bibinfo {author} {\bibfnamefont {L.-H.}\ \bibnamefont {Hu}}, \bibinfo
  {author} {\bibfnamefont {C.}~\bibnamefont {Li}}, \bibinfo {author}
  {\bibfnamefont {G.-Y.}\ \bibnamefont {Wang}}, \bibinfo {author}
  {\bibfnamefont {H.-Y.}\ \bibnamefont {Ma}}, \bibinfo {author} {\bibfnamefont
  {Z.-A.}\ \bibnamefont {Xu}}, \bibinfo {author} {\bibfnamefont {C.-L.}\
  \bibnamefont {Gao}}, \bibinfo {author} {\bibfnamefont {D.-D.}\ \bibnamefont
  {Guan}}, \bibinfo {author} {\bibfnamefont {Y.-Y.}\ \bibnamefont {Li}},
  \bibinfo {author} {\bibfnamefont {C.}~\bibnamefont {Liu}}, \bibinfo {author}
  {\bibfnamefont {D.}~\bibnamefont {Qian}}, \bibinfo {author} {\bibfnamefont
  {Y.}~\bibnamefont {Zhou}}, \bibinfo {author} {\bibfnamefont {L.}~\bibnamefont
  {Fu}}, \bibinfo {author} {\bibfnamefont {S.-C.}\ \bibnamefont {Li}}, \bibinfo
  {author} {\bibfnamefont {F.-C.}\ \bibnamefont {Zhang}}, \ and\ \bibinfo
  {author} {\bibfnamefont {J.-F.}\ \bibnamefont {Jia}},\ }\bibfield  {title}
  {\enquote {\bibinfo {title} {Majorana zero mode detected with spin selective
  {A}ndreev reflection in the vortex of a topological superconductor},}\ }\href
  {\doibase 10.1103/PhysRevLett.116.257003} {\bibfield  {journal} {\bibinfo
  {journal} {Phys. Rev. Lett.}\ }\textbf {\bibinfo {volume} {116}},\ \bibinfo
  {pages} {257003} (\bibinfo {year} {2016})}\BibitemShut {NoStop}%
\bibitem [{\citenamefont {Sun}\ and\ \citenamefont {Jia}(2017)}]{Sun17}%
  \BibitemOpen
  \bibfield  {author} {\bibinfo {author} {\bibfnamefont {H.-H.}\ \bibnamefont
  {Sun}}\ and\ \bibinfo {author} {\bibfnamefont {J.-F.}\ \bibnamefont {Jia}},\
  }\bibfield  {title} {\enquote {\bibinfo {title} {Detection of {M}ajorana zero
  mode in the vortex},}\ }\href {\doibase 10.1038/s41535-017-0037-4} {\bibfield
   {journal} {\bibinfo  {journal} {NPJ Quantum Mater.}\ }\textbf {\bibinfo
  {volume} {2}},\ \bibinfo {pages} {34} (\bibinfo {year} {2017})}\BibitemShut
  {NoStop}%
\bibitem [{\citenamefont {Wu}\ and\ \citenamefont {Zhou}(2017)}]{Wu17}%
  \BibitemOpen
  \bibfield  {author} {\bibinfo {author} {\bibfnamefont {H.-D.}\ \bibnamefont
  {Wu}}\ and\ \bibinfo {author} {\bibfnamefont {T.}~\bibnamefont {Zhou}},\
  }\bibfield  {title} {\enquote {\bibinfo {title} {Vortex pinning by the point
  potential in topological superconductors: A scheme for braiding {M}ajorana
  bound states},}\ }\href {\doibase 10.1103/PhysRevB.96.184508} {\bibfield
  {journal} {\bibinfo  {journal} {Phys. Rev. B}\ }\textbf {\bibinfo {volume}
  {96}},\ \bibinfo {pages} {184508} (\bibinfo {year} {2017})}\BibitemShut
  {NoStop}%
\bibitem [{\citenamefont {November}\ \emph {et~al.}(2019)\citenamefont
  {November}, \citenamefont {Sau}, \citenamefont {Williams},\ and\
  \citenamefont {Hoffman}}]{November19}%
  \BibitemOpen
  \bibfield  {author} {\bibinfo {author} {\bibfnamefont {B.~H.}\ \bibnamefont
  {November}}, \bibinfo {author} {\bibfnamefont {J.~D.}\ \bibnamefont {Sau}},
  \bibinfo {author} {\bibfnamefont {J.~R.}\ \bibnamefont {Williams}}, \ and\
  \bibinfo {author} {\bibfnamefont {J.~E.}\ \bibnamefont {Hoffman}},\
  }\href@noop {} {\enquote {\bibinfo {title} {Scheme for {M}ajorana
  manipulation using magnetic force microscopy},}\ } (\bibinfo {year} {2019}),\
  \Eprint {http://arxiv.org/abs/1905.09792} {arXiv:1905.09792} \BibitemShut
  {NoStop}%
\bibitem [{\citenamefont {Polshyn}\ \emph {et~al.}(2019)\citenamefont
  {Polshyn}, \citenamefont {Naibert},\ and\ \citenamefont
  {Budakian}}]{Polshyn19}%
  \BibitemOpen
  \bibfield  {author} {\bibinfo {author} {\bibfnamefont {H.}~\bibnamefont
  {Polshyn}}, \bibinfo {author} {\bibfnamefont {T.}~\bibnamefont {Naibert}}, \
  and\ \bibinfo {author} {\bibfnamefont {R.}~\bibnamefont {Budakian}},\
  }\bibfield  {title} {\enquote {\bibinfo {title} {Manipulating multivortex
  states in superconducting structures},}\ }\href {\doibase
  10.1021/acs.nanolett.9b01983} {\bibfield  {journal} {\bibinfo  {journal}
  {Nano Lett.}\ }\textbf {\bibinfo {volume} {19}},\ \bibinfo {pages}
  {5476--5482} (\bibinfo {year} {2019})}\BibitemShut {NoStop}%
\bibitem [{\citenamefont {Posske}\ \emph {et~al.}(2019)\citenamefont {Posske},
  \citenamefont {Chiu},\ and\ \citenamefont {Thorwart}}]{Posske19}%
  \BibitemOpen
  \bibfield  {author} {\bibinfo {author} {\bibfnamefont {T.}~\bibnamefont
  {Posske}}, \bibinfo {author} {\bibfnamefont {C.-K.}\ \bibnamefont {Chiu}}, \
  and\ \bibinfo {author} {\bibfnamefont {M.}~\bibnamefont {Thorwart}},\
  }\href@noop {} {\enquote {\bibinfo {title} {Robustly emulating vortex
  {M}ajorana braiding in a finite time},}\ } (\bibinfo {year} {2019}),\ \Eprint
  {http://arxiv.org/abs/1908.03576} {arXiv:1908.03576} \BibitemShut {NoStop}%
\bibitem [{\citenamefont {Liang}\ \emph {et~al.}(2012)\citenamefont {Liang},
  \citenamefont {Wang},\ and\ \citenamefont {Hu}}]{Liang12}%
  \BibitemOpen
  \bibfield  {author} {\bibinfo {author} {\bibfnamefont {Q.-F.}\ \bibnamefont
  {Liang}}, \bibinfo {author} {\bibfnamefont {Z.}~\bibnamefont {Wang}}, \ and\
  \bibinfo {author} {\bibfnamefont {X.}~\bibnamefont {Hu}},\ }\bibfield
  {title} {\enquote {\bibinfo {title} {Manipulation of {M}ajorana fermions by
  point-like gate voltage in the vortex state of a topological
  superconductor},}\ }\href {\doibase 10.1209/0295-5075/99/50004} {\bibfield
  {journal} {\bibinfo  {journal} {EPL}\ }\textbf {\bibinfo {volume} {99}},\
  \bibinfo {pages} {50004} (\bibinfo {year} {2012})}\BibitemShut {NoStop}%
\bibitem [{\citenamefont {Beenakker}(2019)}]{Beenakker19}%
  \BibitemOpen
  \bibfield  {author} {\bibinfo {author} {\bibfnamefont {C.~W.~J.}\
  \bibnamefont {Beenakker}},\ }\href@noop {} {\enquote {\bibinfo {title}
  {Search for non-{A}belian {M}ajorana braiding statistics in
  superconductors},}\ } (\bibinfo {year} {2019}),\ \Eprint
  {http://arxiv.org/abs/1907.06497} {arXiv:1907.06497} \BibitemShut {NoStop}%
\bibitem [{\citenamefont {Stern}\ and\ \citenamefont {Berg}(2019)}]{Stern19}%
  \BibitemOpen
  \bibfield  {author} {\bibinfo {author} {\bibfnamefont {A.}~\bibnamefont
  {Stern}}\ and\ \bibinfo {author} {\bibfnamefont {E.}~\bibnamefont {Berg}},\
  }\bibfield  {title} {\enquote {\bibinfo {title} {Fractional {J}osephson
  vortices and braiding of {M}ajorana zero modes in planar
  superconductor-semiconductor heterostructures},}\ }\href {\doibase
  10.1103/PhysRevLett.122.107701} {\bibfield  {journal} {\bibinfo  {journal}
  {Phys. Rev. Lett.}\ }\textbf {\bibinfo {volume} {122}},\ \bibinfo {pages}
  {107701} (\bibinfo {year} {2019})}\BibitemShut {NoStop}%
\bibitem [{\citenamefont {Beenakker}\ \emph {et~al.}(2019)\citenamefont
  {Beenakker}, \citenamefont {Baireuther}, \citenamefont {Herasymenko},
  \citenamefont {Adagideli}, \citenamefont {Wang},\ and\ \citenamefont
  {Akhmerov}}]{Beenakker19a}%
  \BibitemOpen
  \bibfield  {author} {\bibinfo {author} {\bibfnamefont {C.~W.~J.}\
  \bibnamefont {Beenakker}}, \bibinfo {author} {\bibfnamefont {P.}~\bibnamefont
  {Baireuther}}, \bibinfo {author} {\bibfnamefont {Y.}~\bibnamefont
  {Herasymenko}}, \bibinfo {author} {\bibfnamefont {I.}~\bibnamefont
  {Adagideli}}, \bibinfo {author} {\bibfnamefont {L.}~\bibnamefont {Wang}}, \
  and\ \bibinfo {author} {\bibfnamefont {A.~R.}\ \bibnamefont {Akhmerov}},\
  }\bibfield  {title} {\enquote {\bibinfo {title} {Deterministic creation and
  braiding of chiral edge vortices},}\ }\href {\doibase
  10.1103/PhysRevLett.122.146803} {\bibfield  {journal} {\bibinfo  {journal}
  {Phys. Rev. Lett.}\ }\textbf {\bibinfo {volume} {122}},\ \bibinfo {pages}
  {146803} (\bibinfo {year} {2019})}\BibitemShut {NoStop}%
\bibitem [{\citenamefont {R{\o}ising}\ \emph {et~al.}(2019)\citenamefont
  {R{\o}ising}, \citenamefont {Ilan}, \citenamefont {Meng}, \citenamefont
  {Simon},\ and\ \citenamefont {Flicker}}]{Roising19}%
  \BibitemOpen
  \bibfield  {author} {\bibinfo {author} {\bibfnamefont {H.~S.}\ \bibnamefont
  {R{\o}ising}}, \bibinfo {author} {\bibfnamefont {R.}~\bibnamefont {Ilan}},
  \bibinfo {author} {\bibfnamefont {T.}~\bibnamefont {Meng}}, \bibinfo {author}
  {\bibfnamefont {S.~H.}\ \bibnamefont {Simon}}, \ and\ \bibinfo {author}
  {\bibfnamefont {F.}~\bibnamefont {Flicker}},\ }\bibfield  {title} {\enquote
  {\bibinfo {title} {Finite temperature effects on {M}ajorana bound states in
  chiral $p$-wave superconductors},}\ }\href {\doibase
  10.21468/SciPostPhys.6.5.055} {\bibfield  {journal} {\bibinfo  {journal}
  {SciPost Phys.}\ }\textbf {\bibinfo {volume} {6}},\ \bibinfo {pages} {055}
  (\bibinfo {year} {2019})}\BibitemShut {NoStop}%
\bibitem [{\citenamefont {Baert}\ \emph {et~al.}(1995)\citenamefont {Baert},
  \citenamefont {Metlushko}, \citenamefont {Jonckheere}, \citenamefont
  {Moshchalkov},\ and\ \citenamefont {Bruynseraede}}]{Baert95}%
  \BibitemOpen
  \bibfield  {author} {\bibinfo {author} {\bibfnamefont {M.}~\bibnamefont
  {Baert}}, \bibinfo {author} {\bibfnamefont {V.~V.}\ \bibnamefont
  {Metlushko}}, \bibinfo {author} {\bibfnamefont {R.}~\bibnamefont
  {Jonckheere}}, \bibinfo {author} {\bibfnamefont {V.~V.}\ \bibnamefont
  {Moshchalkov}}, \ and\ \bibinfo {author} {\bibfnamefont {Y.}~\bibnamefont
  {Bruynseraede}},\ }\bibfield  {title} {\enquote {\bibinfo {title} {Composite
  flux-line lattices stabilized in superconducting films by a regular array of
  artificial defects},}\ }\href {\doibase 10.1103/PhysRevLett.74.3269}
  {\bibfield  {journal} {\bibinfo  {journal} {Phys. Rev. Lett.}\ }\textbf
  {\bibinfo {volume} {74}},\ \bibinfo {pages} {3269--3272} (\bibinfo {year}
  {1995})}\BibitemShut {NoStop}%
\bibitem [{\citenamefont {Harada}\ \emph {et~al.}(1996)\citenamefont {Harada},
  \citenamefont {Kamimura}, \citenamefont {Kasai}, \citenamefont {Matsuda},
  \citenamefont {Tonomura},\ and\ \citenamefont {Moshchalkov}}]{Harada96}%
  \BibitemOpen
  \bibfield  {author} {\bibinfo {author} {\bibfnamefont {K.}~\bibnamefont
  {Harada}}, \bibinfo {author} {\bibfnamefont {O.}~\bibnamefont {Kamimura}},
  \bibinfo {author} {\bibfnamefont {H.}~\bibnamefont {Kasai}}, \bibinfo
  {author} {\bibfnamefont {T.}~\bibnamefont {Matsuda}}, \bibinfo {author}
  {\bibfnamefont {A.}~\bibnamefont {Tonomura}}, \ and\ \bibinfo {author}
  {\bibfnamefont {V.~V.}\ \bibnamefont {Moshchalkov}},\ }\bibfield  {title}
  {\enquote {\bibinfo {title} {Direct observation of vortex dynamics in
  superconducting films with regular arrays of defects},}\ }\href {\doibase
  10.1126/science.274.5290.1167} {\bibfield  {journal} {\bibinfo  {journal}
  {Science}\ }\textbf {\bibinfo {volume} {274}},\ \bibinfo {pages} {1167--1170}
  (\bibinfo {year} {1996})}\BibitemShut {NoStop}%
\bibitem [{\citenamefont {Mart\'{\i}n}\ \emph {et~al.}(1997)\citenamefont
  {Mart\'{\i}n}, \citenamefont {V\'elez}, \citenamefont {Nogu\'es},\ and\
  \citenamefont {Schuller}}]{Martin97}%
  \BibitemOpen
  \bibfield  {author} {\bibinfo {author} {\bibfnamefont {J.~I.}\ \bibnamefont
  {Mart\'{\i}n}}, \bibinfo {author} {\bibfnamefont {M.}~\bibnamefont
  {V\'elez}}, \bibinfo {author} {\bibfnamefont {J.}~\bibnamefont {Nogu\'es}}, \
  and\ \bibinfo {author} {\bibfnamefont {I.~K.}\ \bibnamefont {Schuller}},\
  }\bibfield  {title} {\enquote {\bibinfo {title} {Flux pinning in a
  superconductor by an array of submicrometer magnetic dots},}\ }\href
  {\doibase 10.1103/PhysRevLett.79.1929} {\bibfield  {journal} {\bibinfo
  {journal} {Phys. Rev. Lett.}\ }\textbf {\bibinfo {volume} {79}},\ \bibinfo
  {pages} {1929--1932} (\bibinfo {year} {1997})}\BibitemShut {NoStop}%
\bibitem [{\citenamefont {Berdiyorov}\ \emph {et~al.}(2006)\citenamefont
  {Berdiyorov}, \citenamefont {Milo\ifmmode \check{s}\else
  \v{s}\fi{}evi\ifmmode~\acute{c}\else \'{c}\fi{}},\ and\ \citenamefont
  {Peeters}}]{Berdiyorov06}%
  \BibitemOpen
  \bibfield  {author} {\bibinfo {author} {\bibfnamefont {G.~R.}\ \bibnamefont
  {Berdiyorov}}, \bibinfo {author} {\bibfnamefont {M.~V.}\ \bibnamefont
  {Milo\ifmmode \check{s}\else \v{s}\fi{}evi\ifmmode~\acute{c}\else
  \'{c}\fi{}}}, \ and\ \bibinfo {author} {\bibfnamefont {F.~M.}\ \bibnamefont
  {Peeters}},\ }\bibfield  {title} {\enquote {\bibinfo {title} {Novel
  commensurability effects in superconducting films with antidot arrays},}\
  }\href {\doibase 10.1103/PhysRevLett.96.207001} {\bibfield  {journal}
  {\bibinfo  {journal} {Phys. Rev. Lett.}\ }\textbf {\bibinfo {volume} {96}},\
  \bibinfo {pages} {207001} (\bibinfo {year} {2006})}\BibitemShut {NoStop}%
\bibitem [{\citenamefont {Kemmler}\ \emph {et~al.}(2006)\citenamefont
  {Kemmler}, \citenamefont {G\"urlich}, \citenamefont {Sterck}, \citenamefont
  {P\"ohler}, \citenamefont {Neuhaus}, \citenamefont {Siegel}, \citenamefont
  {Kleiner},\ and\ \citenamefont {Koelle}}]{Kemmler06}%
  \BibitemOpen
  \bibfield  {author} {\bibinfo {author} {\bibfnamefont {M.}~\bibnamefont
  {Kemmler}}, \bibinfo {author} {\bibfnamefont {C.}~\bibnamefont {G\"urlich}},
  \bibinfo {author} {\bibfnamefont {A.}~\bibnamefont {Sterck}}, \bibinfo
  {author} {\bibfnamefont {H.}~\bibnamefont {P\"ohler}}, \bibinfo {author}
  {\bibfnamefont {M.}~\bibnamefont {Neuhaus}}, \bibinfo {author} {\bibfnamefont
  {M.}~\bibnamefont {Siegel}}, \bibinfo {author} {\bibfnamefont
  {R.}~\bibnamefont {Kleiner}}, \ and\ \bibinfo {author} {\bibfnamefont
  {D.}~\bibnamefont {Koelle}},\ }\bibfield  {title} {\enquote {\bibinfo {title}
  {Commensurability effects in superconducting {Nb} films with quasiperiodic
  pinning arrays},}\ }\href {\doibase 10.1103/PhysRevLett.97.147003} {\bibfield
   {journal} {\bibinfo  {journal} {Phys. Rev. Lett.}\ }\textbf {\bibinfo
  {volume} {97}},\ \bibinfo {pages} {147003} (\bibinfo {year}
  {2006})}\BibitemShut {NoStop}%
\bibitem [{\citenamefont {de~Souza~Silva}\ \emph {et~al.}(2007)\citenamefont
  {de~Souza~Silva}, \citenamefont {Silhanek}, \citenamefont {Van~de Vondel},
  \citenamefont {Gillijns}, \citenamefont {Metlushko}, \citenamefont {Ilic},\
  and\ \citenamefont {Moshchalkov}}]{deSouzaSilva07}%
  \BibitemOpen
  \bibfield  {author} {\bibinfo {author} {\bibfnamefont {C.~C.}\ \bibnamefont
  {de~Souza~Silva}}, \bibinfo {author} {\bibfnamefont {A.~V.}\ \bibnamefont
  {Silhanek}}, \bibinfo {author} {\bibfnamefont {J.}~\bibnamefont {Van~de
  Vondel}}, \bibinfo {author} {\bibfnamefont {W.}~\bibnamefont {Gillijns}},
  \bibinfo {author} {\bibfnamefont {V.}~\bibnamefont {Metlushko}}, \bibinfo
  {author} {\bibfnamefont {B.}~\bibnamefont {Ilic}}, \ and\ \bibinfo {author}
  {\bibfnamefont {V.~V.}\ \bibnamefont {Moshchalkov}},\ }\bibfield  {title}
  {\enquote {\bibinfo {title} {Dipole-induced vortex ratchets in
  superconducting films with arrays of micromagnets},}\ }\href {\doibase
  10.1103/PhysRevLett.98.117005} {\bibfield  {journal} {\bibinfo  {journal}
  {Phys. Rev. Lett.}\ }\textbf {\bibinfo {volume} {98}},\ \bibinfo {pages}
  {117005} (\bibinfo {year} {2007})}\BibitemShut {NoStop}%
\bibitem [{\citenamefont {Goldberg}\ \emph {et~al.}(2009)\citenamefont
  {Goldberg}, \citenamefont {Segev}, \citenamefont {Myasoedov}, \citenamefont
  {Gutman}, \citenamefont {Avraham}, \citenamefont {Rappaport}, \citenamefont
  {Zeldov}, \citenamefont {Tamegai}, \citenamefont {Hicks},\ and\ \citenamefont
  {Moler}}]{Goldberg09}%
  \BibitemOpen
  \bibfield  {author} {\bibinfo {author} {\bibfnamefont {S.}~\bibnamefont
  {Goldberg}}, \bibinfo {author} {\bibfnamefont {Y.}~\bibnamefont {Segev}},
  \bibinfo {author} {\bibfnamefont {Y.}~\bibnamefont {Myasoedov}}, \bibinfo
  {author} {\bibfnamefont {I.}~\bibnamefont {Gutman}}, \bibinfo {author}
  {\bibfnamefont {N.}~\bibnamefont {Avraham}}, \bibinfo {author} {\bibfnamefont
  {M.}~\bibnamefont {Rappaport}}, \bibinfo {author} {\bibfnamefont
  {E.}~\bibnamefont {Zeldov}}, \bibinfo {author} {\bibfnamefont
  {T.}~\bibnamefont {Tamegai}}, \bibinfo {author} {\bibfnamefont {C.~W.}\
  \bibnamefont {Hicks}}, \ and\ \bibinfo {author} {\bibfnamefont {K.~A.}\
  \bibnamefont {Moler}},\ }\bibfield  {title} {\enquote {\bibinfo {title} {Mott
  insulator phases and first-order melting in
  {Bi$_2$Sr$_2$CaCu$_2$O$_{8+\delta}$} crystals with periodic surface holes},}\
  }\href {\doibase 10.1103/PhysRevB.79.064523} {\bibfield  {journal} {\bibinfo
  {journal} {Phys. Rev. B}\ }\textbf {\bibinfo {volume} {79}},\ \bibinfo
  {pages} {064523} (\bibinfo {year} {2009})}\BibitemShut {NoStop}%
\bibitem [{\citenamefont {Kemmler}\ \emph {et~al.}(2009)\citenamefont
  {Kemmler}, \citenamefont {Bothner}, \citenamefont {Ilin}, \citenamefont
  {Siegel}, \citenamefont {Kleiner},\ and\ \citenamefont {Koelle}}]{Kemmler09}%
  \BibitemOpen
  \bibfield  {author} {\bibinfo {author} {\bibfnamefont {M.}~\bibnamefont
  {Kemmler}}, \bibinfo {author} {\bibfnamefont {D.}~\bibnamefont {Bothner}},
  \bibinfo {author} {\bibfnamefont {K.}~\bibnamefont {Ilin}}, \bibinfo {author}
  {\bibfnamefont {M.}~\bibnamefont {Siegel}}, \bibinfo {author} {\bibfnamefont
  {R.}~\bibnamefont {Kleiner}}, \ and\ \bibinfo {author} {\bibfnamefont
  {D.}~\bibnamefont {Koelle}},\ }\bibfield  {title} {\enquote {\bibinfo {title}
  {Suppression of dissipation in {Nb} thin films with triangular antidot arrays
  by random removal of pinning sites},}\ }\href {\doibase
  10.1103/PhysRevB.79.184509} {\bibfield  {journal} {\bibinfo  {journal} {Phys.
  Rev. B}\ }\textbf {\bibinfo {volume} {79}},\ \bibinfo {pages} {184509}
  (\bibinfo {year} {2009})}\BibitemShut {NoStop}%
\bibitem [{\citenamefont {Gutierrez}\ \emph {et~al.}(2009)\citenamefont
  {Gutierrez}, \citenamefont {Silhanek}, \citenamefont {Van~de Vondel},
  \citenamefont {Gillijns},\ and\ \citenamefont {Moshchalkov}}]{Gutierrez09}%
  \BibitemOpen
  \bibfield  {author} {\bibinfo {author} {\bibfnamefont {J.}~\bibnamefont
  {Gutierrez}}, \bibinfo {author} {\bibfnamefont {A.~V.}\ \bibnamefont
  {Silhanek}}, \bibinfo {author} {\bibfnamefont {J.}~\bibnamefont {Van~de
  Vondel}}, \bibinfo {author} {\bibfnamefont {W.}~\bibnamefont {Gillijns}}, \
  and\ \bibinfo {author} {\bibfnamefont {V.~V.}\ \bibnamefont {Moshchalkov}},\
  }\bibfield  {title} {\enquote {\bibinfo {title} {Transition from turbulent to
  nearly laminar vortex flow in superconductors with periodic pinning},}\
  }\href {\doibase 10.1103/PhysRevB.80.140514} {\bibfield  {journal} {\bibinfo
  {journal} {Phys. Rev. B}\ }\textbf {\bibinfo {volume} {80}},\ \bibinfo
  {pages} {140514} (\bibinfo {year} {2009})}\BibitemShut {NoStop}%
\bibitem [{\citenamefont {Lib\'al}\ \emph {et~al.}(2009)\citenamefont
  {Lib\'al}, \citenamefont {Reichhardt},\ and\ \citenamefont
  {Reichhardt}}]{Libal09}%
  \BibitemOpen
  \bibfield  {author} {\bibinfo {author} {\bibfnamefont {A.}~\bibnamefont
  {Lib\'al}}, \bibinfo {author} {\bibfnamefont {C.~J.~Olson}\ \bibnamefont
  {Reichhardt}}, \ and\ \bibinfo {author} {\bibfnamefont {C.}~\bibnamefont
  {Reichhardt}},\ }\bibfield  {title} {\enquote {\bibinfo {title} {Creating
  artificial ice states using vortices in nanostructured superconductors},}\
  }\href {\doibase 10.1103/PhysRevLett.102.237004} {\bibfield  {journal}
  {\bibinfo  {journal} {Phys. Rev. Lett.}\ }\textbf {\bibinfo {volume} {102}},\
  \bibinfo {pages} {237004} (\bibinfo {year} {2009})}\BibitemShut {NoStop}%
\bibitem [{\citenamefont {Hoffmann}\ \emph {et~al.}(2012)\citenamefont
  {Hoffmann}, \citenamefont {Prieto}, \citenamefont {Metlushko},\ and\
  \citenamefont {Schuller}}]{Hoffmann12}%
  \BibitemOpen
  \bibfield  {author} {\bibinfo {author} {\bibfnamefont {A.}~\bibnamefont
  {Hoffmann}}, \bibinfo {author} {\bibfnamefont {P.}~\bibnamefont {Prieto}},
  \bibinfo {author} {\bibfnamefont {V.}~\bibnamefont {Metlushko}}, \ and\
  \bibinfo {author} {\bibfnamefont {I.~K.}\ \bibnamefont {Schuller}},\
  }\bibfield  {title} {\enquote {\bibinfo {title} {Superconducting vortex
  pinning with magnetic dots: Does size and magnetic configuration matter?}}\
  }\href {\doibase 10.1007/s10948-012-1647-5} {\bibfield  {journal} {\bibinfo
  {journal} {J. Supercond. Novel Mag.}\ }\textbf {\bibinfo {volume} {25}},\
  \bibinfo {pages} {2187--2191} (\bibinfo {year} {2012})}\BibitemShut {NoStop}%
\bibitem [{\citenamefont {Swiecicki}\ \emph {et~al.}(2012)\citenamefont
  {Swiecicki}, \citenamefont {Ulysse}, \citenamefont {Wolf}, \citenamefont
  {Bernard}, \citenamefont {Bergeal}, \citenamefont {Briatico}, \citenamefont
  {Faini}, \citenamefont {Lesueur},\ and\ \citenamefont
  {Villegas}}]{Swiecicki12}%
  \BibitemOpen
  \bibfield  {author} {\bibinfo {author} {\bibfnamefont {I.}~\bibnamefont
  {Swiecicki}}, \bibinfo {author} {\bibfnamefont {C.}~\bibnamefont {Ulysse}},
  \bibinfo {author} {\bibfnamefont {T.}~\bibnamefont {Wolf}}, \bibinfo {author}
  {\bibfnamefont {R.}~\bibnamefont {Bernard}}, \bibinfo {author} {\bibfnamefont
  {N.}~\bibnamefont {Bergeal}}, \bibinfo {author} {\bibfnamefont
  {J.}~\bibnamefont {Briatico}}, \bibinfo {author} {\bibfnamefont
  {G.}~\bibnamefont {Faini}}, \bibinfo {author} {\bibfnamefont
  {J.}~\bibnamefont {Lesueur}}, \ and\ \bibinfo {author} {\bibfnamefont
  {Javier~E.}\ \bibnamefont {Villegas}},\ }\bibfield  {title} {\enquote
  {\bibinfo {title} {Strong field-matching effects in superconducting
  {YBa$_2$Cu$_3$O$_{7-\delta}$} films with vortex energy landscapes engineered
  via masked ion irradiation},}\ }\href {\doibase 10.1103/PhysRevB.85.224502}
  {\bibfield  {journal} {\bibinfo  {journal} {Phys. Rev. B}\ }\textbf {\bibinfo
  {volume} {85}},\ \bibinfo {pages} {224502} (\bibinfo {year}
  {2012})}\BibitemShut {NoStop}%
\bibitem [{\citenamefont {Latimer}\ \emph {et~al.}(2013)\citenamefont
  {Latimer}, \citenamefont {Berdiyorov}, \citenamefont {Xiao}, \citenamefont
  {Peeters},\ and\ \citenamefont {Kwok}}]{Latimer13}%
  \BibitemOpen
  \bibfield  {author} {\bibinfo {author} {\bibfnamefont {M.~L.}\ \bibnamefont
  {Latimer}}, \bibinfo {author} {\bibfnamefont {G.~R.}\ \bibnamefont
  {Berdiyorov}}, \bibinfo {author} {\bibfnamefont {Z.~L.}\ \bibnamefont
  {Xiao}}, \bibinfo {author} {\bibfnamefont {F.~M.}\ \bibnamefont {Peeters}}, \
  and\ \bibinfo {author} {\bibfnamefont {W.~K.}\ \bibnamefont {Kwok}},\
  }\bibfield  {title} {\enquote {\bibinfo {title} {Realization of artificial
  ice systems for magnetic vortices in a superconducting {MoGe} thin film with
  patterned nanostructures},}\ }\href {\doibase 10.1103/PhysRevLett.111.067001}
  {\bibfield  {journal} {\bibinfo  {journal} {Phys. Rev. Lett.}\ }\textbf
  {\bibinfo {volume} {111}},\ \bibinfo {pages} {067001} (\bibinfo {year}
  {2013})}\BibitemShut {NoStop}%
\bibitem [{\citenamefont {Wang}\ \emph {et~al.}(2013)\citenamefont {Wang},
  \citenamefont {Latimer}, \citenamefont {Xiao}, \citenamefont {Divan},
  \citenamefont {Ocola}, \citenamefont {Crabtree},\ and\ \citenamefont
  {Kwok}}]{Wang13}%
  \BibitemOpen
  \bibfield  {author} {\bibinfo {author} {\bibfnamefont {Y.~L.}\ \bibnamefont
  {Wang}}, \bibinfo {author} {\bibfnamefont {M.~L.}\ \bibnamefont {Latimer}},
  \bibinfo {author} {\bibfnamefont {Z.~L.}\ \bibnamefont {Xiao}}, \bibinfo
  {author} {\bibfnamefont {R.}~\bibnamefont {Divan}}, \bibinfo {author}
  {\bibfnamefont {L.~E.}\ \bibnamefont {Ocola}}, \bibinfo {author}
  {\bibfnamefont {G.~W.}\ \bibnamefont {Crabtree}}, \ and\ \bibinfo {author}
  {\bibfnamefont {W.~K.}\ \bibnamefont {Kwok}},\ }\bibfield  {title} {\enquote
  {\bibinfo {title} {Enhancing the critical current of a superconducting film
  in a wide range of magnetic fields with a conformal array of nanoscale
  holes},}\ }\href {\doibase 10.1103/PhysRevB.87.220501} {\bibfield  {journal}
  {\bibinfo  {journal} {Phys. Rev. B}\ }\textbf {\bibinfo {volume} {87}},\
  \bibinfo {pages} {220501} (\bibinfo {year} {2013})}\BibitemShut {NoStop}%
\bibitem [{\citenamefont {Ray}\ \emph {et~al.}(2014)\citenamefont {Ray},
  \citenamefont {Reichhardt},\ and\ \citenamefont {Reichhardt}}]{Ray14b}%
  \BibitemOpen
  \bibfield  {author} {\bibinfo {author} {\bibfnamefont {D.}~\bibnamefont
  {Ray}}, \bibinfo {author} {\bibfnamefont {C.}~\bibnamefont {Reichhardt}}, \
  and\ \bibinfo {author} {\bibfnamefont {C.~J.~Olson}\ \bibnamefont
  {Reichhardt}},\ }\bibfield  {title} {\enquote {\bibinfo {title} {Pinning,
  ordering, and dynamics of vortices in conformal crystal and gradient pinning
  arrays},}\ }\href {\doibase 10.1103/PhysRevB.90.094502} {\bibfield  {journal}
  {\bibinfo  {journal} {Phys. Rev. B}\ }\textbf {\bibinfo {volume} {90}},\
  \bibinfo {pages} {094502} (\bibinfo {year} {2014})}\BibitemShut {NoStop}%
\bibitem [{\citenamefont {Trastoy}\ \emph {et~al.}(2014)\citenamefont
  {Trastoy}, \citenamefont {Malnou}, \citenamefont {Ulysse}, \citenamefont
  {Bernard}, \citenamefont {Bergeal}, \citenamefont {Faini}, \citenamefont
  {Lesueur}, \citenamefont {Briatico},\ and\ \citenamefont
  {Villegas}}]{Trastoy14}%
  \BibitemOpen
  \bibfield  {author} {\bibinfo {author} {\bibfnamefont {J.}~\bibnamefont
  {Trastoy}}, \bibinfo {author} {\bibfnamefont {M.}~\bibnamefont {Malnou}},
  \bibinfo {author} {\bibfnamefont {C.}~\bibnamefont {Ulysse}}, \bibinfo
  {author} {\bibfnamefont {R.}~\bibnamefont {Bernard}}, \bibinfo {author}
  {\bibfnamefont {N.}~\bibnamefont {Bergeal}}, \bibinfo {author} {\bibfnamefont
  {G.}~\bibnamefont {Faini}}, \bibinfo {author} {\bibfnamefont
  {J.}~\bibnamefont {Lesueur}}, \bibinfo {author} {\bibfnamefont
  {J.}~\bibnamefont {Briatico}}, \ and\ \bibinfo {author} {\bibfnamefont
  {J.~E.}\ \bibnamefont {Villegas}},\ }\bibfield  {title} {\enquote {\bibinfo
  {title} {Freezing and thawing of artificial ice by thermal switching of
  geometric frustration in magnetic flux lattices},}\ }\href {\doibase
  10.1038/NNANO.2014.158} {\bibfield  {journal} {\bibinfo  {journal} {Nature
  Nanotechnol.}\ }\textbf {\bibinfo {volume} {9}},\ \bibinfo {pages} {710--715}
  (\bibinfo {year} {2014})}\BibitemShut {NoStop}%
\bibitem [{\citenamefont {Sadovskyy}\ \emph {et~al.}(2017)\citenamefont
  {Sadovskyy}, \citenamefont {Wang}, \citenamefont {Xiao}, \citenamefont
  {Kwok},\ and\ \citenamefont {Glatz}}]{Sadovskyy17}%
  \BibitemOpen
  \bibfield  {author} {\bibinfo {author} {\bibfnamefont {I.~A.}\ \bibnamefont
  {Sadovskyy}}, \bibinfo {author} {\bibfnamefont {Y.~L.}\ \bibnamefont {Wang}},
  \bibinfo {author} {\bibfnamefont {Z.-L.}\ \bibnamefont {Xiao}}, \bibinfo
  {author} {\bibfnamefont {W.-K.}\ \bibnamefont {Kwok}}, \ and\ \bibinfo
  {author} {\bibfnamefont {A.}~\bibnamefont {Glatz}},\ }\bibfield  {title}
  {\enquote {\bibinfo {title} {Effect of hexagonal patterned arrays and defect
  geometry on the critical current of superconducting films},}\ }\href
  {\doibase 10.1103/PhysRevB.95.075303} {\bibfield  {journal} {\bibinfo
  {journal} {Phys. Rev. B}\ }\textbf {\bibinfo {volume} {95}},\ \bibinfo
  {pages} {075303} (\bibinfo {year} {2017})}\BibitemShut {NoStop}%
\bibitem [{\citenamefont {Zechner}\ \emph {et~al.}(2017)\citenamefont
  {Zechner}, \citenamefont {Jausner}, \citenamefont {Haag}, \citenamefont
  {Lang}, \citenamefont {Dosmailov}, \citenamefont {Bodea},\ and\ \citenamefont
  {Pedarnig}}]{Zechner17}%
  \BibitemOpen
  \bibfield  {author} {\bibinfo {author} {\bibfnamefont {G.}~\bibnamefont
  {Zechner}}, \bibinfo {author} {\bibfnamefont {F.}~\bibnamefont {Jausner}},
  \bibinfo {author} {\bibfnamefont {L.~T.}\ \bibnamefont {Haag}}, \bibinfo
  {author} {\bibfnamefont {W.}~\bibnamefont {Lang}}, \bibinfo {author}
  {\bibfnamefont {M.}~\bibnamefont {Dosmailov}}, \bibinfo {author}
  {\bibfnamefont {M.~A.}\ \bibnamefont {Bodea}}, \ and\ \bibinfo {author}
  {\bibfnamefont {J.~D.}\ \bibnamefont {Pedarnig}},\ }\bibfield  {title}
  {\enquote {\bibinfo {title} {Hysteretic vortex-matching effects in
  high-${T}_{c}$ superconductors with nanoscale periodic pinning landscapes
  fabricated by {H}e ion-beam projection},}\ }\href {\doibase
  10.1103/PhysRevApplied.8.014021} {\bibfield  {journal} {\bibinfo  {journal}
  {Phys. Rev. Applied}\ }\textbf {\bibinfo {volume} {8}},\ \bibinfo {pages}
  {014021} (\bibinfo {year} {2017})}\BibitemShut {NoStop}%
\bibitem [{\citenamefont {Xue}\ \emph {et~al.}(2018)\citenamefont {Xue},
  \citenamefont {Ge}, \citenamefont {He}, \citenamefont {Zharinov},
  \citenamefont {Moshchalkov}, \citenamefont {Zhou}, \citenamefont {Silhanek},\
  and\ \citenamefont {Van~de Vondel}}]{Xue18}%
  \BibitemOpen
  \bibfield  {author} {\bibinfo {author} {\bibfnamefont {C.}~\bibnamefont
  {Xue}}, \bibinfo {author} {\bibfnamefont {J.-Y.}\ \bibnamefont {Ge}},
  \bibinfo {author} {\bibfnamefont {A.}~\bibnamefont {He}}, \bibinfo {author}
  {\bibfnamefont {V.~S.}\ \bibnamefont {Zharinov}}, \bibinfo {author}
  {\bibfnamefont {V.~V.}\ \bibnamefont {Moshchalkov}}, \bibinfo {author}
  {\bibfnamefont {Y.~H.}\ \bibnamefont {Zhou}}, \bibinfo {author}
  {\bibfnamefont {A.~V.}\ \bibnamefont {Silhanek}}, \ and\ \bibinfo {author}
  {\bibfnamefont {J.}~\bibnamefont {Van~de Vondel}},\ }\bibfield  {title}
  {\enquote {\bibinfo {title} {Tunable artificial vortex ice in nanostructured
  superconductors with a frustrated kagome lattice of paired antidots},}\
  }\href {\doibase 10.1103/PhysRevB.97.134506} {\bibfield  {journal} {\bibinfo
  {journal} {Phys. Rev. B}\ }\textbf {\bibinfo {volume} {97}},\ \bibinfo
  {pages} {134506} (\bibinfo {year} {2018})}\BibitemShut {NoStop}%
\bibitem [{\citenamefont {Ge}\ \emph {et~al.}(2018)\citenamefont {Ge},
  \citenamefont {Gladilin}, \citenamefont {Tempere}, \citenamefont {Devreese},\
  and\ \citenamefont {Moshchalkov}}]{Ge18}%
  \BibitemOpen
  \bibfield  {author} {\bibinfo {author} {\bibfnamefont {J.-Y.}\ \bibnamefont
  {Ge}}, \bibinfo {author} {\bibfnamefont {V.~N.}\ \bibnamefont {Gladilin}},
  \bibinfo {author} {\bibfnamefont {J.}~\bibnamefont {Tempere}}, \bibinfo
  {author} {\bibfnamefont {J.~T.}\ \bibnamefont {Devreese}}, \ and\ \bibinfo
  {author} {\bibfnamefont {V.~V.}\ \bibnamefont {Moshchalkov}},\ }\bibfield
  {title} {\enquote {\bibinfo {title} {Tunable and switchable magnetic dipole
  patterns in nanostructured superconductors},}\ }\href {\doibase
  10.1038/s41467-018-05045-3} {\bibfield  {journal} {\bibinfo  {journal}
  {Nature Commun.}\ }\textbf {\bibinfo {volume} {9}},\ \bibinfo {pages} {2576}
  (\bibinfo {year} {2018})}\BibitemShut {NoStop}%
\bibitem [{\citenamefont {Wang}\ \emph
  {et~al.}(2018{\natexlab{b}})\citenamefont {Wang}, \citenamefont {Ma},
  \citenamefont {Xu}, \citenamefont {Xiao}, \citenamefont {Snezhko},
  \citenamefont {Divan}, \citenamefont {Ocola}, \citenamefont {Pearson},
  \citenamefont {J{\' a}nko},\ and\ \citenamefont {Kwok}}]{Wang18a}%
  \BibitemOpen
  \bibfield  {author} {\bibinfo {author} {\bibfnamefont {Y.-L.}\ \bibnamefont
  {Wang}}, \bibinfo {author} {\bibfnamefont {X.}~\bibnamefont {Ma}}, \bibinfo
  {author} {\bibfnamefont {J.}~\bibnamefont {Xu}}, \bibinfo {author}
  {\bibfnamefont {Z.-L.}\ \bibnamefont {Xiao}}, \bibinfo {author}
  {\bibfnamefont {A.}~\bibnamefont {Snezhko}}, \bibinfo {author} {\bibfnamefont
  {R.}~\bibnamefont {Divan}}, \bibinfo {author} {\bibfnamefont {L.~E.}\
  \bibnamefont {Ocola}}, \bibinfo {author} {\bibfnamefont {J.~E.}\ \bibnamefont
  {Pearson}}, \bibinfo {author} {\bibfnamefont {B.}~\bibnamefont {J{\' a}nko}},
  \ and\ \bibinfo {author} {\bibfnamefont {W.-K.}\ \bibnamefont {Kwok}},\
  }\bibfield  {title} {\enquote {\bibinfo {title} {Switchable geometric
  frustration in an artificial-spin-ice-superconductor heterosystem},}\ }\href
  {\doibase 10.1038/s41565-018-0162-7} {\bibfield  {journal} {\bibinfo
  {journal} {Nature Nanotechnol.}\ }\textbf {\bibinfo {volume} {13}},\ \bibinfo
  {pages} {560} (\bibinfo {year} {2018}{\natexlab{b}})}\BibitemShut {NoStop}%
\bibitem [{\citenamefont {Georgiev}(2006)}]{Georgiev06}%
  \BibitemOpen
  \bibfield  {author} {\bibinfo {author} {\bibfnamefont {L.~S.}\ \bibnamefont
  {Georgiev}},\ }\bibfield  {title} {\enquote {\bibinfo {title} {Topologically
  protected gates for quantum computation with non-{A}belian anyons in the
  {P}faffian quantum {H}all state},}\ }\href {\doibase
  10.1103/PhysRevB.74.235112} {\bibfield  {journal} {\bibinfo  {journal} {Phys.
  Rev. B}\ }\textbf {\bibinfo {volume} {74}},\ \bibinfo {pages} {235112}
  (\bibinfo {year} {2006})}\BibitemShut {NoStop}%
\bibitem [{\citenamefont {Georgiev}(2008)}]{Georgiev08}%
  \BibitemOpen
  \bibfield  {author} {\bibinfo {author} {\bibfnamefont {L.~S.}\ \bibnamefont
  {Georgiev}},\ }\bibfield  {title} {\enquote {\bibinfo {title} {Towards a
  universal set of topologically protected gates for quantum computation with
  {P}faffian qubits},}\ }\href {\doibase 10.1016/j.nuclphysb.2007.07.016}
  {\bibfield  {journal} {\bibinfo  {journal} {Nucl. Phys. B}\ }\textbf
  {\bibinfo {volume} {789}},\ \bibinfo {pages} {552--590} (\bibinfo {year}
  {2008})}\BibitemShut {NoStop}%
\bibitem [{\citenamefont {Yang}\ \emph {et~al.}(2016)\citenamefont {Yang},
  \citenamefont {Stano}, \citenamefont {Klinovaja},\ and\ \citenamefont
  {Loss}}]{Yang16}%
  \BibitemOpen
  \bibfield  {author} {\bibinfo {author} {\bibfnamefont {G.}~\bibnamefont
  {Yang}}, \bibinfo {author} {\bibfnamefont {P.}~\bibnamefont {Stano}},
  \bibinfo {author} {\bibfnamefont {J.}~\bibnamefont {Klinovaja}}, \ and\
  \bibinfo {author} {\bibfnamefont {D.}~\bibnamefont {Loss}},\ }\bibfield
  {title} {\enquote {\bibinfo {title} {Majorana bound states in magnetic
  skyrmions},}\ }\href {\doibase 10.1103/PhysRevB.93.224505} {\bibfield
  {journal} {\bibinfo  {journal} {Phys. Rev. B}\ }\textbf {\bibinfo {volume}
  {93}},\ \bibinfo {pages} {224505} (\bibinfo {year} {2016})}\BibitemShut
  {NoStop}%
\bibitem [{\citenamefont {G\"ung\"ord\"u}\ \emph {et~al.}(2018)\citenamefont
  {G\"ung\"ord\"u}, \citenamefont {Sandhoefner},\ and\ \citenamefont
  {Kovalev}}]{Gungordu18}%
  \BibitemOpen
  \bibfield  {author} {\bibinfo {author} {\bibfnamefont {U.}~\bibnamefont
  {G\"ung\"ord\"u}}, \bibinfo {author} {\bibfnamefont {S.}~\bibnamefont
  {Sandhoefner}}, \ and\ \bibinfo {author} {\bibfnamefont {A.~A.}\ \bibnamefont
  {Kovalev}},\ }\bibfield  {title} {\enquote {\bibinfo {title} {Stabilization
  and control of {M}ajorana bound states with elongated skyrmions},}\ }\href
  {\doibase 10.1103/PhysRevB.97.115136} {\bibfield  {journal} {\bibinfo
  {journal} {Phys. Rev. B}\ }\textbf {\bibinfo {volume} {97}},\ \bibinfo
  {pages} {115136} (\bibinfo {year} {2018})}\BibitemShut {NoStop}%
\bibitem [{\citenamefont {Rex}\ \emph {et~al.}(2019)\citenamefont {Rex},
  \citenamefont {Gornyi},\ and\ \citenamefont {Mirlin}}]{Rex19}%
  \BibitemOpen
  \bibfield  {author} {\bibinfo {author} {\bibfnamefont {S.}~\bibnamefont
  {Rex}}, \bibinfo {author} {\bibfnamefont {I.~V.}\ \bibnamefont {Gornyi}}, \
  and\ \bibinfo {author} {\bibfnamefont {A.~D.}\ \bibnamefont {Mirlin}},\
  }\href@noop {} {\enquote {\bibinfo {title} {Majorana bound states in magnetic
  skyrmions imposed onto a superconductor},}\ } (\bibinfo {year} {2019}),\
  \Eprint {http://arxiv.org/abs/1904.04177} {arXiv:1904.04177} \BibitemShut
  {NoStop}%
\bibitem [{\citenamefont {Reichhardt}\ \emph {et~al.}(2018)\citenamefont
  {Reichhardt}, \citenamefont {Ray},\ and\ \citenamefont
  {Reichhardt}}]{Reichhardt18}%
  \BibitemOpen
  \bibfield  {author} {\bibinfo {author} {\bibfnamefont {C.}~\bibnamefont
  {Reichhardt}}, \bibinfo {author} {\bibfnamefont {D.}~\bibnamefont {Ray}}, \
  and\ \bibinfo {author} {\bibfnamefont {C.~J.~O.}\ \bibnamefont
  {Reichhardt}},\ }\bibfield  {title} {\enquote {\bibinfo {title}
  {Nonequilibrium phases and segregation for skyrmions on periodic pinning
  arrays},}\ }\href {\doibase 10.1103/PhysRevB.98.134418} {\bibfield  {journal}
  {\bibinfo  {journal} {Phys. Rev. B}\ }\textbf {\bibinfo {volume} {98}},\
  \bibinfo {pages} {134418} (\bibinfo {year} {2018})}\BibitemShut {NoStop}%
\bibitem [{\citenamefont {Hanneken}\ \emph {et~al.}(2016)\citenamefont
  {Hanneken}, \citenamefont {Kubetzka}, \citenamefont {von Bergmann},\ and\
  \citenamefont {Wiesendanger}}]{Hanneken16}%
  \BibitemOpen
  \bibfield  {author} {\bibinfo {author} {\bibfnamefont {C.}~\bibnamefont
  {Hanneken}}, \bibinfo {author} {\bibfnamefont {A.}~\bibnamefont {Kubetzka}},
  \bibinfo {author} {\bibfnamefont {K.}~\bibnamefont {von Bergmann}}, \ and\
  \bibinfo {author} {\bibfnamefont {R.}~\bibnamefont {Wiesendanger}},\
  }\bibfield  {title} {\enquote {\bibinfo {title} {Pinning and movement of
  individual nanoscale magnetic skyrmions via defects},}\ }\href {\doibase
  10.1088/1367-2630/18/5/055009} {\bibfield  {journal} {\bibinfo  {journal}
  {New J. Phys.}\ }\textbf {\bibinfo {volume} {18}},\ \bibinfo {pages} {055009}
  (\bibinfo {year} {2016})}\BibitemShut {NoStop}%
\bibitem [{\citenamefont {Menezes}\ \emph {et~al.}(2019)\citenamefont
  {Menezes}, \citenamefont {Neto}, \citenamefont {Silva},\ and\ \citenamefont
  {Milo\ifmmode \check{s}\else \v{s}\fi{}evi\ifmmode~\acute{c}\else
  \'{c}\fi{}}}]{Menezes19}%
  \BibitemOpen
  \bibfield  {author} {\bibinfo {author} {\bibfnamefont {R.~M.}\ \bibnamefont
  {Menezes}}, \bibinfo {author} {\bibfnamefont {J.~F.~S.}\ \bibnamefont
  {Neto}}, \bibinfo {author} {\bibfnamefont {C.~C. de~Souza}\ \bibnamefont
  {Silva}}, \ and\ \bibinfo {author} {\bibfnamefont {M.~V.}\ \bibnamefont
  {Milo\ifmmode \check{s}\else \v{s}\fi{}evi\ifmmode~\acute{c}\else
  \'{c}\fi{}}},\ }\bibfield  {title} {\enquote {\bibinfo {title} {Manipulation
  of magnetic skyrmions by superconducting vortices in
  ferromagnet-superconductor heterostructures},}\ }\href {\doibase
  10.1103/PhysRevB.100.014431} {\bibfield  {journal} {\bibinfo  {journal}
  {Phys. Rev. B}\ }\textbf {\bibinfo {volume} {100}},\ \bibinfo {pages}
  {014431} (\bibinfo {year} {2019})}\BibitemShut {NoStop}%
\bibitem [{\citenamefont {Cheng}\ \emph {et~al.}(2009)\citenamefont {Cheng},
  \citenamefont {Lutchyn}, \citenamefont {Galitski},\ and\ \citenamefont
  {Das~Sarma}}]{Cheng09}%
  \BibitemOpen
  \bibfield  {author} {\bibinfo {author} {\bibfnamefont {M.}~\bibnamefont
  {Cheng}}, \bibinfo {author} {\bibfnamefont {R.~M.}\ \bibnamefont {Lutchyn}},
  \bibinfo {author} {\bibfnamefont {V.}~\bibnamefont {Galitski}}, \ and\
  \bibinfo {author} {\bibfnamefont {S.}~\bibnamefont {Das~Sarma}},\ }\bibfield
  {title} {\enquote {\bibinfo {title} {Splitting of {M}ajorana-{F}ermion modes
  due to intervortex tunneling in a ${p}_{x}+i{p}_{y}$ superconductor},}\
  }\href {\doibase 10.1103/PhysRevLett.103.107001} {\bibfield  {journal}
  {\bibinfo  {journal} {Phys. Rev. Lett.}\ }\textbf {\bibinfo {volume} {103}},\
  \bibinfo {pages} {107001} (\bibinfo {year} {2009})}\BibitemShut {NoStop}%
\bibitem [{\citenamefont {Cheng}\ \emph {et~al.}(2010)\citenamefont {Cheng},
  \citenamefont {Lutchyn}, \citenamefont {Galitski},\ and\ \citenamefont
  {Das~Sarma}}]{Cheng10}%
  \BibitemOpen
  \bibfield  {author} {\bibinfo {author} {\bibfnamefont {M.}~\bibnamefont
  {Cheng}}, \bibinfo {author} {\bibfnamefont {R.~M.}\ \bibnamefont {Lutchyn}},
  \bibinfo {author} {\bibfnamefont {V.}~\bibnamefont {Galitski}}, \ and\
  \bibinfo {author} {\bibfnamefont {S.}~\bibnamefont {Das~Sarma}},\ }\bibfield
  {title} {\enquote {\bibinfo {title} {Tunneling of anyonic {M}ajorana
  excitations in topological superconductors},}\ }\href {\doibase
  10.1103/PhysRevB.82.094504} {\bibfield  {journal} {\bibinfo  {journal} {Phys.
  Rev. B}\ }\textbf {\bibinfo {volume} {82}},\ \bibinfo {pages} {094504}
  (\bibinfo {year} {2010})}\BibitemShut {NoStop}%
\bibitem [{\citenamefont {Hastings}\ \emph {et~al.}(2003)\citenamefont
  {Hastings}, \citenamefont {Reichhardt},\ and\ \citenamefont
  {Reichhardt}}]{Hastings03}%
  \BibitemOpen
  \bibfield  {author} {\bibinfo {author} {\bibfnamefont {M.~B.}\ \bibnamefont
  {Hastings}}, \bibinfo {author} {\bibfnamefont {C.~J.~Olson}\ \bibnamefont
  {Reichhardt}}, \ and\ \bibinfo {author} {\bibfnamefont {C.}~\bibnamefont
  {Reichhardt}},\ }\bibfield  {title} {\enquote {\bibinfo {title} {Ratchet
  cellular automata},}\ }\href {\doibase 10.1103/PhysRevLett.90.247004}
  {\bibfield  {journal} {\bibinfo  {journal} {Phys. Rev. Lett.}\ }\textbf
  {\bibinfo {volume} {90}},\ \bibinfo {pages} {247004} (\bibinfo {year}
  {2003})}\BibitemShut {NoStop}%
\bibitem [{\citenamefont {Milosevic}\ \emph {et~al.}(2007)\citenamefont
  {Milosevic}, \citenamefont {Berdiyorov},\ and\ \citenamefont
  {Peeters}}]{Milosevic07}%
  \BibitemOpen
  \bibfield  {author} {\bibinfo {author} {\bibfnamefont {M.~V.}\ \bibnamefont
  {Milosevic}}, \bibinfo {author} {\bibfnamefont {G.~R.}\ \bibnamefont
  {Berdiyorov}}, \ and\ \bibinfo {author} {\bibfnamefont {F.~M.}\ \bibnamefont
  {Peeters}},\ }\bibfield  {title} {\enquote {\bibinfo {title} {Fluxonic
  cellular automata},}\ }\href {\doibase 10.1063/1.2813047} {\bibfield
  {journal} {\bibinfo  {journal} {Appl. Phys. Lett.}\ }\textbf {\bibinfo
  {volume} {91}},\ \bibinfo {pages} {212501} (\bibinfo {year}
  {2007})}\BibitemShut {NoStop}%
\end{thebibliography}%
\end{document}